\documentclass[12pt]{article}
\newcommand{\micron}{$\mu$m}
\usepackage{newtxtext,newtxmath}
\usepackage{multirow} 

\usepackage{graphicx}
\usepackage{afterpage}

\usepackage[letterpaper,margin=1in]{geometry}

\renewenvironment{abstract}
	{\quotation}
	{\endquotation}

\date{}

\makeatletter
\renewcommand{\fnum@figure}{\textbf{Figure \thefigure}}
\renewcommand{\fnum@table}{\textbf{Table \thetable}}
\makeatother

\usepackage{scicite}

\usepackage{url}

\def\scititle{
	Resolving the black hole sphere of influence in a hyper-luminous obscured quasar at
redshift 4.6
}
\title{\bfseries \boldmath \scititle}

\author{
	Mai Liao$^{1,2,3\ast}$,
	Roberto J.Assef$^{2}$,
	Chao-Wei Tsai$^{1,4,5}$,
    Manuel Aravena$^{2}$,\and
    Rom\'an Fern\'andez Aranda$^{6}$,
    Andrew W. Blain$^{7}$, 
    Tanio D\'iaz-Santos$^{8}$,\and
    Peter Eisenhardt$^{9}$,
    Jorge Gonz\'alez-L\'opez$^{10,11}$,
    Hyunsung D. Jun$^{12,13}$,\and
    Xiaofeng Li$^{14}$,
    Guodong Li$^{15}$,
    Lee R.Martin$^{7}$, Ana Posses$^{16}$,\and
    Devika Shobhana$^{2}$, Manuel Solimano$^{6}$,
    Daniel Stern$^{9}$,\and
    Andrey Vayner$^{17,18}$,
    Jingwen Wu$^{1,5}$, Dejene Zewdie$^{19}$ \and
	\small$^{1}$National Astronomical Observatories, Chinese Academy of Sciences, 20A Datun Road, Beijing,100101, China.\and
    \small$^{2}$Instituto de Estudios Astrof\'isicos Facultad de Ingenier\'ia y Ciencias, Universidad Diego Portales, Av. Ej\'ercito \and \small Libertador 441, Santiago, 8370191, Chile. \and
    \small$^{3}$Chinese Academy of Sciences South America Center for Astronomy, Chinese Academy of Sciences, 20A Datun \and 
    \small Road,  Beijing, 100101, China.\and
    \small$^{4}$Institute for Frontiers in Astronomy and Astrophysics, Beijing Normal University, Beijing, 102206, China.\and
    \small$^{5}$University of Chinese Academy of Sciences, Beijing, 100049, China.\and
    \small$^{6}$Centro de Astrobiología (CAB), INTA-CSIC, Carretera de Ajalvir km 4, Torrejón de Ardoz 28850, Madrid, Spain. \and
    \small$^{7}$Physics \& Astronomy, University of Leicester, Leicester, LE1 7RH, UK.\and
    \small$^{8}$ Institute of Astrophysics, Foundation for Research and Technology-Hellas (FORTH), Heraklion, 70013, Greece. \and
    \small$^{9}$Jet Propulsion Laboratory, California Institute of Technology, Pasadena, CA 91109, USA.\and
    \small$^{10}$Instituto de Astrof\'isica, Facultad de F\'isica, Pontificia Universidad Cat\'olica de Chile, Av. Vicuña Mackenna 4860, \and
     \small782-0436 Macul, Santiago, Chile.\and
    \small$^{11}$ Las Campanas Observatory, Carnegie Institution for Science, Ra\'ul Bitr\'an 1200, La Serena, Chile. \and
    \small$^{12}$Department of Physics, Northwestern College, 101 7th St SW, Orange
City, IA 51041, USA.\and
    \small$^{13}$School of Physics, Korea Institute for Advanced Study, 85 Hoegiro,
Dongdaemun-gu, Seoul, 02455,\and \small Republic of Korea.\and
    \small$^{14}$School of Computer Science and Information Engineering, Changzhou 
Institute of Technology, Changzhou,\and \small 213002, China.\and
    \small$^{15}$Kavli Institute for Astronomy and Astrophysics, Peking University,
Beijing, 100871, China.\and
\small$^{16}$ Department of Physics and Astronomy and George P. and Cynthia Woods Mitchell Institute for Fundamental \and
\small Physics and Astronomy, Texas A\&M University, 4242 TAMU, College Station, TX 77843-4242, USA \and
\small$^{17}$IPAC, California Institute of Technology, 1200 E. California Boulevard, Pasadena, 91125, CA, USA. \and
\small$^{18}$Florida Gulf Coast University, 10501 FGCU Blvd. South, Fort Myers, 33965, FL, USA. \and
\small$^{19}$Centre for Space Research, North-West University, Potchefstroom, 2520, South Africa.\and
\small$^\ast$Corresponding author. Email: mai.liao@mail.udp.cl\and
}

\begin{document} 

\maketitle

\begin{abstract} \bfseries \boldmath
Supermassive black holes (SMBHs) imprint gravitational signatures on the matter within their sphere of influence (SoI). Nuclear gas dynamics can hence be used to accurately measure the mass of an SMBH, yet such measurements remain elusive in the early Universe. We report the first dynamical measurement of an SMBH mass at  $z >$ 2, based on high spatial resolution observations of the [C{~\sc ii}] 157.7$\rm \mu $m and CO (12-11) 216.93$\rm \mu $m emission lines that resolve the SoI in an obscured quasar at $z$ = 4.6. The radial profile of the velocity dispersion reveals a clear Keplerian rise, 
requiring the presence of an approximately $6$ $\times$ $\rm 10^{9}~\rm M_{\odot}$ SMBH. We propose that heavily obscured quasars allow tracers like [C{~\sc ii}] to survive in the inner regions, and may be ideal targets for increasing dynamical SMBH mass estimates in the early Universe.
\end{abstract}

\noindent
\section*{Introduction}
SMBH mass measurements are crucial for understanding the growth and evolution of galaxies. The most direct way to measure the mass of an SMBH is by observing its gravitational effects over nearby gas and stars. In active galactic nuclei (AGN), we can observe the emission of gas close to the SMBH that is typically moving at $>$ 1000 km$~\rm s^{-1}$ due to the SMBH's gravitational pull in the so-called broad-line region (BLR). The BLR location depends on the brightness of the SMBH's accretion disk, but is commonly between tens to thousands of light days from it. Recent technological advances have enabled directly resolving the BLR using near-infrared (NIR) interferometry in a limited number of objects \cite{2018Natur.563..657G,2020A&A...643A.154G,2021A&A...654A..85G,2024Natur.627..281A,2024A&A...684A.167G}, but traditionally this has been done using a technique called reverberation mapping (RM) \cite{2000ApJ...533..631K,2004ApJ...613..682P}. A key result of RM studies has been the identification of a direct relation between the accretion disk luminosity and the size of the BLR \cite{2013ApJ...767..149B}, which allow SMBH masses in AGN across the Universe to be estimated from a single spectrum. These estimates are referred to as "single-epoch" black hole mass ($M_{\rm BH}$) estimates \cite{2011ApJS..194...45S}. However, the critical assumption that these empirical estimates are universally applicable has been put into question by a number of studies \cite{2023ApJ...956..127M}. In inactive galaxies, on the other hand, the most direct method to measure the mass of the SMBH is through the motion of the stars and the gas within the region of the galaxy where dynamics are dominated by the SMBH's gravity rather than by the galaxy's baryonic or dark matter components. This region is called the sphere of influence (SoI, $R_{\rm SoI} = G$$M_{\rm BH}/\sigma_{\star}^2$) and it is typically between a few and hundreds of parsecs depending on the relative masses of the SMBH and the galaxy.
Recent NIR interferometric observations have attempted to estimate the SMBH mass using dynamics within the BLR in local AGN as well as in a luminous AGN at $z \sim 2$, which can then be compared to estimates based on the single-epoch method \cite{2024Natur.627..281A,2024A&A...684A.167G}. 
These studies have found that while in general the agreement is good, the BLR radius ($R_{\rm BLR}$) for the hydrogen emission could be smaller than expected from the canonical $R-L_{\rm 5100 \AA}$ relation at high luminosities, implying that single-epoch SMBH mass measurements in such objects could be systematically overestimated \cite{2024A&A...684A.167G}.

Recently, a few studies have attempted these dynamical measurements in distant luminous AGN, called quasars, at the early Universe ($z >$ 6) by using the typical brightest far-infrared (FIR) emission line, [C{~\sc ii}], observed at the high spatial resolutions (200--400 pc) accessible by ALMA. However, none were able to find an unequivocal signal for the SMBH, which let them conclude the SoI was smaller than expected due to substantial gas content in the central resolution
element \cite{2019ApJ...874L..30V,2022ApJ...927...21W}. Alternatively, the [C{~\sc ii}] gas near the SMBH could be over-ionized \cite{2015A&A...580A...5L, 2023ApJ...956..127M}, inhibiting the line emission in the proximity of the SMBH.
These null results imply that the ideal object in which to attempt these measurements at high redshift would be a very luminous, highly obscured quasar, with an expected large SMBH mass. The large luminosity would increase the signal to noise ratio (SNR) of the emission line, the large obscuration would make over-ionization of carbon atoms less likely around the nucleus, and the large SMBH would make the SoI more spatially extended. Here we present observations high spatial resolution observations of [C~{\sc ii}], as well as the dense-gas tracer CO (12-11), for such an object.

\section*{Results}
\subsection*{High-resolution observations of WISE J224607.6--052634.9}
WISE J224607.6--052634.9 (W2246--0526) is one of the most luminous obscured quasar currently known \cite{2015ApJ...805...90T}, with $L_{\rm bol} = $ 3.6 $\times~10^{14}$ $\rm L_{\odot}$ and $\sim$15 mag of dust obscuration at $V$-band. The galaxy is at a redshift of $z = $ 4.6 \cite{2016ApJ...816L...6D, 2018ApJ...868...15T,2018Sci...362.1034D}, when the age of the Universe was about 1.3 billion years. 
W2246--0526 is in a multiple merger system with dusty streams connecting three companions to the central galaxy \cite{2018ApJ...868...15T}, and its interstellar medium (ISM) is globally dominated by turbulence, possibly with a weak rotational component \cite{2016ApJ...816L...6D}.
Its black hole mass has been estimated to be 4.0$^{+6.0}_{-2.4}$ $\times~10^{9}~\rm M_{\odot}$ by \cite{2018ApJ...868...15T} based on the single-epoch method applied to the width of the Mg II 2800\AA\ emission line (see Section \ref{mass_mgii}). Assuming the observed [C{~\sc ii}] velocity dispersion of 250 $\rm km~s^{-1}$ found by \cite{2016ApJ...816L...6D} using low spatial-resolution ($\sim$2 kpc, 0.35$^{\prime\prime}$) ALMA observations is representative of the stellar dispersion velocity along the line of sight of the galaxy, the diameter of the SoI is predicted to be $550^{+825}_{-330}$ pc ($0.084^{+0.13}_{-0.05}$$^{\prime\prime}$). While the errorbars are large, spatially resolving the SoI is within the reach of modern observatories, inspiring us to obtain higher spatial resolution observations. We present here ALMA high spatial resolution observations of the [C~{\sc ii}] and CO (12-11) emission of W2246--0526, with respective beam FWHMs of 0.089$^{\prime\prime}$ $\times$ 0.053$^{\prime\prime}$ and 0.038$^{\prime\prime}$ $\times$ 0.030$^{\prime\prime}$ in their uniform-weighted data cubes.

Fig. \ref{M0} shows the intensity of the continuum-subtracted [C{~\sc ii}] and CO (12-11) emission for different antenna weightings that trade between total SNR and spatial resolution. More details about these observations and the data processing are presented in Section \ref{ALMA_data_reduction} and Table \ref{cube_info}.

\subsection*{Analysis of [C{~\sc ii}] and CO (12-11) Kinematics}
Analysis of the moment maps, position-velocity (PV) diagrams and channel maps of [C{~\sc ii}] (see Section \ref{dispersion_system}) shows the ISM of W2246-0526 is detected in high velocity dispersion, with significant outflow components within a radius of $\sim$ 1.5 kpc (0.23$^{\prime\prime}$).
To isolate the contribution of the host dispersion from the outflows in the [C{~\sc ii}] dynamics, we start by spatially splitting the data cubes in a series of elliptical rings (see Section \ref{spectra_fitting} for details). We use the uniform-weighted data cube within the central $\sim$500 pc, and the natural-weighted data cube for larger scales. We model the integrated line emission in each of these regions as a combination of Gaussian functions, which is found to be sufficient given the data quality. We adopt one Gaussian function centered at the systemic velocity of the host to model the dynamics of the non-outflowing gas. We then add Gaussian functions with different velocity offsets to model the outflowing gas.
The total number of Gaussian functions was determined by successively adding functions until no improvements were identified in the $\chi^2$ per degree of freedom statistic.
No outflows are fitted to the central resolution element as their contribution is not detected.

CO (12-11) does not show significant outflow components in moment and channel maps (see Section \ref{co_kinematics}). We use uniform-weighted data cube within the central $\sim$220 pc, and three data cubes for larger scales (see Table S1). Through the same spectro-spatial modeling applied to [C~{\sc ii}], only a weak outflow component is recoved for CO (12-11). As in [C~{\sc ii}], no outflows are detected within the central resolution element.

A more detailed account of spectro-spatial modeling, as well as the best-fits to the spectra, are presented in Section \ref{spectra_fitting}. The detailed properties of the spatially resolved [C{~\sc ii}] outflowing gas will be explored elsewhere.

\subsection*{SMBH mass and SoI diameter measurements}
Fig. \ref{fig: dispersion_profile} (top panel) shows the joint radial profile of the best-fit host velocity dispersion (i.e., without the contribution from outflows) of the [C{~\sc ii}] and CO (12-11) data. Both emission lines show a consistent radial profile out to ~520 pc (0.08$^{\prime\prime}$), indicating that both [C{~\sc ii}] and CO (12-11) are co‑spatial and trace the same dynamical component. This is expected in X-ray dominated regions (XDR, \cite{2022ARA&A..60..247W}), which \cite{2024A&A...682A.166F} determined is the case in the nucleus of W2246--0526. The combined radial dispersion profile starts to diverge at larger angular scales (see Section 10 and Fig. \ref{full combined dispersion profile}), where heating likely starts being dominated by photodissociation regions (PDRs) for [C{~\sc ii}] and shocks for CO (12-11).

The combined dispersion drops with increasing radius up to $\sim~$0.85~kpc (0.13$^{\prime\prime}$) and levels off at larger distances from the center. This behavior is qualitatively consistent with the SMBH's gravity dominating within the central $\sim \rm 0.85~kpc$ region. To estimate the SMBH mass from the observed velocity dispersion curve we consider both the gravitational pull of the SMBH and of the mass distribution of the host galaxy (which could potentially contribute significantly to the observed velocity dispersions, particularly at larger radii), as well as the smoothing effect of the ALMA beam and the ring integration on the observations. The details of our model are discussed in Section \ref{dispersion fitting}. The joint fit yields an SMBH mass of $M_{\mathrm{BH}} = 6.5^{+0.6}_{-0.7} \times 10^{9}\,\mathrm{M}_{\odot}$ and a total host dynamical mass of $M_{\mathrm{host}} = 1.3^{+0.6}_{-0.5} \times 10^{10}\,\mathrm{M}_{\odot}$. The intrinsic dispersion profile contributed by the best-fit SMBH and host dynamical masses is separately shown in the middle panel of Fig.~\ref{fig: dispersion_profile}, and the residuals between best‑fit total model and the observations are shown in the bottom panel of Fig. \ref{fig: dispersion_profile}. We caution the reader that the error-bars on the SMBH mass here could be somewhat underestimated due to unaccounted systematics. We consider a number of different scenarios and assumptions in Section \ref{dispersion fitting}, and find that the best-fit SMBH mass could be as large as $~7.8~\pm~0.4$ $\times$ $\rm 10^{9}~\rm M_{\odot}$, corresponding to a scenario where the host galaxy contribution is neglected, or as little as $5.2^{+1.0}_{-0.9
}$ $\times$ $\rm 10^{9}~\rm M_{\odot}$ if the modeling is carried out by dropping the central beam points of the dispersion profile (see Tables \ref{All_models_fitting} and \ref{All_models_fitting_no_central_beam}). We note as well that the best-fit host
mass is surprisingly low when considering the best-fit SMBH mass. While at face value this
may seem odd, it may imply that other stellar mass components dominate at larger scales
than those probed by our ALMA data. Note that the SMBH mass is primarily set by the Keplerian rise and is largely unaffected by the specific host model (see Sections \ref{Host galaxy mass issues} and \ref{enclosed mass}).
A detailed account for the dynamical host mass's reliability and a comparison
with estimates obtained from other indicators, most notably from JWST NIRSpec and MIRI
observations, are presented in methods section.
Finally, we also note that consistent results of the dynamical SMBH mass are obtained we analyze each emission line independently.
Only using [C{~\sc ii}] yields $M_{\mathrm{BH}} = 6.6^{+0.8}_{-0.7} \times 10^{9}\,\mathrm{M}_{\odot}$, while CO (12‑11) alone yields $M_{\mathrm{BH}} = 6.9\pm0.8 \times 10^{9}\,\mathrm{M}_{\odot}$ (see Section~\ref{dispersion fitting} for details). 

The centrally enhanced dispersion profile is naturally explained by the action of an SMBH.  
This interpretation is strongly supported by the inability of alternative scenarios to account for the central rise. We consider a concentrated gas component to be unlikely given that the dust continuum observations around the [C{~\sc ii}] emission indicate a gas mass of only 1.3--2.0 $\times~10^{9}~\rm M_{\odot}$ within the [C{~\sc ii}] central beam, 
well below what is needed to account for the central [C{~\sc ii}] and CO (12-11) dispersions (see Section \ref{dust mass} and Table \ref{dust_con_T_gas_mass}).
We also consider whether a compact stellar component could explain this rise by considering a dual S\'ersic profile host without an SMBH. While a satisfactory model is obtained with an improbably concentrated central component (effective radius of $R_{\rm e}$ = 50$_{-50}^{+60}$ pc, S\'ersic index of $n$ = 5.8$_{-3.0}^{+3.6}$), it is statistically disfavored when compared to the model assuming the SMBH gravity (see Section \ref{Host-only Model} and Table \ref{All_models_summary}). A nuclear star cluster is also statistically disfavored, and furthermore it would also have to be more massive than any such cluster ever observed before (see Section \ref{Nuclear star cluster} and Table \ref{All_models_summary}). Such concentration is also reminiscent of little red dots (LRDs,\cite{2024ApJ...963..129M}) which can have effective radius below 100 pc \cite{2025ApJ...983...60C}. A number of articles have discussed about the AGN nature of LRDs (e.g., \cite{2026Natur.653.1017J}) and some have compared them to luminous obscured quasars \cite{2024MNRAS.533.2948S,2025ApJ...992..117B}. While it remains plausible that the nucleus of W2246--0526 resembles an LRD, we consider this unlikely given that LRDs lack the extended massive host of this object.

The combined dispersion profile keeps rising steeply towards the center in a Keplerian manner, which would be inconsistent with outflows or non-gravitational motions like turbulence dominating central kinematics. Furthermore, the morphology of the nuclear CO (12-11) emission 
strongly supports the gas being dispersion supported as it is best-fit by a S\'ersic profile with a steep index of 3.69$^{+0.55}_{-0.50}$ (see Section \ref{turbulence}). Turbulence dominated regions, on the contrary, tend to have flatter S\'ersic indices between 0.5 and 1 \cite{2014ApJ...781...11E,2024A&A...684A..24P,2024A&A...689A.347G}.
It is likely that the energy injection from the AGN is sufficient to maintain in a dispersion dominated distribution, particularly considering that \cite{2025ApJ...989..230V} determined the AGN activity in W2246-0526 has been ongoing for over 10 myr, as otherwise the gas would quickly coalesce into a rotating disk. Something similar is observed in the nuclear regions of nearby Ellipticals, where the sporadic AGN activity maintains the dispersion-dominated gas kinematics as proposed in \cite{2006AJ....131.1961V}.

Combined with the bolometric luminosity estimate of \cite{2018ApJ...868...15T}, the dynamical black hole mass of $M_{\rm BH}$ =  $6.5^{+0.6}_{-0.7}$ $\times$ $\rm 10^{10}~\rm M_{\odot}$ implies a super Eddington ratio of $1.54^{+0.19}_{-0.13}$, which is consistent with that found by \cite{2018ApJ...868...15T}. 
The derived corresponding intrinsic velocity dispersion fields of the best-fit SMBH and host components, shown in the bottom panel of Fig. \ref{fig: dispersion_profile}, imply the black hole gravity dominates over the host gravity within an SoI radius of $1.37^{+1.96}_{-0.39}$ kpc ($0.21^{+0.30}_{-0.06}$$^{\prime\prime}$). 
Note the best-fit diameter of the SoI is significantly larger than our our initial expectation. The mismatch is due to the assumptions about the host galaxy. The spatially global velocity dispersion measured by \cite{2016ApJ...816L...6D} based on low spatial resolution observations may be contaminated by outflows or turbulence driven by the interaction between the AGN outflows and host ISM. Additionally, as shown in Fig. 2, the best-fit host velocity dispersion flattens at a value close to 120 km~s$^{-1}$ instead of the assumed 250 km~s$^{-1}$, which by itself would result in a factor of 4 larger estimated size for the SoI. 

Now, as mentioned earlier, our dynamical model yields a surprisingly low mass for the host galaxy. If we assume it is underestimated and instead take a host model built from the stellar and dust-based gas mass profiles (see Sections \ref{stellar mass} and \ref{dust mass}) along with our best-fit SMBH mass, the SoI radius could be as small as $386_{-39}^{+46}$ pc ($0.059_{-0.007}^{+0.006}$$^{\prime\prime}$), which would still be resolved by our observations (see details in Section \ref{enclosed mass}).
To our knowledge, this is the first time the black hole SoI has been spatially resolved beyond $z = 2$ \cite{2024Natur.627..281A}. While \cite{2022A&A...668A.121S} constrained the SMBH mass of a $z$ $\sim$ 6 quasar through its [C{~\sc ii}] kinematics, they were not able to resolve the SoI which is key to assess the robustness of the measurement.
\section*{Discussion}
\subsection*{Implications for SMBH-galaxy co-evolution}

We have presented the first dynamical measurement of an SMBH mass at $z$ $>$ 2 by  resolving the black hole SoI.
The experience with W2246-0526 contrasts with the lack of results in similar experiments focusing on type 1 quasars at $z > 6$ discussed earlier, and suggests that [C{~\sc ii}] may be over-ionized near the SMBHs in unobscured objects. As shown by \cite{2015A&A...580A...5L}, when the ISM is exposed to the strong X-ray radiation field of an AGN with $L_{\rm X-ray} > 10^{43}~\rm erg~s^{-1}$, C$^{+}$ could be significantly transferred into C$^{2+}$ or higher ionization states, thereby depressing [C{~\sc ii}] emission within a few hundred pc from the SMBH. Even though W2246--0526 is very luminous, with an inferred X-ray luminosity of $L_{\rm 2-10~keV} = 6.4~\times~10^{45}~\rm erg~s^{-1}$ based on the mid-infrared to X-ray relation of \cite{2015ApJ...807..129S}, its heavy obscuration may be shielding the [C{~\sc ii}] gas from the AGN radiation. Although no direct X-ray detection exists to date, recently \cite{2024A&A...682A.166F}  estimated $N_{\rm H}$ of at least $10^{24}~\rm cm^{-2}$ based on the ratios of several far-IR emission line. Furthermore, X-ray studies for other similar heavily obscured quasars show that these objects are typically close to, or above, the Compton-thick limit (e.g., \cite{2014ApJ...794..102S,2015A&A...574L...9P,2016ApJ...819..111A,2020ApJ...897..112A,2017ApJ...835..105R,2018MNRAS.474.4528V,2026A&A...705A..90V}).
This scenario should be mirrored by other similar heavily obscured quasars \cite{2012ApJ...755..173E,2012ApJ...756...96W,2015MNRAS.447.3368B,2017MNRAS.464.3431H,2022ApJ...934..101A}, making them ideal targets for dynamically measuring their SMBH masses through high-resolution observations of their [C{~\sc ii}] emission.

Our SMBH mass estimate for W2246-0526 is consistent within the errorbars of that based on the Mg II emission line through the single-epoch method obtained by \cite{2018ApJ...868...15T}.
This consistency suggests the Mg II emission in W2246--0526 originates in the BLR, and it reaches the observer through scattering of the central engine emission, which has been observed in other similar obscured systems \cite{2016ApJ...819..111A,2020ApJ...897..112A,2022ApJ...934..101A,2018MNRAS.479.4936A,2024MNRAS.533.2948S}.
It also suggests the Mg II based single-epoch empirical relations may be reliable \cite{2009ApJ...707.1334W,2009ApJ...699..800V,2011ApJS..194...45S}, and suitable for (hyper-) luminous quasars at high redshift whose black hole masses have been suspected to be overestimated if based on the single-epoch method \cite{2023ApJ...956..127M}. We further find that if we estimate the H$\beta$ line width from the Mg II line width measured for W2246-0526 by \cite{2018ApJ...868...15T}, our dynamical $M_{\rm BH}$ estimates imply hydrogen emission $R_{\rm BLR}$ values that are largely consistent with the expectations from the best-fit relation of \cite{2024A&A...684A.167G} for high luminosity AGN and also the canonical relation of \cite{2013ApJ...767..149B} (see Section \ref{5100}).
The uncertainties are unfortunately too large to robustly differentiate between them using our observations.
Therefore, W2246-0526 provides a glimpse of assessing the reliability of the Mg II single-epoch spectral method, which underpins much of our current understanding of SMBHs and galaxy co-evolution in the early Universe. Focusing on obscured objects could
increase the number of dynamical SMBH
mass measurements in the early Universe and test the reliability of the single-epoch spectral method for an statistically meaningful sample.
Heavily obscured quasars have been associated with a specific evolutionary stage
where the SMBHs have high accretion rates and create winds that will eventually clear paths through much of the obscuring gas and dust. After this blowout phase, they may become regular quasars with a traditional torus \cite{2015ApJ...805...90T,2018ApJ...868...15T,2024ApJ...971...40L}.
W2246--0526 and similar objects \cite{2012ApJ...755..173E,2012ApJ...756...96W,2015MNRAS.447.3368B,2017MNRAS.464.3431H,2022ApJ...934..101A}, represent the most extreme obscured systems and possibly trace the peak of SMBH growth and radiative feedback into the host galaxy. The super-Eddington ratio implied by our SMBH mass estimate, coupled with the multiple outflows observed in our [C{~\sc ii}] data and by \cite{2025ApJ...989..230V} in [O{~\sc iii}] support this scenario for W2246--0526.
In Fig. \ref{fig:BH-host}, we show that W2246--0526 and other similar heavily obscured quasars (with $M_{\rm BH}$ estimated based on Mg II) have significantly overmassive SMBHs compared to the local early-type galaxy-$M_{\rm BH}$ relation \cite{2020ARA&A..58..257G}. Similar results have been found for hyperluminous optical quasars ($L_{\rm bol} \gtrsim $ $10^{13}$ $\rm L_{\odot}$) at $z \sim$ 2 \cite{2017A&A...598A.122B,2024Natur.627..281A}, and for luminous quasars at $z >6$ \cite{2016ApJ...816...37V}. If W2246-0526 is to evolve into the local relation by $z$ = 0, then a complex interplay between the growth of the SMBH and the host galaxy is needed, as the currently observed feedback must be insufficient to stop star-formation at this point. While a high AGN feedback efficiency of 5\% was found by \cite{2025ApJ...989..230V} based on [O{~\sc iii}] outflows, they also found a lack of photoionization by the central quasar beyond 1 kpc, suggesting a more nuanced interaction between the AGN and the host galaxy on larger scales that may require recurrent luminous AGN episodes as suggested by \cite{2018Sci...362.1034D} and \cite{2024ApJ...971...40L}.
Alternatively, W2246--0526 and other similar objects may evolve into outliers of the local relation at $z$ = 0. Separating between these scenarios is critical for understanding the long-term effects of radiative AGN feedback, as our dynamical estimates of $M_{\rm BH}$ in W2246-0526 suggest the offsets in this plane are real and not a by-product of poor SMBH mass estimates.

\section*{Materials \& Methods}
Detailed discussion of the materials and methods used in this paper are presented in the supplementary materials. Here we provide a brief overview of most important. We provide details of the JWST (Section 4) and ALMA (Section 6) observations used and of their reduction. Stellar and gas mass estimates for the host galaxy are presented in Section 5 and Section 7. 
We also provide details on the creation of ALMA data cubes, moment maps and position‑velocity diagrams used throughout this study (Section 8 and Section 9). The specifics of the spectra modeling of the data cubes are presented in Section 10. The method used for dynamically modeling the observed radial dispersion profile and for measuring the SMBH mass is presented in Section 11. Different combinations of data and model component are also explored in that Section. An extended discussion of the robustness of our SMBH mass measurements are presented in Section 12, while Section 13 and Section 14 present a discussion of potential issue affecting our measurements of host mass while showing they do not affect our SMBH mass measurements. Finally, Section 15 presents a discussion on why radiation pressure and turbulence are unlikely affect our results.

\clearpage

\begin{figure} [tb!]
\centering
\includegraphics[width=1.0\columnwidth  ]{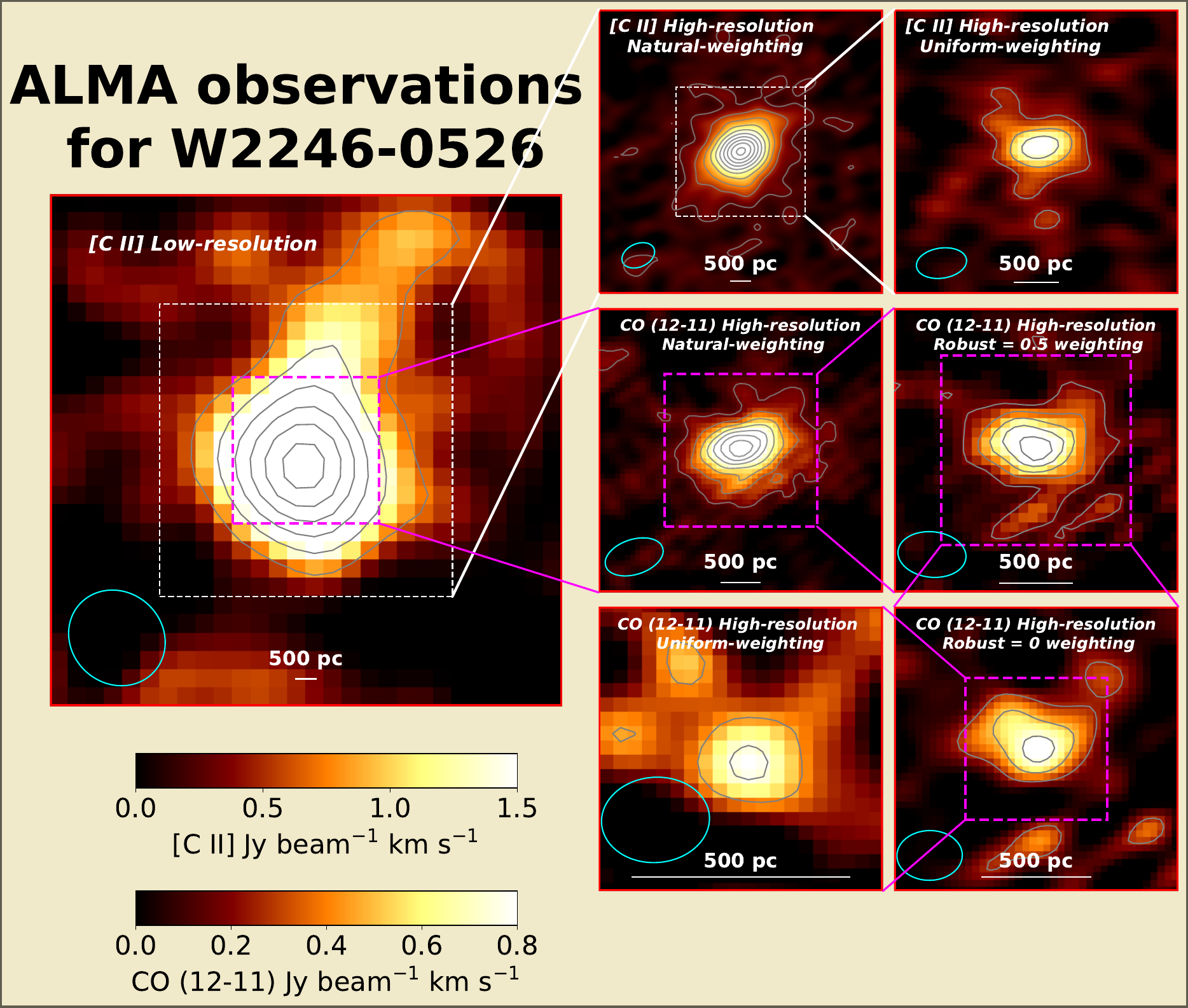}
\caption{\textbf{W2246--0526 ALMA [C{~\sc ii}] 157.7$\rm \mu $m and CO (12-11) 216.93 $\rm \mu $m images}.  The left panel shows the low-resolution [C{~\sc ii}] map of \cite{2016ApJ...816L...6D} and the six panels on the right are organized as a 3‑row by 2‑column grid, illustrating the data used in this work for different antenna weightings (see details in Table S1).
The synthesized ALMA beam for each observation is shown as a cyan ellipse in the lower left corner of each map. 
The contours of the [C{~\sc ii}] emission are drawn at 3$\sigma$ and increase by powers of 3 ($\sigma$ = 0.235 ($\it left$), 0.085 ($\it right‑grid~ top‑left$), 0.146 ($\it right‑grid~ top‑right$), 0.054 ($\it right‑grid~ middle‑left$), 0.061 ($\it right‑grid~ middle‑right$), 0.085 ($\it right‑grid~ bottom‑right$), and 0.146 ($\it right‑grid~ bottom‑left$) Jy $\rm km~s^{-1} beam^{-1}$). The boxes with white dashed lines are 1$^{\prime\prime}$ $\times$ 1$^{\prime\prime}$ ($\it left$) and 0.5$^{\prime\prime}$ $\times$ 0.5$^{\prime\prime}$ ($\it right‑grid~ top‑left$). The boxes with magenta dashed lines are 0.6$^{\prime\prime}$ $\times$ 0.6$^{\prime\prime}$ ($\it left$), 0.3$^{\prime\prime}$ $\times$ 0.3$^{\prime\prime}$ ($\it right‑grid~ middle‑left$), 0.2$^{\prime\prime}$ $\times$ 0.2$^{\prime\prime}$ ($\it right‑grid~ middle‑right$), and 0.1$^{\prime\prime}$ $\times$ 0.1$^{\prime\prime}$ ($\it right‑grid~ bottom‑right$). }
\label{M0}
\end{figure}

\begin{figure} [htp!]
\centering
\includegraphics[width=1.0\columnwidth]{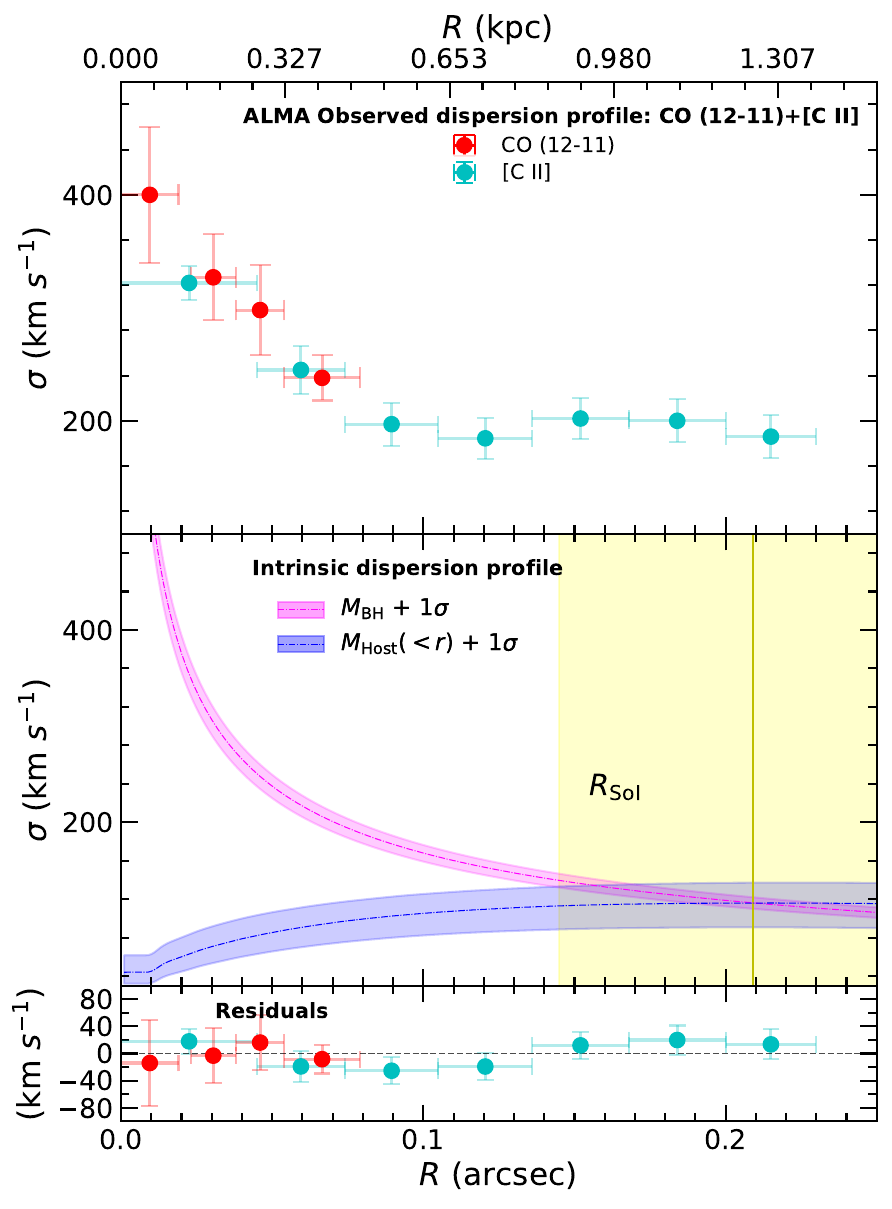}
\end{figure}

\afterpage{
  \begin{figure}[htp!]
    \caption[]{{\bf Velocity dispersion radial profile.} $\it Top$: the combined dispersion profile from CO (12-11) (red circles) and [C~{\sc ii}] (cyan circles).
The radii are along the major-axis with the values corresponding to the distance of the middle position of each ring from the center.
$\it Middle$: the intrinsic velocity dispersion fields of the best-fit SMBH dynamical mass (6.5$^{+0.6}_{-0.7}$ $\times$ $\rm 10^{9}~\rm M_{\odot}$, the magenta line and shaded region) and the best-fit host galaxy dynamical mass $M_{\rm host}~(< r)$ (the blue line and shaded region), respectively.
The radius of the SoI obtained from the dynamical model is shown by the yellow line. The yellow shaded regions show the 1$\sigma$ confidence intervals. $\it Bottom$: the residual between the observed outflow-free host dispersion profile and the best-fit dynamical total dispersion model with considering the beam smoothing effect, including contributions from SMBH mass and host mass (see Section 11 for details).}
\label{fig: dispersion_profile}
  \end{figure}
}

\clearpage

\begin{figure} [hbp!]
\centering
\includegraphics[width=1.\columnwidth]{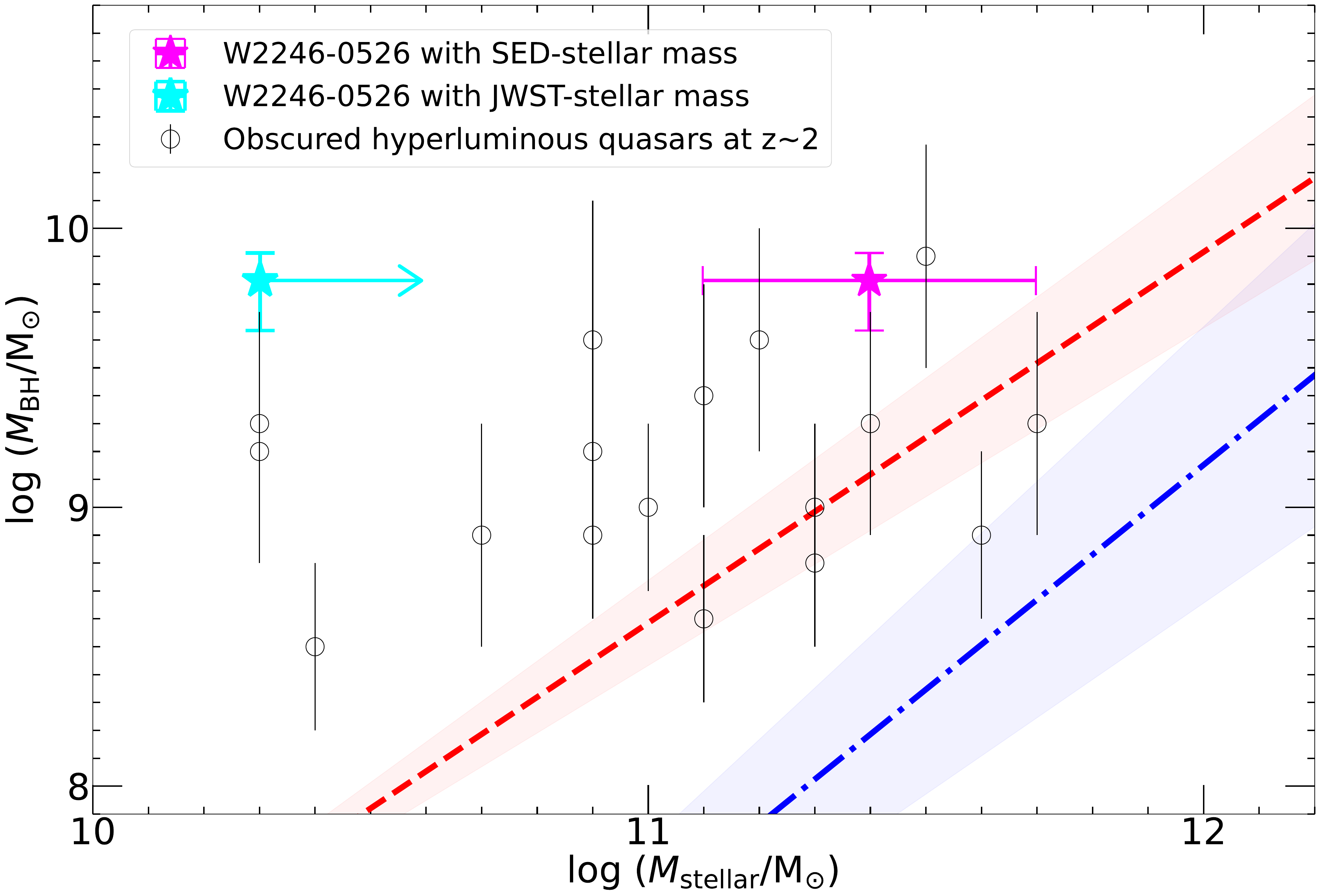}
\caption{\textbf{Black hole mass vs host galaxy mass relation.} The dashed red and dotted-dash blue lines are recent local relationships for early- and late-type galaxies from \cite{2020ARA&A..58..257G}, respectively. The cyan and magenta stars show W2246--0526, with stellar mass estimated by JWST NIRSpec and MIRI observations ($>$ 2 $\times$ 10$^{10}$ $\rm M_{\odot}$) and 2.5 $\times$ 10$^{11}$ $\rm M_{\odot}$ (with a factor of 2 for the errorbars) by \cite{2018Sci...362.1034D}, respectively.
The y-axis error-bar shows the range of plausible $M_{\rm BH}$ estimates for W2246--0526 discussed in Section of SMBH mass and SoI diameter measurements in the main text.
The black open circles show related heavily obscured hyperluminous quasars at $z \sim 2$ with Mg II-based black hole masses from \cite{2024ApJ...971...40L}. The stellar mass for all sources except W2246--0526 are derived from SED fitting and have typical uncertainties of a factor of 2 (see \cite{2024ApJ...971...40L} for details).}
\label{fig:BH-host}
\end{figure}

\clearpage

\bibliography{science_template} 
\bibliographystyle{sciencemag}

%
%
%
%
%
%


\section*{Acknowledgments}
ALMA is a partnership of ESO (representing its
member states), NSF (USA) and NINS (Japan), together with NRC (Canada) and NSC and ASIAA (Taiwan), in cooperation with the Republic of Chile. The Joint ALMA Observatory is operated by ESO, AUI/NRAO and NAOJ.

\paragraph*{Funding:}

This work was supported by the National Natural Science Foundation of China (Nos.12588202, 11988101, 11973051, 12041302 to M.L. and C.-W.T.), the New Cornerstone Science Foundation grant (to C.-W.T.), the National Science Foundation of China (No. 12203014 to X.-F.L.), the International Partnership Program of Chinese Academy of Sciences (No.114A11KYSB20210010 to C.-W.T. and M.L.), the National Aeronautics and Space Administration (80NM0018D0004 to P.E. and D.St.), Fondo Nacional de Desarrollo Cient\'ifico, Tecnol\'ogico y de Innovaci\'on Tecnol\'ogica (number 1231718 to M.L. and R.J.A.; number 1252054 to M.A. and J.G.L.), Agencia Nacional de Investigaci\'on y Desarrollo (FB210003 to M.A., J.G.L., R.J.A. and M.L.), the Rubin-Chile Fund grant (R.J.A and M.L.), Agencia Nacional de Investigaci\'on y Desarrollo (NCN2024\_112 to J.G.L. and M.A), Agencia Nacional de Investigaci\'on y Desarrollo (\#21221511 to M.S.), and European Union’s Horizon 2020 research and
innovation programme (DistantDust, No. 101117541 to R.F.A).

\paragraph*{Author contributions:}
Conceptualization: R.J.A. and M.L.
Data curation: R.J.A. and M.L.
Formal analysis: R.J.A. and M.L.
Funding acquisition: C.T., R.J.A. and M.L.
Investigation: A.W.B., R.F.A., M.A., C.T., R.J.A., M.L., D.Z., J.W., A.V., D.St., M.S., D.Sh., A.P., L.R.M., G.L., X.L., H.J., J.G., P.E., and T.D.S.
Methodology: A.W.B., R.F.A., M.A., C.T., R.J.A., M.L., A.V., D.St., H.J., J.G., P.E., and T.D.S.
Project administration: R.J.A.
Resources: R.J.A.
Software: R.J.A.
Supervision: R.J.A.
Validation: R.J.A. and M.L.
Visualization: M.L.
Writing‑original draft: M.L.
Writing‑review \& editing: A.W.B., R.J.A., M.L. and D.St.

\paragraph*{Competing interests:}
There are no competing interests to declare.
\paragraph*{Data, Code and Materials Availability:}
All data and code needed to evaluate and reproduce the results in the paper are present in the paper and/or the Supplementary Materials. This study did not generate new materials.
The ALMA high-resolution [C{~\sc ii}] (project code: 2016.1.00184.S) and high-resolution CO (12-11) (project code: 2025.1.01515.S), VLA CO (2-1) (project code: 17B-312), and JWST/NIRSpec/MIRI (project code: 1712) data in this work is publicly available. All ALMA data can be accessed at the ALMA data archive of https://almascience.eso.org/aq/, and VLA data at the NRAO data archive of https://data.nrao.edu/portal/\#/. All JWST data can be accessed at https://archive.stsci.edu/missions-and-data/jwst.
The code for implementing dynamically estimating the SMBH mass is archived on Zenodo \cite{roberto_2026_22879557}. 
The GALFIT software used in this work is available from the Astrophysics Source Code Library at https://ascl.net/1104.010. The WebbPSF software for JWST PSF simulation is available at https://ascl.net/1504.001.

\newpage


\renewcommand{\thefigure}{S\arabic{figure}}
\renewcommand{\thetable}{S\arabic{table}}
\renewcommand{\theequation}{S\arabic{equation}}
\renewcommand{\thepage}{S\arabic{page}}
\setcounter{figure}{0}
\setcounter{table}{0}
\setcounter{equation}{0}
\setcounter{page}{1} 


\begin{center}
\section*{Supplementary Materials for\\ \scititle}

    Mai Liao$^{1,2,3\ast}$,
	Roberto J.Assef$^{2}$,
	Chao-Wei Tsai$^{1,4,5}$,
    Manuel Aravena$^{2}$,\\
    Rom\'an Fern\'andez Aranda$^{6}$,
    Andrew W. Blain$^{7}$, 
    Tanio D\'iaz-Santos$^{8}$,\\
    Peter Eisenhardt$^{9}$,
    Jorge Gonz\'alez-L\'opez$^{10,11}$,
    Hyunsung D. Jun$^{12,13}$,\\
    Xiaofeng Li$^{14}$,
    Guodong Li$^{15}$,
    Lee R.Martin$^{7}$, Ana Posses$^{16}$,\\
    Devika Shobhana$^{2}$, Manuel Solimano$^{6}$,
    Daniel Stern$^{9}$,\\
    Andrey Vayner$^{17,18}$,
    Jingwen Wu$^{1,5}$, Dejene Zewdie$^{19}$\\
    \small$^\ast$Corresponding author. Email: mai.liao@mail.udp.cl
\end{center}

\subsubsection*{This PDF file includes:}
Materials and Methods\\
Figures \ref{JWST_plots} to \ref{LF_cii}\\
Tables S1 to S7\\


\newpage



\section{Cosmology}
Throughout the paper, we adopt a flat $\Lambda$CDM cosmology with $H_0 = 70\, \mathrm{km\,s^{-1}\, Mpc^{-1}}$ and $\Omega_\mathrm{m} = 0.3$. At $z$ = 4.6, this cosmology implies an angular scale of 6.5 kpc/$^{\prime\prime}$.

\section{Single-epoch SMBH mass estimate based on Mg II} \label{mass_mgii}
In \cite{2018ApJ...868...15T}, the width of the Mg II line is well fitted by one symmetrical and broad Gaussian component with FWHM = 3300 $\pm$ 600 $\rm km~s^{-1}$. Using $L_{\rm 3000\AA}$ = 2.6 $\times$ 10$^{47} \rm erg~s^{-1}$ estimated from the bolometric luminosity obtained by assuming the unobsured mean AGN SED template of \cite{2006ApJS..166..470R}, necessary due to the high optical extinction, \cite{2018ApJ...868...15T} estimated a black hole mass of 4.0$^{+6.0}_{-2.4}$ $\times~10^{9}~\rm M_{\odot}$ using the local calibration of \cite{2009ApJ...707.1334W}. The error range includes the 1$\sigma$ systematic uncertainties from both the single-epoch calibration (0.35 dex) and the Mg II profile fitting (0.05 dex). We note that consistent black hole mass estimates of 4.0$^{+3.9}_{-2.0}$ and 7.8$^{+7.2}_{-3.9}$ $\times~10^{9}~\rm M_{\odot}$ are found using the alternative Mg II-based single-epoch mass estimate recipes from \cite{2009ApJ...699..800V} and \cite{2011ApJS..194...45S}, respectively.

\section{The broad line region radius} \label{5100}
We derive log ($L_{\rm 5100\AA}$/$\rm L_{\odot}$) = 13.88 $\pm~0.15$ based on the bolometric luminosity $L_{\rm bol}$ from \cite{2018ApJ...868...15T} and the scaling relation between $L_{\rm bol}$ and the rest-frame continuum luminosity at  5100\AA~derived for highly obscured luminous quasars in \cite{2024ApJ...971...40L}.

The 5100\AA~continuum luminosity corresponds to log ($R_{\rm BLR}$/ld) = 3.37 $\pm~ 0.15$ for the canonical $R_{\rm BLR}-L_{\rm 5100 \AA}$ relation from \cite{2013ApJ...767..149B}, while in the revised $R_{\rm BLR}-L_{\rm 5100 \AA}$ relation for high luminosity AGN in \cite{2024A&A...684A.167G} it corresponds to log ($R_{\rm BLR}$/ld) = 2.96 $\pm~ 0.67$.
Our dynamical SMBH mass estimate implies log ($R_{\rm BLR}$/ld) = 3.12~$\pm~ 0.20$ by using the standard virial relation $M_{\rm BH} = f(R_{\rm BLR})\Delta v^{2}/G$ with log $<f>$ = 0.05 $\pm ~0.12$ \cite{2015ApJ...801...38W} and $\Delta v$ = FWHM$_{\rm H\beta}$ = 4714 $\pm$ 857 $\rm km~s^{-1}$. We scale the Mg II FWHM to the ${\rm H\beta}$ FWHM using the scaling factor from \cite{2016MNRAS.460..187M} as ${\rm H\beta}$ and ${\rm H\alpha}$ are heavily affected by the outflows in W2246--0526 \cite{2025ApJ...989..230V}. Unfortunately, we are not able to discriminate between the two relations given the precision of our measurements.
\section{JWST observations} \label{JWST observations}

We make use of the JWST NIRSpec and MIRI observations of W2246--0526 (project code: 1712) that were presented by \cite{2025ApJ...989..230V,2025arXiv251009870V}. The data reduction for the JWST NIRSpec data sets is described in \cite{2025ApJ...989..230V}, while that for the MIRI imaging data is described in \cite{2025arXiv251009870V}.

The NIRSPEC observations were obtained with a combination of the G235H and G395H gratings, covering the rest-frame 0.3--0.94\micron wavelength range. 
To perform the PSF subtraction and identify the stellar continuum, we used an empirical PSF from the recent Cycle 2 calibration program 3399, which observed a bright PSF star suitable for PSF subtraction (technical paper is in preparation). This PSF subtraction approach differs from that in \cite{2025ApJ...989..230V}, where the PSF was derived from the data itself using the \texttt{q3dfit} software. Following the methodology from \cite{2026ApJ..1006...76C}, we collapsed the data cube and PSF star along the spectral direction over the same line-free wavelength ranges for NIRSpec. A combined S\'ersic + PSF model is then fitted to the collapsed white-light image of W2246--0526 using Galfit \cite{2011ascl.soft04010P}. In this initial fit, all parameters are left free. We then fix all parameters except the PSF and S\'ersic flux, and re-fit the entire spectral range of each grating, binning the data spectrally by a factor of 3 to improve SNR. The wavelength-dependent flux of the S\'ersic component is finally extracted and attributed to the extended host galaxy. We find that the point-source emission from the quasar contributes a significant fraction of the total light (30-50\%) throughout the spectrum, indicating this emission likely stems from a combination of transmitted and scattered central engine light.

\begin{figure}[h]
\centering
\includegraphics[width=1.\textwidth]
{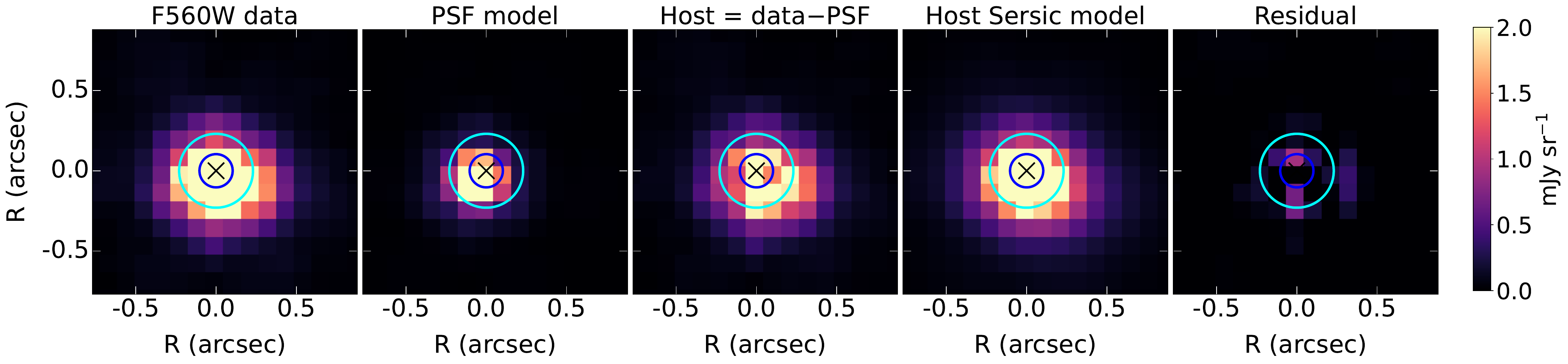}
\caption{PSF subtraction for JWST MIRI F560W imaging of W2246--0526, to measure the extended host emission. The blue, cyan circles are for the size of FWHM PSF (0.207$^{\prime\prime}$) at F560W, and the size of effective radius (2~$\times$ 0.23$^{\prime\prime}$) derived from the Galfit. The black cross marks the nominal position of the central quasar.}
\label{JWST_plots} 
\end{figure}

For MIRI imaging at F560W and F770W (rest-frame 0.94 -- 1.4\micron), we estimate the PSF in each filter at the precise position of W2246-0526 in the MIRI focal plane using the ``stpsf" package \cite{2014SPIE.9143E..3XP}. We then fit the data with a combination of a point-source and S\'ersic profile using Galfit to extract the extended host galaxy flux in each filter. For the F770W observations, only the flux of the PSF and the S\'ersic profile are allowed to vary; all other parameters are fixed to the F560W filter since the galaxy is marginally resolved at this longer wavelength. Fig. \ref{JWST_plots} shows the results of this modeling for the MIRI F560W imaging. The host galaxy is well fit by a single S\'ersic profile with $R_{\rm e}$ = 1.5 $\pm ~0.13$ kpc (0.23 $^{\prime\prime}$ $\pm~ 0.02^{\prime\prime}$) and $n$ = 1.64 $\pm~0.35$.

Interestingly, the best-fit positions of the point source and the S\'ersic profile in the JWST/MIRI F560W observations are offset by $\sim$1 kpc (1.3 pixels), suggesting the SMBH is not located at the geometric center of light of the host. This potential offset could be due to an offset either
in the mass distribution or just in the light distribution due to differential reddening. We explore the potential implications of the quasar being offset from the mass center of the galaxy in Section \ref{Offset stellar distribution}. We note that the emission peaks in our high‑resolution ALMA [C{~\sc ii}] and CO (12‑11) data coincide with the JWST point source position within the uncertainties, indicating the gas emission is centered at the quasar position. Specifically, we align the JWST observations using Gaia and measure distances to the MIRI peaks of 0.02$''$ $\pm$ 0.042$''$ (0.025$''$ $\pm$ 0.040$''$) and 0.031$''$ $\pm$ 0.040$''$ (0.031$''$ ± 0.038$''$) for [C{~\sc ii}]  and CO (12-11) peaks in the natural-weighting (uniform-weighting) intensity maps. Each total positional error budget includes the uncertainties in the ALMA absolute astrometric accuracy (https://help.almascience.org/kb/articles/what-is-the-absolute-astrometric-accuracy-of-alma), the centroiding errors of the ALMA emission peaks, the JWST PSF-fitting centroid error, and the geometric distortion uncertainties of MIRI imaging (https://jwst-docs.stsci.edu/jwst-calibration-status/miri-calibration-status/miri-imaging-calibration status\#gsc.tab=0), as well as the uncertainties due to the Gaia alignment.

\section{Stellar mass} \label{stellar mass} 

\begin{figure}[h]
    \centering
    \includegraphics[width=1.\linewidth]{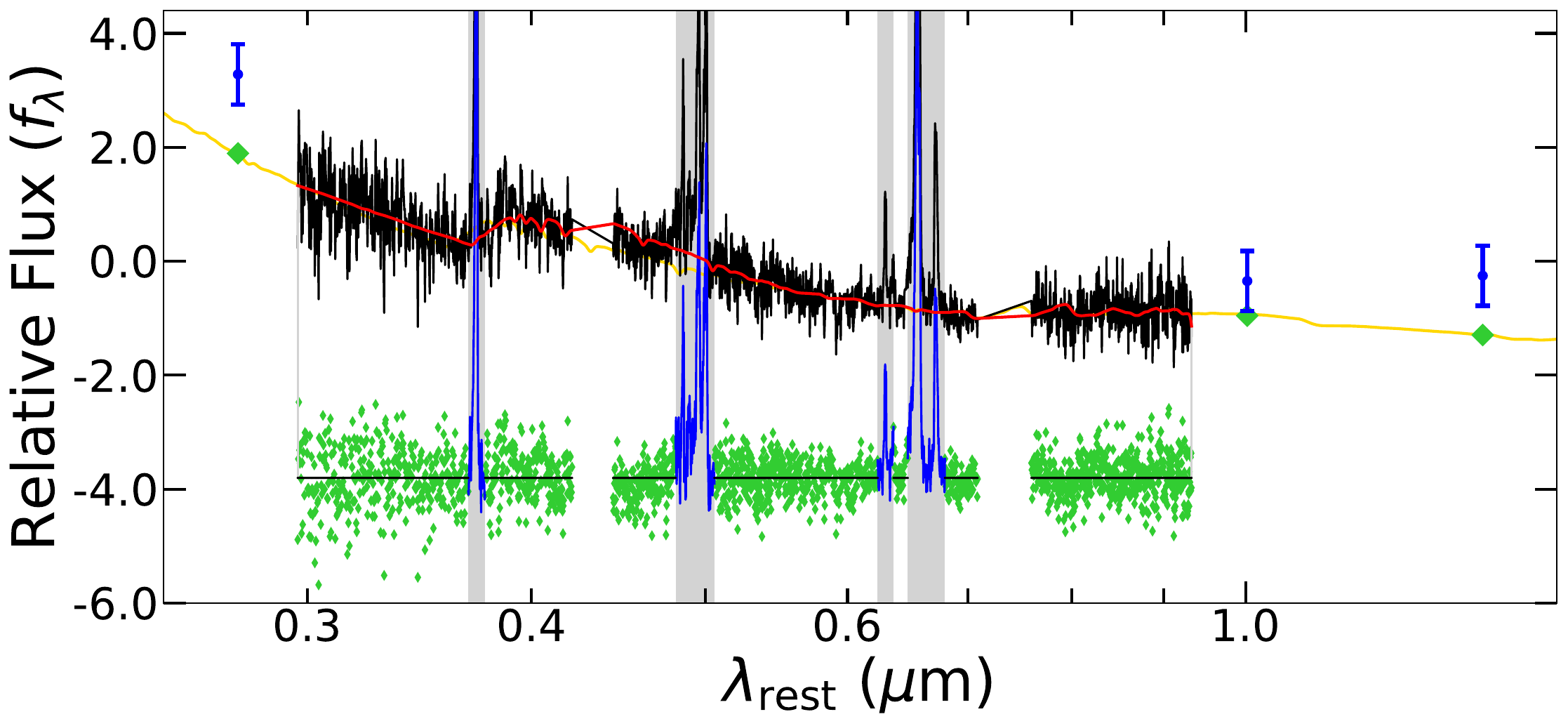}
    \caption{Fitting the stellar continuum of W2246-0526 using stellar population synthesis modelling with pPXF. We present the best-fit model to the combined spectrophotometry (black: NIRSpec) and photometry (blue points: HST F160W, MIRI F560W and F770W) in yellow line, while the red line shows the best-fit model only to the spectrophotometry. The HST F160W and MIRI F770W data points are slightly above the best-fit yellow line, due to unsolved quasar contribution in the former and the uncertainties for PSF subtraction in the latter (See section 8.1).
    Grey regions show wavelength ranges that are excluded from the fit, where strong emission lines are present from the extended host galaxy.}
    \label{fig:ppxf_fit}
\end{figure}

\begin{figure}
\centering
\includegraphics[width=1.0\textwidth]
{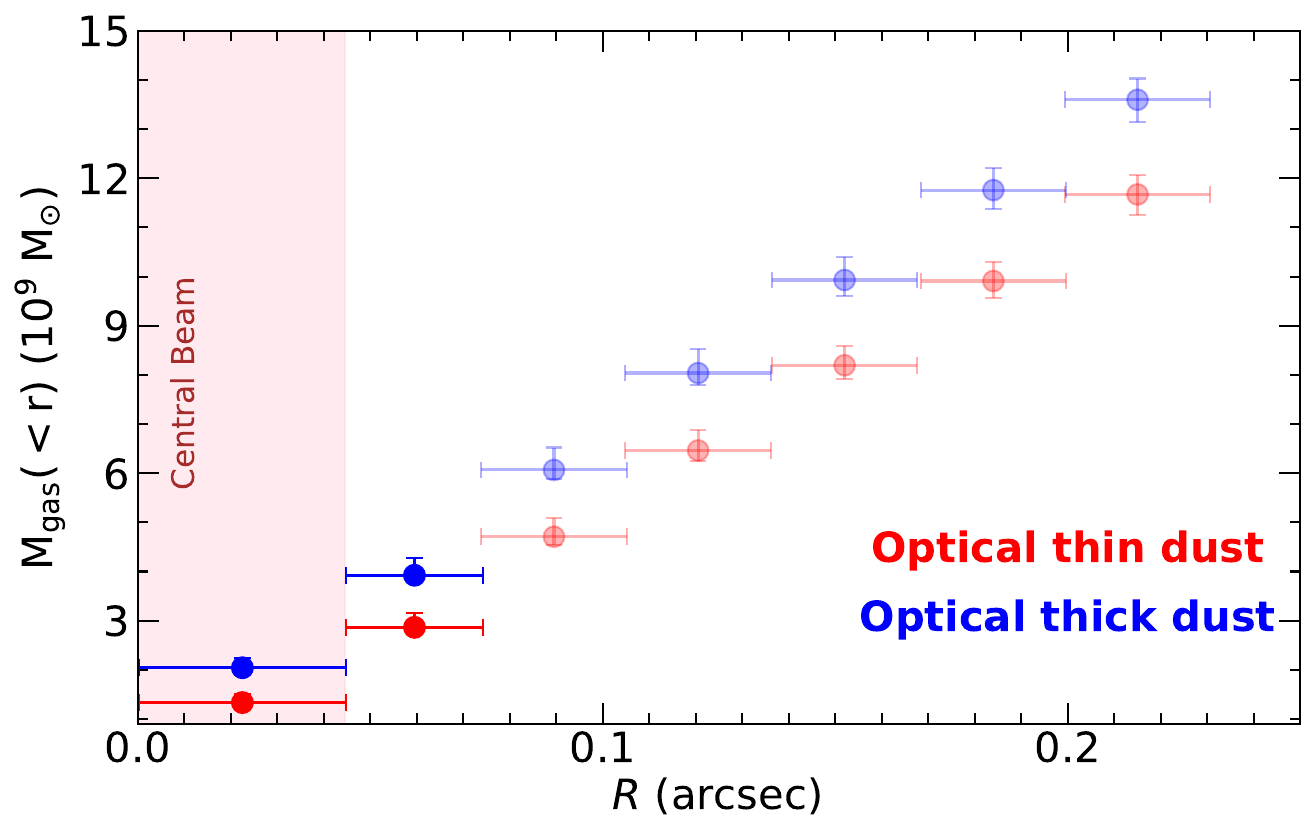}
\caption{The enclosed gas mass profile created from the dust continuum emission by using ALMA high-resolution dust observations for both optical thick and thin dust. The vertical shaded area shows the size of the beam in the [C~{\sc ii}] uniform-weighted data cube.}
\label{dust_gas_profile} 
\end{figure}

We use pPXF \cite{Cappellari2023} to fit the spectrophotometry of the best-fit S\'ersic profile model to the NIRSPEC and MIRI observations, along with the photometry from archival HST F160W imaging (see \cite{2016ApJ...816L...6D}). We mask all bright emission lines in the NIRSPEC data and fit only the continuum emission (see Fig. \ref{fig:ppxf_fit}). We use the fsps v3.2 model templates by \cite{2009ApJ...699..486C} and \cite{2010ApJ...712..833C} with MIST isochrones and a Chabrier IMF. The star formation history and metallicity are reconstructed by finding the linear combination of single-age, single-metallicity templates that best match the data. We consider a grid of stellar population ages and metallicities spanning $10^{6-9}$ years and $[Z/H]=-1.75-0.25$, respectively. More details about the spectral fitting will be presented in a forthcoming paper.

The total stellar mass, estimated from the contribution of each fsps template used, is approximately 2$~\times~$10$^{10}$ $\rm M_{\odot}$.
We note, however, that the best-fit model from pPXF has an unusually large star formation rate (SFR) of 3000 $\pm$ 1000 M$_{\odot}$~yr$^{-1}$ averaged over 28 myr. Such activity could dominate the observed stellar light and outshine an older, more massive stellar component. Furthermore, it could signify that there is a significant contribution from light of the obscured central engine scattered in sufficiently extended spatial scales so as to not be captured by the PSF subtraction. Such a contribution would make the observed SED artificially blue, leading to an overestimation of the SFR and an underestimation of the host stellar mass. We note that scattered-light is commonly observed in other similarly obscured, hyperluminous quasars \cite{2016ApJ...819..111A,2020ApJ...897..112A,2022ApJ...934..101A,2024ApJ...971...40L}, and recently \cite{2025A&A...702A.124A} has shown a case where this scattering is occurring in scales of $\sim$10~kpc.  Because of these reasons, we consider this stellar mass estimate to be a lower limit on the true value, and can only conclude that the host stellar mass is likely $M_{\rm Host}>2\times 10^{10}~\rm M_{\odot}$.

This estimate is significantly lower than that of \cite{2018Sci...362.1034D}, who estimated an stellar mass of $2.5$ $\times$ $\rm 10^{11}~\rm M_{\odot}$. They obtained this value by first estimating the brightness of the host galaxy in the rest-frame $H$-band relying on AGN/host SED modeling and decomposition of spatially unresolved photometry. Then, by adopting a combination of stellar ages (100 and 500~myr) and exponentially declining star-formation histories (with characteristic times of 100~myr and 1~gyr), they estimated a range of stellar mass-to-light ratios in the $H$-band that resulted in a range of stellar masses between $1.7\times 10^{11}~\rm M_{\odot}$ and $3.4\times 10^{11}~\rm M_{\odot}$ (see their Methods section for further details).

If we extrapolate the best-fit pPXF model to longer wavelenghts, we estimate a rest-frame $H$-band flux of about 7 $\times$ 10$^{-20}~\rm erg~s^{-1}~cm^{-2}~\AA^{-1}$, which is a factor of $\sim$2 times lower than that estimated by \cite{2018Sci...362.1034D} through the SED decomposition. Combined with the differences in star-formation histories, they account for the large discrepancy between the two stellar mass estimates. Combining all these issues, we consider it likely that the stellar mass of the host galaxy in W2246--0526 is within the range of $2\times 10^{10}~\rm M_{\odot}$ and $3\times 10^{11}~\rm M_{\odot}$.

\section{ALMA high-resolution observations  and data reduction} \label{ALMA_data_reduction}
\subsection{[C{~\sc ii}] }
Observations of the [C{~\sc ii}] 157.7$\rm \mu $m emission line of W2246--0526 were obtained during ALMA Cycle-4 (ID: 2016.1.00184.S) on UT 2017 July 28 using 45 antennas. 
The data were collected using the band 7 receiver in two observing sessions with a total on-source integration time of 80 min (divided into two 40 min sessions).
The [C{~\sc ii}] line was placed close to the center of the reference spectral window, while the other three spectral windows were used to detect the underlying continuum emission bluewards of the line.

The data were reduced using the Common Astronomy Software Applications (CASA) package version 4.7.2 \cite{2007ASPC..376..127M}. We subtracted the continuum from the calibrated visibilities using the task $\tt uvcontsub$ with a zeroth-order polynomial fitted to the continuum spectral window (SPW) immediately next to the one containing the emission line. We created two cubes for the line emission, one using uniform-weighting of the antennas and the other using natural-weighting of the antennas. Both are useful as the former’s higher spatial resolution allows us to probe the kinematics closer to the SMBH, and
the latter’s higher SNR allows us to better study the outer regions.
Both cubes were created using the task $tclean$ with a pixel size of 0.01$^{\prime\prime}$ and a channel velocity width of 30 $\rm km~s^{-1}$. They were cleaned down to a level of 2$\sigma$, where $\sigma$ is the average rms per channel of the [C{~\sc ii}] spectral window, calculated using a source-free rectangular region in the respective dirty images. The synthesized beams have FWHM of 0.089$^{\prime\prime}$ $\times$ 0.053$^{\prime\prime}$ (position angle PA: -81.7$^{\circ}$) in the uniform-weighted cube, and 0.13$^{\prime\prime}$ $\times$ 0.09$^{\prime\prime}$ (PA: -66$^{\circ}$) in the natural-weighted cube. The synthesized beam size for each cube is listed in Table \ref{cube_info}.

The average noise of the data cubes are 0.4 and 0.20 mJy $\rm beam^{-1}$ per channel for the uniform- and natural-weighted cubes, respectively. The systematic velocity we use throughout is based on the improved redshift of $z$ = 4.6019 $\pm$ 0.0001 derived by fitting a Gaussian function to the central beam emission of the uniform-weighted cube.

\subsection{CO (12-11)}

Observations of the CO (12-11) 216.93$\rm \mu $m emission line of W2246--0526 were obtained during ALMA Cycle-12 (ID: Project ID: 2025.1.01515.S) on UT 2025 December 9 using 45 antennas. The data were collected using the band 6 receiver in two observing sessions with a total on-source integration time of 90 min (divided into two 45 min sessions).
The CO (12-11) line was placed close to the center of the reference spectral window, while the other three spectral windows were used to detect the underlying continuum emission bluewards of the line.

We reduced the data using the CASA package version 6.5.1 \cite{2007ASPC..376..127M}. Following the same continuum-subtraction procedure as for [C~{\sc ii}], we created four CO(12-11) data cubes with different antenna weightings to optimise the balance between spatial resolution and sensitivity. Specifically, the uniform-weighted cube (robust = -2, beam FWHM: $0.038''\times0.030''$) provides the highest resolution (twice that of the uniform-weighted [C~{\sc ii}] cube) to resolve the innermost kinematics. However, only the central beam of this cube has sufficient SNR for our analysis. The natural-weighted cube (robust = 2, beam FWHM: $0.118''\times0.069''$) maximises sensitivity to trace the extended outer emission, but has lower spatial resolution,  comparable with that of natural-weighting [C~{\sc ii}] data cube. To map the spatially resolved gas kinematics between the uniform-weighted and natural-weighted CO (12-11) data cubes, two intermediate cubes were also created, one with robust = 0 (beam FWHM: $0.046''\times0.035''$) and one with robust = 0.5 (beam FWHM: $0.072''\times0.048''$).

All four cubes were created with a channel velocity width of 30 $\rm km~\rm s^{-1}$ and their synthesized beam sizes and pixel sizes are listed in Table \ref{cube_info}.
The average RMS of each data cube is 0.47, 0.24, 0.18 and 0.16 mJy $\rm beam^{-1}$ per channel for the uniform-weighting, robust = 0 weighting, robust = 0.5 weighting, and natural-weighting cubes, respectively.

\section{Gas mass estimates} \label{Gas mass estimates}
\subsection{Dust-based gas mass} \label{dust mass}

Recently \cite{2025A&A...695L..15F} carried out a detailed ALMA multi-band modeling of the continuum dust emission in W2246--0526. Within their 1$^{\prime\prime}$ FWHM beam, they found the dust temperature to be about 110~K at the position of the quasar and to drop to about 40~K at a distance of 1.2$^{\prime\prime}$ (8~kpc) from it, suggesting heating from the quasar stops being dominant at those distances. They estimated a dust mass of 8.4$_{-1.9}^{+2.0}$ $\times$~10$^{7}$ M$_{\odot}$ within the 1$^{\prime\prime}$ FWHM beam centered at the position of the quasar, which is significantly lower than the 5.6 $\times$ 10$^{8}$ $\rm M_{\odot}$ estimated within the same aperture by \cite{2018Sci...362.1034D} based on ALMA observations at 212\micron ~and assuming a slightly lower temperature of 100~K. This discrepancy is primarily ascribed by \cite{2025A&A...695L..15F} to the different treatment of the optical depth, as they fit for it while \cite{2018Sci...362.1034D} assumed the medium to be optically thin. Furthermore, \cite{2025A&A...695L..15F} assumed a dust mass opacity about a factor 2 larger than that assumed by \cite{2018Sci...362.1034D}. Using the gas-to-dust ratio of 200 adopted in \cite{2025A&A...695L..15F} results in a gas mass of 1.7 $\pm$ 0.4 $~\times~$10$^{10}$ M$_{\odot}$.

While this estimate is useful to compare to the global results of our dynamical modeling (see Section \ref{dispersion fitting}), we can use the high-resolution dust continuum from our ALMA observations to construct a radial profile in the same physical scales probed by [C{~\sc ii}]. To do this we first compute the radial profile of the dust continuum emission using the same rings used for the [C{~\sc ii}] dynamical analysis (see Section \ref{spectra_fitting} for further details), and fit it using a S\'ersic profile that considers the beam smearing in the same manner as done for the [C{~\sc ii}] modeling in the main text and explained in detail in Section \ref{dispersion fitting}. In order to compute a dust mass, we need to assume a temperature profile that considers the distance to the quasar rather than the average temperature of 110~K computed by \cite{2025A&A...695L..15F} within the central 1$^{\prime\prime}$. Specifically, we compute the average temperature $\langle T \rangle$ within each ring (see Section \ref{spectra_fitting} and Table \ref{cii_fitting_table}) weighted by the flux of the intrinsic (i.e., without beam smearing) S\'ersic profile to then obtain the dust mass using the same procedure as \cite{2025A&A...695L..15F}. As we lack sufficient information to fit for the temperature profile $T$($r$) we consider two prescriptions that should encompass its plausible values. First, we assume an scenario where the dust is distributed in optically and geometrically thick clouds, as in the torus, and adopt the temperature that corresponds to the non-illuminated side of the clouds as estimated from the $T_{\rm min}$ value in the Section 3.1.2 of \cite{2008ApJ...685..147N}, namely
\begin{equation}
T (r) = 400\left(\frac{1}{\rho}\right)^{0.42}, 
\end{equation}
\noindent where $\rho$ = $r/R_{\rm d}$, $R_{\rm d}$ = 14.43 pc is the dust sublimation radius, and $L_{\rm 45}$ = $L_{\rm bol}/(10^{45}~\rm erg~\rm s^{-1})$. Alternatively, we also consider an scenario where the dust is optical thin and the dust temperature is simply the equilibrium temperature with respect to the quasar, namely
\begin{equation}
T (r) = \frac{L_{\rm bol}^{1/4}}{(16\pi \sigma_{\rm SB})^{1/4} r^{1/2}} (1-B)^{1/4}
\end{equation}
\noindent where $\sigma_{\rm SB}$ is the Stefan-Boltzmann constant and $B$ is the albedo of the dust particles. We assume $B=0.5$  which is appropriate for ISM dust particles at UV/optical wavelengths \cite{2003ApJ...598.1017D}. The $\langle T \rangle$ estimates along with the dust mass estimates for the two scenarios are summarized in Table \ref{dust_con_T_gas_mass}. Finally, we estimate the gas mass by adopting the same dust-to-gas ratio of 200 of \cite{2025A&A...695L..15F}. The enclosed gas mass profiles are presented in Fig. \ref{mass_budget}. We find the gas mass within the central beam is only 1.3--2.0$~\times~$10$^{9}$ M$_{\odot}$, while the spatially integrated total mass is 1.2--1.4$~\times~$10$^{10}$ M$_{\odot}$ within the 0.23$^{\prime\prime}$ radius of the area probed by our ALMA data.

\subsection{CO-based gas mass} \label{CO mass}
We can also estimate the gas mass using the CO (2-1) observations from the VLA B configuration, which has a 0.25$^{\prime\prime}$ FWHM beam and an rms of 0.05 mJy~$\rm beam^{-1}$ per 100 km~s$^{-1\rm}$ (VLA/17B-312) as discussed in \cite{2025A&A...703A.216H}. Within the 0.25$''$ FWHM beam of the observations, we find a CO (2-1) luminosity of 5.9 $\times$ 10$^{6}$ L$_{\odot}$. Assuming the $\alpha_{\rm CO}$ $=$ 1.57 $\pm$ 0.9 [K~km$~\rm s^{-1}~\rm pc^{2}$]$^{-1}$ value for this transition in W2246--0526 computed by \cite{2025A&A...703A.216H} we estimate a gas mass of 2.4 $\pm$ 1.4 $\times$ 10$^{10}$ M$_{\odot}$. This is broadly consistent with that estimated in the previous section for a similar-sized region using the dust continuum from our ALMA high resolution observations.

\subsection{[C{~\sc ii}]-based gas mass} \label{cii mass}

In principle, it should also be possible to estimate the gas mass from the intensity of the [C~{\sc ii}] emission. If we use the standard conversion factor of $\alpha_{\rm [C~{\sc ii}]}$ = 30 M$_{\odot}$/L$_{\odot}$ from \cite{2018MNRAS.481.1976Z}, we find gas masses of 6.6 $\times$ 10$^{10}$ M$_{\odot}$ within a 0.13$^{\prime\prime}$ radius and 1.1 $\times$ 10$^{11}$ M$_{\odot}$within a 0.23$^{\prime\prime}$ radius of the center, which are significantly larger than those estimated from the dust continuum and the CO (2-1) emission. However, it is unclear whether this value of $\alpha_{\rm [C~{\sc ii}]}$ is appropriate for our source, as it was empirically calibrated for main sequence galaxies whose physical conditions may differ greatly from those in W2246--0526. Additionally, $\alpha_{\rm [C~{\sc ii}]}$ has a very large dispersion, which significantly increases the uncertainty of these estimates. For example, a value of $\alpha_{\rm [C~{\sc ii}]}$ = 7 M$_{\odot}$/L$_{\odot}$ is consistent with the lower boundary of the range of conversion factors observed by \cite{2018MNRAS.481.1976Z} (also see below for a physical interpretation of such a low value), and would result in much smaller masses of 1.7 $\times$ 10$^{10}$ M$_{\odot}$  within a 0.13$^{\prime\prime}$  radius and 2.8 $\times$ 10$^{10}$ M$_{\odot}$ within a 0.23$^{\prime\prime}$ radius respectively, which are comparable to those derived from the other indicators.

To assess the plausibility of this low $\alpha_{\rm [C~{\sc ii}]}$ = 7 M$_{\odot}$/L$_{\odot}$, we start by noting that $\alpha_{\rm [C~{\sc ii}]}$ can be expressed as (see eqn. (19) of \cite{2025A&ARv..33....4D}):
\[
\alpha_{\mathrm{[C~ii]}} = \frac{M_{\mathrm{\rm H}}}{L_{\mathrm{[C ii]}}} \propto (2e^{91.3/T_{\rm ex}}+4) \cdot \frac{1}{[\mathrm{C/H}]} \cdot \frac{1}{f_{\mathrm{C}^+}},
\]
where \(T_{\mathrm{ex}}\) is the excitation temperature with \(T_{\mathrm{ex}}\) = \(T_{\mathrm{kin}}\) (kinetic
temperature) for gas density much larger than the critical value (i.e., $n \gg n_{\rm crit}$), \(\mathrm{[C/H]}\) is the carbon abundance relative to hydrogen, and \(f_{\mathrm{C}^+}\) is the fraction of carbon that is singly ionised in C element. For W2246--0526, we adopt a solar metallicity, i.e., \(\mathrm{[C/H]}\) = 1.6 $\times 10^{-4}$, and \(T_{\mathrm{ex}}\) = \(T_{\mathrm{kin}}\) = 400 K (as inferred from the high-J CO excitation; \cite{2025A&A...703A.216H}),
a value of \(\alpha_{\mathrm{[C~ii]}} \sim 7\,\rm M_{\odot}/L_{\odot}\) can be achieved under \(f_{\mathrm{C}^+}\) $\sim$ 0.5. The model calculations in \cite{2015A&A...580A...5L} (their Fig. 4) show that under conditions resembling the dense, warm gas in the nucleus of W2246--0526, such a \(f_{\mathrm{C}^+}\) corresponds to an X-ray flux of 1.5 $\times$ 10$^{4}~\rm erg\rm~s^{-1}\rm cm^{-2}$.  Using the MIR-predicted X-ray luminosity of W2246--0526 of 6.4 $\times$ 10$^{45} \rm erg~s^{-1}$, this flux level would be reached at a distance of only 60 pc. This is significantly smaller than the physical size of the [C~{\sc ii}] central beam, indicating that the required X-ray flux can be achieved well within [C~{\sc ii}] central beam. Therefore, \(f_{\mathrm{C}^+}\) for the spatial scales we studied tend to be higher due to less X-ray flux, and \(\alpha_{\mathrm{[C~ii]}} \sim 7\,\rm M_{\odot}/L_{\odot}\) should be considered a conservative upper limit.
We caution that, the gas density in the nuclear region can significantly exceed the critical density of the [C~{\sc ii}] transition, makes [C~{\sc ii}] luminosity saturated thus potentially increase \(\alpha_{\mathrm{[C~ii]}}\). However, the low \(\alpha_{\mathrm{[C~ii]}}\) discussed above is already derived under the condition \(n \gg n_{\mathrm{crit}}\). We therefore consider that a low \(\alpha_{\mathrm{[C~ii]}}\) is physically justified in the extreme environment of W2246--0526.

\section{Observed [C{~\sc ii}] kinematics} \label{dispersion_system}

\subsection{Moment maps} \label{moments}
Fig. \ref{M0} shows the integrated intensity (the zeroth moment) of the high-resolution [C{~\sc ii}] observations, while Fig. \ref{cii_M1_M2} shows the line of sight mean velocity (the first moment) and the velocity dispersion (the second moment) for both data cubes. 
Both the first and second moments are derived using a 1$\sigma$ clipping to avoid biasing them due to noise. While clipping at a higher significance could further avoid noise, the second moment is very sensitive to this and doing so would result in underestimated dispersion values.

For the natural-weighted cube, a two-dimensional (2D) elliptical Gaussian fit to the zeroth moment with a deconvolved size of 0.25$^{\prime\prime}$ $\times$ 0.19$^{\prime\prime}$ at PA = 138$^{\circ}$, has a significant residuals, indicating that the [C{~\sc ii}] emission is spatially extended. We measure a frequency-integrated flux of 10.6 $\pm$ 0.7 Jy $\rm km~s^{-1}$ using a 0.6$^{\prime\prime}$ diameter aperture on the zeroth moment map. The first and second moments of this  cube show ranges that are consistent with those
found by \cite{2016ApJ...816L...6D} in lower spatial resolution observations. However, the first moment of our observations shows significant structure, with redshifted emission  spreading in three different directions.
If we instead focus on the cube created using uniform-weighting, a 2D elliptical Gaussian fitting to its zeroth moment shows a deconvolved size of 0.156$^{\prime\prime}$ $\times$ 0.128$^{\prime\prime}$ (PA = 24$^{\circ}$) and a total flux of 4.45 $\pm~0.5$ Jy $\rm km~s^{-1}$ within a 0.13$^{\prime\prime}$ diameter aperture. Its first moment has a velocity range between $-$60 and 150 $\rm km~s^{-1}$ which is narrower than that of natural-weighted cube ($-$200--100 $\rm km~s^{-1}$), with a high positive velocity region towards the south. Its second moment shows a similar range to the natural-weighted cube, but is more centrally peaked.

\subsection{PV diagrams} \label{PV}
Fig. \ref{cii_M1_M2} shows that the kinematics of the [C{~\sc ii}] emission line are dispersion dominated, as the width of the line is significantly larger than the velocity shifts throughout the source. This is consistent with the results of \cite{2016ApJ...816L...6D}. However both maps show significant structure, so we use position-velocity (PV) diagrams to further elucidate the nature of all the components.

Fig. \ref{fig:PV} shows 
the PV diagrams extracted from the natural-weighted cube. Clearly, the kinematic structure along the minor-axis is roughly symmetric as expected for a dispersion dominated system. The PV diagram along the major-axis, however, shows additional asymmetric structures which could be due to a combination of low velocity rotation and high velocity outflows. 
To isolate these structures, we subtract the dispersion dominated component, which we model using a 2D Gaussian fit to the minor-axis PV diagram since this component should be roughly axi-symmetric.
As illustrated in the middle panel of Fig. \ref{fig:2D}, our simple model provides a good representation of the observed kinematics along the minor-axis. However,  
the subtraction leaves substantial residuals along the major-axis as shown in the right panel of Fig. \ref{fig:2D}. The low velocity clumps ($|v|  < 200~\rm km~\rm s^{-1}$) could indicate a neglected, weak rotation component, as they appear on both sides with opposite velocity signs. The peak emission is at about +100 $\rm km~s^{-1}$ and -100 $\rm km~s^{-1}$, and we find the spatially integrated residuals at this velocity to have flux densities of 0.4 and 0.6 mJy, respectively.
The kinematics of the low velocity residual clumps are roughly consistent with the expected rotation curve for a galaxy with a stellar mass matching that of W2246--0526 (i.e., $2.5\times10^{11}$ $\rm M_{\odot}$ within a 2$^{\prime\prime}$ radius, \cite{2018Sci...362.1034D}), assuming an effective radius of 1~kpc and a Sersic profile index of $n=3$, and a central SMBH of mass $4.0\times 10^9~\rm M_{\odot}$ \cite{2018ApJ...868...15T} (see the right panel of Fig. \ref{fig:2D}).
The higher velocity ($\sim$ 200--400 $\rm km~s^{-1}$) residual clumps, however, only appear on one side. They may be related to AGN-driven outflows, inflows, or imprints from a recent gravitational interaction. They could also be left-overs from a cannibalized satellite galaxy.
We believe that they are more likely related to AGN outflows, as we would not expect coherent structures of inflowing gas at such high velocities. In fact, the dusty streamers connecting W2246-0526 already show the effects of tidal disruption at much larger scales \cite{2018Sci...362.1034D}. Moreover, the high velocity [C{~\sc ii}] gas (200--600 $\rm km~s^{-1}$) is highly asymmetric and collimated as shown on the channel maps of Fig. \ref{fig:cii_channel_maps}. This is more readily explained by directional outflows than by tidal streams or gravitational interactions. 
We note that, AGN-driven massive outflows detected in [O{~\sc iii}] 5007\AA ~are a common occurrence in the similar hyperluminous obscured quasars \cite{2020ApJ...905...16F, 2020ApJ...888..110J} and in particular in W2246-0526 itself \cite{2025ApJ...989..230V}.

\subsection{Channel maps} \label{channels maps}
The channel maps of the natural-weighted cube (see Fig. \ref{fig:cii_channel_maps}) show that the gas between -240 and 240 $\rm km~s^{-1}$ is more symmetrically distributed around the central peak emission of the integrated line emission. In contrast, the faster moving gas show asymmetry.
The observed blue-shifted gas (V $<$ -300 $\rm km~s^{-1}$) becomes progressively more disturbed and offset from the central peak with increasing speed.
The gas in the 240 to 720 $\rm km~s^{-1}$ range displays a rounder morphology with a smaller offset but is significantly  disturbed towards the southeast, northeast and southwest. 
We further isolate the high-velocity gas by separately collapsing the channels with velocities above $>$ 240 $\rm km~s^{-1}$ and below $<$ $-$420 $\rm km~s^{-1}$. As shown in Fig. \ref{fig: outflows with the first moment}, this fast moving gas seems to account for most of the spatial structures observed in the first moment of the line emission (Fig. \ref{cii_M1_M2}), suggesting the first moment is dominated by outflows rather than rotation.
Fig. \ref{fig:cii_channel_maps} and Fig. \ref{fig: outflows with the first moment} also indicate the red-outflows and blue-outflows are not distributed symmetrically: the blue outflow appears only in one direction and significantly more collimated. 

The higher velocities towards the southern edge of the first moment in the uniform-weighted cube (see the top-right panel of Fig. \ref{cii_M1_M2}) also points towards outflowing gas in the innermost scales. This high velocity component may correspond to the base of the red outflows seen in the natural-weighted cube.

\section{Observed CO (12-11) kinematics} \label{co_kinematics}

\subsection{Moment maps}
Fig. \ref{M0} shows the zeroth moment (i.e., intensity map) of the high-resolution CO (12-11) observations, while Fig. \ref{CO_M1} and Fig. \ref{CO_M2} shows the first and second moments of the four data cubes (see Table \ref{cube_info}). As for [C{~\sc ii}], the first and second moments were constructed using 1$\sigma$ clipping.
The deconvolved sizes for the CO (12-11) emission were obtained by fitting the intensity maps with 2D elliptical Gaussians. We find sizes of 
 0.13$^{\prime\prime}$ $\times$ 0.094$^{\prime\prime}$ at PA = 120$^{\circ}$, 0.077$^{\prime\prime}$ $\times$ 0.053$^{\prime\prime}$ at PA = 90$^{\circ}$, 0.056$^{\prime\prime}$ $\times$ 0.026$^{\prime\prime}$ at PA = 82$^{\circ}$ for robust = 0 weighting, robust = 0.5 weighting, and natural-weighted cubes, respectively. We find spatially integrated fluxes  of 0.8 $\pm$ 0.15, 1.5 $\pm$ 0.2, 2.4 $\pm$ 0.2 and 2.8 $\pm$ 0.17 Jy $\rm km~s^{-1}$ using a 0.06$^{\prime\prime}$, 0.14$^{\prime\prime}$, 0.25$^{\prime\prime}$ and
 0.4$^{\prime\prime}$ diameter apertures on the intensity maps derived for the uniform-weighted, robust = 0 weighting, robust = 0.5 weighting, and natural-weighted data cubes, respectively.

All four CO (12-11) data cubes globally exhibit a similar velocity field, which is also similar to that of the uniform‑weighted [C{~\sc ii}] cube, indicating that both lines in these cubes trace the same kinematic components. The second moment maps of CO (12-11) show a clear centrally peaked dispersion that increases with spatial resolution, qualitatively consistent with a black hole Keplerian rise.

\subsection{Channel maps}

In Fig. \ref{CO_channel_maps}, we show the channel maps for CO (12-11) constructed from the natural-weighting data cube to maximize SNR. In contrast to the channel maps of [C{~\sc ii}] (Fig. \ref{fig:cii_channel_maps}), CO (12-11) does not exhibit comparable high‑velocity asymmetric gas structures. Instead it shows the gas is symmetrically distributed around the peak of the integrated line emission. We further explore this in Section \ref{spectra_fitting}.

\section{Spectro-Spatial Modeling} \label{spectra_fitting}
To spatially isolate the host dispersion and [C{~\sc ii}] outflows, we model the combined spectra from W2246--0526's central region and its extended regions using successive rings (see Table \ref{cii_fitting_table}). We fit the line emission in each annular region over a 0.4 $\mu$m range that includes the [C{~\sc ii}] line center (i.e., between 157.3 and 158.1 $\mu$m at rest-frame).
We assume that the outflowing gas and host dispersion can be modeled by simple Gaussian functions with different central velocities and dispersions. 
Specifically, for the uniform-weighted cube, we consider a central ellipse equal to the beam of the observations (0.089$^{\prime\prime}$ $\times$ 0.053$^{\prime\prime}$, PA of -81.7 deg) and the first concentric successive ring ($U_{\rm R1}$ hereafter) with a width 1/3 the size of the beam in each axis. We do not include the second ring for the uniform-weighted cube since it has only partial pixels with $>$ 3$\sigma$ detections in the zeroth moment map. To ensure the spatial independency of our measurements between the uniform- and natural-weighted data cubes, for the natural-weighted cube, we only consider spatial regions outside $U_{\rm R1}$ with five successive rings (see Table \ref{cii_fitting_table}). Each ring has a width 1/8 the deconvolved size of the galaxy in each axis (0.25$^{\prime\prime}$ $\times$ 0.19$^{\prime\prime}$, see details in section \ref{moments}) with the same PA (138 deg). The outer ring of the final ellipse has major and minor axes of 0.46$^{\prime\prime}$ and 0.35$^{\prime\prime}$, respectively.

We adopt one Gaussian function to account for the dynamics of the host dispersion, with the centroid fixed at the expected wavelength of [C{~\sc ii}] given the redshift of the source. We then successively add Gaussian components to account for the red- and blue-outflow contributions and test if they are required by checking whether $\chi^{2}_{\rm d.o.f.}$ is significantly improved.
We require outflows to be offset by at least 200 km$~\rm s^{-1}$ from the systemic velocity of the host. We find that a single blue outflow component is required in the natural-weighted cube and none are required in the uniform-weighted cube.
No outflow component is needed in the central beam of the uniform-weighted cube (see Fig. \ref{cii_ring_fitting_uniform}).
For the red-outflows, we find that one component is sufficient for the uniform-weighted cube, while multiple Gaussians are required for the natural-weighted cube, likely related to the complex morphology observed in Fig. \ref{fig:cii_channel_maps} and Fig. \ref{fig: outflows with the first moment}. Based on inspection of their emission in the channel maps, we find one component centered at $\sim$ 350 $\rm km~s^{-1}$ and another centered at $\sim$ 600 $\rm km~s^{-1}$.
A minimum width of 40 $\rm km~s^{-1}$ is used to avoid fitting noise spikes.

The best-fit models for each ring and central beam are shown in Fig.~\ref{cii_ring_fitting_uniform} and Fig.~\ref{cii_ring_fitting_natutal}, while the best-fit parameters are presented in Table \ref{cii_fitting_table}. To estimate the uncertainties in each parameter we generated 200 mock spectra created by resampling the observed spectra according to the uncertainty, and use the same fitting routines as above. Note we also included outflow components as free parameters when deriving errors of the host dispersion for the central beam of the uniform-weighted data cube to account for their potential presence even if formally undetected. We take the standard deviation of the distribution of each best-fit parameter as its uncertainty.

We applied the same ring‑fitting procedure to the CO (12‑11) data cubes. 
Specifically, for the uniform-weighted cube of CO (12-11), we considered a central ellipse equal to the beam of the observations (0.038$^{\prime\prime}$ $\times$ 0.03$^{\prime\prime}$, PA of -85.9 deg). No outer rings are considered in this cube because no pixels exceed a 3$\sigma$ detection outside the central beam.
For robust = 0 weighting cube, we considered two rings outside its central beam (denoted as $R0_{\rm R1}$ and $R0_{\rm R2}$). The ring width was set to one third of its beam size in each axis (0.046$^{\prime\prime}$ $\times$ 0.035$^{\prime\prime}$) with PA of -89 deg. For robust = 0.5 weighting cube, we considered one ring outside its central beam (denoted as $R0.5_{\rm R1}$), with the ring width equal to one third of its beam size in each axis (0.072$^{\prime\prime}$ $\times$ 0.048$^{\prime\prime}$) with PA of 84 deg. For the natural‑weighted cube (robust = 2), we considered two rings outside the spatial regions of $R0.5_{\rm R1}$, denoted as $NN_{\rm R1}$ and $NN_{\rm R2}$. The ring widths were defined as 1/4 its beam size in each axis (0.118$^{\prime\prime}$ $\times$ 0.069$^{\prime\prime}$) with PA of -72 deg. $NN_{\rm R2}$, i.e., the outermost ring of CO (12-11) emission, has major and minor axes of 0.28$^{\prime\prime}$ and 0.18$^{\prime\prime}$, respectively. Note, here the ratios of adopted ring widths to the beam sizes are same as those applied for [C{~\sc ii}].

Fig. \ref{CO_ring_fitting1} and Fig. \ref{CO_ring_fitting2}  shows the ring fitting for CO (12-11) emission with Gaussian functions for the host dispersion and outflows, obtained using the same priors used to fit the [C{~\sc ii}] emission. Notably, the CO (12‑11) fits require only a weak blue‑shifted component centered at $\sim$ -550 $\rm km~s^{-1}$ in the outer rings, while no red‑shifted outflows are presented in all the concentric rings. The best‑fit parameters are listed in Table \ref{co_ring_fitting_table}. 

Fig.~\ref{full combined dispersion profile} presents the radial profile of the host velocity dispersion derived from [C~{\sc ii}] and CO (12-11). The two tracers show remarkable consistency out to a radius of 520~pc ($\sim$0.08$^{\prime\prime}$), confirming that within this region both lines trace the same dynamically cold gas in the X-ray dominated region (XDR; \cite{2022ARA&A..60..247W}). Beyond this radius, the profiles diverge as that the [C~{\sc ii}] dispersion levels off, while the CO (12-11) dispersion shows a rise, suggesting heating likely starts being dominated by photodissociation regions (PDRs) for [C{~\sc ii}] and shocks for CO (12-11). To avoid contamination from shocks, we exclude the outermost two CO (12-11) data points from the dynamical modelling for the gravitational components of SMBH and host masses described in Section~\ref{dispersion fitting}.

\section{SMBH mass measurement} \label{dispersion fitting}

The SMBH mass can be measured from the observed velocity dispersion profile of [C{~\sc ii}] and CO (12-11) presented in Fig. \ref{fig: dispersion_profile}. Doing so, however, is not trivial, as we must also account for the gravitation pull of all other components of the host galaxy, as well as the smoothing effects of the beam and the spatial integration in each ring (see Section \ref{spectra_fitting}). In this section we describe the procedure used to estimate the SMBH mass from the profile in Fig. \ref{fig: dispersion_profile}. 

\subsection{Methodology}
Following \cite{2018ApJ...854...97D,2023ApJ...956..127M}, we assume that the intrinsic velocity dispersion at a given distance from the center, $r$, is given by 
\begin{equation}\label{dispersion model}
    \sigma(r) = \sqrt{\frac{2}{3}~\frac{G [M_{\rm BH} + M_{\rm Host}(<r)]}{r}}, 
\end{equation}
\noindent where $M_{\rm BH}$ is the SMBH mass, $M_{\rm Host}(<r)$ is the host galaxy mass within $r$, and $G$ is the gravitational constant. We assume that the non-SMBH mass of the host galaxy follows a S\'ersic profile with index $n_{\rm Host}$, effective radius $R_e^{\rm Host}$ and total dynamical mass $M_{\rm Host}^{\rm Total}$ $\equiv$ $M_{\rm Host}$ ($<\infty$).

Let $I_R(r)$ be the total intensity radial profile of the [C {\sc ii}] emission, which was shown by \cite{2021A&A...654A..37D} to be consistent with a 2D Gaussian with a de-convolved circularized effective radius of 1.2~kpc. The intrinsic intensity at a given velocity $v$ with respect to the systemic redshift and at radius $r$, $I_{\rm int}(r,v)$ can be therefore written as
\begin{equation}\label{intrisic spectrum}
    I_{\rm int}(r,v) \propto I_R(r)\ \exp{\left\{-\frac{1}{2}\left(\frac{v}{\sigma(r)}\right)^2\right\}},
\end{equation}

However, since the observations are smoothed by the beam, the observed intensity, at radius $r_{\rm o}$, would actually be given by
\begin{equation}\label{eq:Iobs}
    I_{\rm obs}(r_{\rm o},v)\propto \int_{0}^{2\pi} \int_{0}^{\infty} I_{\rm int}(r,v)\ \exp{\left\{-\frac{1}{2}\left(\frac{D(r,r_{\rm o},\theta)}{\sigma_{\rm Beam}}\right)^2\right\}}\ r\ dr\ d\theta,
\end{equation}
\noindent where $\sigma_{\rm Beam}$ is the standard deviation of the beam, transformed from the geometric mean of FWHM axes of the respective beams given in prior sections, and $D(r,r_{\rm o},\theta) = (r^2 + r^{2}_{\rm o} - 2r r_{\rm o}\cos{\theta})^{1/2}$ is the Euclidean distance between two points in polar coordinates. We note that equation (\ref{eq:Iobs}) can be conveniently rewritten in terms of the zeroth order modified Bessel function of the first kind, $\mathcal{I}_0$, such that 
\begin{equation}\label{observed spectrum}
    I_{\rm obs}(r_{\rm o},v)\propto \int_{0}^{\infty} I_{\rm int}(r,v)\ \exp{\left\{-\frac{1}{2}\left(\frac{r^{2}_{\rm o}+r^2}{\sigma_{\rm Beam}^2}\right)\right\}}\ \mathcal{I}_0\left(\frac{r_{\rm o} r}{\sigma_{\rm Beam}^2}\right)\ r\ dr,
\end{equation}

Finally, the measured line velocity profile spatially integrated within one of the rings defined in Section \ref{spectra_fitting}, $I_{\rm M}$, is given by 
\begin{equation} \label{ring spectrum}
    I_{\rm M}(v)\propto \int_{r_{\rm in}}^{r_{\rm out}} I_{\rm obs}(r_{\rm o},v)\ r_{\rm o}\ dr_{\rm o},
\end{equation}
\noindent where $r_{\rm in}$ and $r_{\rm out}$ are the inner and outer radii of the rings in question, respectively. Since the rings are slightly elliptical, we consider the geometric average between the respective major and minor axes as the values of $r_{\rm in}$ and $r_{\rm out}$.

We estimate the SMBH mass by finding the values of $M_{\rm BH}$, $M_{\rm Host}^{\rm Total}$, $n_{\rm Host}$ and $R_e^{\rm Host}$ that best-fit the observed velocity dispersion profile, shown in Fig. \ref{fig: dispersion_profile}. For this we run a Markov-Chain Monte Carlo process using the software {\tt{emcee}} \cite{2013PASP..125..306F} with 32 walkers and 5000 iterations. For each combination of parameters we compute the measured line velocity profile, $I_{\rm M}(v)$, in each ring, and fit a Gaussian function to it to obtain an accurate representation of the measured velocity dispersion values. We take the median of the distribution of each parameter as its best-fit value and the 16-to-84 percentile range of each distribution as the uncertainty. We have made the routines used to fit the observations as well as a notebook showing all the results discussed below publicly available (see details in Section of Data, Code and Materials Availability). 

\subsection{[C{~\sc ii}] and CO (12-11) joint fit }\label{joint_disperison_fitting}
Firstly, we consider the case for fitting the joint dispersion profile of [C{~\sc ii}] and CO (12-11). As shown in Fig. \ref{fig:BH_host_combined_dis}, the best-fit model has $\chi^2 = 8.71$ ($\chi^2_{\rm d.o.f.} = 1.24$) with $M_{\rm BH}$ = $6.2^{+0.9}_{-1.0}$ $\times$ $\rm 10^{9}~\rm M_{\odot}$ and $M_{\rm Host}^{\rm Total} = 3.6^{+8.1}_{-2.2}$ $\times$ $\rm 10^{10}~\rm M_{\odot}$. Note that because the distribution of $n_{\rm Host}$ is so asymmetric, the minimum $\chi^2$ solution ($\chi^2$ = 6.55, $\chi^2_{\rm d.o.f.} = 0.94$) is significantly offset. 
More importantly, as shown in Fig. \ref{fig:BH_host_combined_dis}, we find $n_{\rm Host}$ and $R_e^{\rm Host}$ are largely unconstrained. To address this, we limit $n_{\rm Host}$ and $R_e^{\rm Host}$ to vary between 0.5--2 and 1.1--1.6~kpc, respectively. 
These are the ranges spanned by the parameters of the best-fit S\'ersic profiles to the HST/WFC3 F160W \cite{2016ApJ...816L...6D} and JWST/MIRI F560W imaging (see Section \ref{JWST observations}), and to the [C~{\sc ii}] integrated emission in \cite{2021A&A...654A..37D}.
We then obtain a best-fit model with $\chi^2 = 7.63$ ($\chi^2_{\rm d.o.f.} = 1.09$), suggesting our model provides an acceptable representation of the data.
Fig. \ref{fig: BH_host_combined_fixed_dis} shows the corresponding corner plot with the distribution of each parameter. The values of  $M_{\rm BH}$ = $6.5^{+0.6}_{-0.7}$ $\times$ $\rm 10^{9}~\rm M_{\odot}$ and $M_{\rm Host}^{\rm Total} = 1.3^{+0.6}_{-0.5}$ $\times$ $\rm 10^{10}~\rm M_{\odot}$ are well constrained. The intrinsic velocity dispersion profile of the specific contributions from the AGN and the host, as well as the residuals between the best-fit total model and observations, are shown in the middle and bottom panels of Fig. \ref{fig: dispersion_profile}, respectively. The best-fit SMBH and host masses, as well as the corresponding values of $\chi^2$ and $\chi^2_{\rm d.o.f.}$, are listed in Table \ref{All_models_fitting}. Clearly, the host parameter priors have a negligible effect on black hole mass measurements since $M_{\rm BH}$ is already well constrained before using the host priors. and the its value would not change with and without those priors.

For completeness we also present in Fig. \ref{BH_only} and Table \ref{All_models_fitting} a fit where the host-galaxy is neglected and only the gravity from a SMBH is considered. The goodness-of-fit is acceptable, with $\chi^2 = 18.56$ and $\chi^2_{\rm d.o.f.} = 1.86$, but Fig. \ref{BH_only} shows that an extended host component is needed to model the data. However, we note that if we only consider the eight central data points, the fit is very good, with $\chi^2 = 4.54$ and $\chi^2_{\rm d.o.f.} = 0.65$, implying that those points are consistent with being located within the SoI.

As an independent test, we also carry out below the dynamical modelling for [C{~\sc ii}]-only and CO (12-11)-only dispersion points, respectively.

\subsection{[C{~\sc ii}]-only fit}\label{cii_disperison_fitting}
We carry out the same modelling by considering the gravity of both the SMBH and the host galaxy as done in the joint fit when using [C{~\sc ii}]-only dispersion points (seven cyan data points shown Fig. \ref{fig: dispersion_profile}). Using the same informative host priors, we obtain a best-fit model with $\chi^2 = 7.20$ ($\chi^2_{\rm d.o.f.} = 2.40$) (see Table \ref{All_models_fitting}), and $M_{\rm BH}$ =  $6.6^{+0.8}_{-0.7}$ $\times$ $\rm 10^{9}~\rm M_{\odot}$, as well as $M_{\rm Host}^{\rm Total} = 1.2\pm0.6$ $\times$ $\rm 10^{10}~\rm M_{\odot}$ (See Fig. \ref{model_plot_only_cii} and Fig.\ref{model_plot_only_cii_dis}).

\subsection{CO (12-11)-only fit}\label{co_disperison_fitting}
Since all four CO (12‑11) dispersion data points are located on the rising part of the dispersion profile shown in Fig. \ref{fig: dispersion_profile}, we fit them only considering the gravity from a SMBH. The goodness-of-fit is good with $\chi^2 = 0.44$ and $\chi^2_{\rm d.o.f.} = 0.15$ (see Table \ref{All_models_fitting}).
The resulting black hole mass is $M_{\rm BH}$ =  $6.9\pm0.8$ $\times$ $\rm 10^{9}~\rm M_{\odot}$ (See Fig. \ref{model_plot_only_co}), highly consistent with measurements from the joint and [C{~\sc ii}]‑only fit.

\section{Robustness of the SMBH mass measurements}\label{robustness}
We consider in this section alternative scenarios to explain the observed central velocity dispersion rise that do not involve the presence of an SMBH. Overall, we find that none of the models we consider here provide a satisfactory alternative to the SMBH interpretation. Table \ref{All_models_summary} summarizes the main reasons ruling out these alternative models.

\subsection{Host-only Model}\label{Host-only Model}

We study here whether a host galaxy without an SMBH can account for the central rise in the combined velocity dispersion. Firstly, we consider a host galaxy described by a single S\'ersic profile. The results are shown in the Fig. \ref{only_host_single} and Fig.~\ref{only_host_single_dis}.
We find that the best-fit values have a significantly larger $\chi^2$ of 16.08 ($\chi^2_{\rm d.o.f.} = 2.01$) compared to the model, with the residuals being most prominent at the center.
Furthermore, we note that the best-fit values of $n_{\rm Host}$ = 9.2$^{+0.6}_{-1.1}$ and $R_e^{\rm Host}$ = 0.7$^{+0.3}_{-0.1}$ kpc imply an unrealistically compact host (see Fig. \ref{only_host_single_dis}). If we apply the same priors as for our preferred model, we find a much worse fit with $\chi^2$ = 105.24 and $\chi^2_{\rm d.o.f.} = 13.15$.
We next consider whether a host described by a double S\'ersic profile (here we only apply the same priors as for our preferred model to the extended host component) can provide a good fit to the data. The results are shown in the Fig. \ref{only_host_double} and Fig. \ref{only_host_double_dis}. We find this model requires an even more compact component at the center, with $n_{\rm Host}$ = 5.8$_{-3.7}^{+3.0}$ and $R_e^{\rm Host}$ = 0.05$_{-0.03}^{+0.05}$ kpc, which is sensible as it starts approaching a central point-like mass within our spatial resolution. We find, however, that this model also results in a worse statistical description of the data, with $\chi^2$ of 10.72 and $\chi^2_{\rm d.o.f}$ of 2.14.

\subsection{Nuclear star cluster} \label{Nuclear star cluster}
We also consider whether a Milky-Way-like nuclear star cluster (NSC) could account for the central rise in the combined radial dispersion profile. We consider the model described by \cite{2014A&A...566A..47S} where the mass density profile $\rho(r)$ of the stellar mass follows a broken power law that transitions between $\rho(r) \propto$ $r^{-2}$ and $\rho(r) \propto$ $r^{-3}$ at a radius $R_{\rm c}$. We note that $R_{\rm c}$ is 5 pc for the Milky Way NSC.
To model the dispersion profile we let $R_{\rm c}$ as a free parameter. The best-fit model is shown in Fig. \ref{star_cluster}, and the best-fit parameter values are shown in Fig. \ref{star_cluster_corner} and Table \ref{All_models_fitting}. The best-fit model has a significantly larger $\chi^2$ of 10.42 ($\chi^2_{\rm d.o.f.} = 1.74$) than the preferred model with an SMBH. Moreover, the best-fit model has an NSC mass of $0.3 \pm 0.1$ $\times$ $\rm 10^{10}~\rm M_{\odot}$, which is larger than that of the most massive NSCs known at $\rm 10^{9}~\rm M_{\odot}$, and much larger than their typical mass of $\rm 10^{7}~\rm M_{\odot}$ \cite{2016MNRAS.457.2122G,2024Natur.632..513A}.

\subsection{Importance of the innermost velocity dispersion measurement} \label{Importance of the innermost velocity dispersion measurement}
We have argued that the presence of the SMBH is necessary to reproduce the central rise of the host combined radial velocity dispersion profile, which is obtained by removing the contribution of outflows to the [C~{\sc ii}] and CO (12-11) line distribution (see Section \ref{spectra_fitting}). However, as discussed in Section \ref{spectra_fitting} and shown Fig. \ref{cii_ring_fitting_uniform} and Fig.~\ref{CO_ring_fitting1}, we are unable to identify a contribution from outflows to the line profile in the central beams of the uniform-weighting [C~{\sc ii}] and CO (12-11) data cubes. This is in contrast to the procedure in all other rings, where blue and/or red outflows are detected and disentangled from the host velocity dispersion. If unrecognized outflows were significant in those central beam profiles, they could lead us to an overestimation of the host velocity dispersion in that region. In this section we study the robustness of our results on those data points by masking them and re-fitting our dynamical models. We summarize the best-fit results in Table \ref{All_models_fitting_no_central_beam}.

The Fig. \ref{no_central_beam_model_BH+host} and Fig. \ref{no_central_beam_model_BH+host_dis} show the results for our preferred model, which consists of an SMBH and a host galaxy described by a single S\'ersic profile. We find that removing the central points of the velocity dispersion profile results in values of $M_{\rm BH}$ and $M_{\rm Host}^{\rm Total}$ that are consistent within 1$\sigma$ of those obtained by fitting all the velocity dispersion data points, albeit with much larger error bars.

Moreover, we note that a host-only model with a single S\'ersic profile (Fig. \ref{no_central_beam_model_only_host} and Fig.~\ref{no_central_beam_model_only_host_dis}) results in a larger $\chi^{2}$ and $\chi_{\rm d.o.f}^{2}$ compared to when including the SMBH component, with the residuals being most prominent at the center (see Fig. \ref{no_central_beam_model_only_host}), and an unrealistically compact host with $n_{\rm Host}$ = 8.5$^{+1.1}_{-1.8}$, indicating that the presence of the latter is statistically and physically disfavored even when removing the central beam.

\section{Host galaxy mass issues} \label{Host galaxy mass issues}
As discussed in the main text and in Section Section~\ref{dispersion fitting}, our dynamical modeling of the [C{~\sc ii}] kinematics finds a total host mass of $M_{\rm Host}^{\rm Total} =  $~1.3$^{+0.6}_{-0.5}$ $\times$ $\rm 10^{10}~\rm M_{\odot}$. This value is consistent with those estimated from the dust continuum emission and the CO (2-1) observations in Section \ref{Gas mass estimates}, which would in principle suggest a large gas fraction. However, our estimate of $M_{\rm Host}^{\rm Total}$ is below the stellar mass estimate of 2--30 $\times$ 10$^{10}~\rm M_{\odot}$ discussed in Section \ref{stellar mass}, where this broad range is estimated from a combination of our modeling of JWST NIRSPEC and MIRI observations with the pPXF code and the analysis of \cite{2018Sci...362.1034D}. In this section we discuss the source of this seeming discrepancy. An important aspect to keep in mind throughout this discussion is that the spatial resolution of the different data sets used here vary significantly, making it difficult to connect all the relevant scales. The PSFs of the HST (FWHM of 0.15$^{\prime\prime}$), JWST NIRSpec (0.20$^{\prime\prime}$), and MIRI (0.207$^{\prime\prime}$) data are only a factor of $\sim$2 smaller than the 0.46$^{\prime\prime}$ diameter of the area probed by our high spatial resolution ALMA observations and, perhaps more importantly, are comparable to our estimated diameter of the SoI (0.42$_{-0.12}^{+0.60}$$^{\prime\prime}$).

The first issue we consider is that our ALMA observations may not be a good probe of the total host mass. There are reasons that the central mass distribution within a few hundred pc of the galaxy center could be hollowed out compared to the outskirts. Such effects in massive BCGs are expected from SMBH mergers, which can expel a large fraction of the stars in the nuclear regions \cite{2016ApJ...833..168M,2013ApJ...773..100K,2018ApJ...864..113R}, and \cite{2018Sci...362.1034D} has shown that there is significant merger activity in W2246--0526. It is also possible that a significant fraction of the stellar mass is in a flatter S\'ersic index component, such as a disk, whose contribution is more important in the UV/NIR imaging/IFU scales than in the scales of our ALMA data. Note that we can obtain a good fit to the [C{~\sc ii}] dispersion profile and match the nominal stellar mass of 2.5 $\times$ 10$^{11}$ $\rm M_{\odot}$ from \cite{2018Sci...362.1034D} as long as we allow for a larger $R_{\rm e}$ of about 8 kpc (see Fig. \ref{BH_host_fix_host} and Fig. \ref{fix_tanio_mass_corner}).
While that would be inconsistent with the $R_{\rm e}$ measured in the space-based HST ($R_{\rm e}$ = 1.12 kpc, \cite{2016ApJ...816L...6D}) and JWST imaging ($R_{\rm e}$ = 1.5 kpc, see Section 8.1), it highlights that there is room within the parameters assumed to significantly increase the best-fit host mass from the dynamical analysis.

Additionally, there are two other important aspects that could affect our estimates of the host mass. One is the possibility of a significant rotational component in (or close to) the plane of the sky that is neglected in our analysis. Additionally, the potential offset suggested by the JWST observations (see Section \ref{Offset stellar distribution}) between the center of the host galaxy and the position of the SMBH can also have an important effect. Below we explore each of these cases separately, and show that while each of them could account for at least part of the observed discrepancy in the host mass, neither has a significant effect over the estimated SMBH mass (see Table \ref{All_models_fitting}), highlighting the robustness of our measurement.

\subsection{Rotation} \label{Rotation}
As mentioned earlier, it is possible that there could be a substantial rotational component to the [C~{\sc ii}] kinematics as long as it is in, or close to, the plane of the sky. Neglecting this component could result in a significantly underestimate of the host mass in our dynamical analysis. To gain insights into such a scenario, we consider a simple model in which the ratio of the rotational velocity ($V_{\rm Rot}$) to the dispersion velocity component caused by the host galaxy itself ($\sigma_{\rm Rot}$), usually denoted using the letter $Q$, is constant throughout the galaxy. We can then rewrite equation \ref{dispersion model} as:

\begin{equation}\label{dispersion model_new}
    \sigma_{\rm Total} (r) = \sqrt{\sigma_{\rm BH}^2 (r) + \sigma_{\rm Host}^2 (r)}, 
\end{equation}

\noindent where $\sigma_{\rm Total} (r)$ is the observed velocity dispersion at distance $r$ from the SMBH, $\sigma_{\rm BH}(r)$ is the dispersion caused by the SMBH gravity and $\sigma_{\rm Host}(r)$ is that caused by the host gravity. We can then rewrite this equation as:

\begin{equation}\label{rotate_dispersion}
    \sigma(r) = \sqrt{\sigma_{\rm BH}^2 (r) + \left( \frac{V_{\rm Rot} (r)}{Q} \right)^2} = \sqrt{\frac{2}{3} \, \frac{GM_{\rm BH}}{r} + \left( \frac{1}{Q} \right)^2 \frac{GM_{\rm Host}(< r)}{r}}.
\end{equation}

\noindent Comparing equation S1 with equation \ref{rotate_dispersion}, it is clear that the shape of the best-fit model is not be affected, with the only difference being that the host mass would instead be $M_{\rm Host}^{\rm Total}$ = 1.3 $\times$ 10$^{10}$ $\rm M_{\rm \odot}$ $\times$ 2 Q$^{2}$ / 3. Matching the stellar mass range of 2 $\times$ 10$^{10}$ $\rm M_{\odot}$ to $3$ $\times$ $\rm 10^{11}~\rm M_{\odot}$ would require $Q$ values in the range of 1.5--5.7. Since $\sigma_{\rm Host}$ flattens out at about 120 km s$^{-1}$ (see the bottom panel of Fig. 2), the rotational velocity in the outskirts would then be in the range of 180 to 680 km s$^{-1}$. Since \cite{2016ApJ...816L...6D} detected a potential rotational component in the galaxy outskirts of $\sim$100 km s$^{-1}$, this would imply an inclination between 9 and 34 degrees, very close to the plane of the sky as expected. Although this approximation is quite coarse, it suggests a missing rotation component due to projection effects could be a feasible explanation for the discrepancy between the dynamical host mass and stellar mass estimates.

\subsection{Offset stellar distribution} \label{Offset stellar distribution}

As discussed in Section \ref{JWST observations}, our morphological decomposition of the emission in the JWST/MIRI F560W observations suggests there could be an offset of $\sim$1 kpc (1.3 pixels) between the AGN and the geometric center of light of the host. If this offset is real and corresponds to an offset in the mass distribution (i.e., not just in the light distribution due to differential reddening), then the dynamical host mass obtained in Section \ref{dispersion fitting} could be underestimated.

Following a similar analysis to that carried out in Section \ref{dispersion fitting}, the intrinsic velocity dispersion at a given distance from the center of stellar mass distribution, $r_{\rm g}$, would be given by 
\begin{equation}
    \sigma_{\rm Host}(r_{\rm g}) = \sqrt{\frac{2}{3}~\frac{G [M_{\rm Host}(<r_{\rm g})]}{r_{\rm g}}}, 
\end{equation}

Equation S6 can be further written at a given distance from the center of SMBH, $r$, as:

\begin{equation}
    \sigma_{\rm Host}(r, \theta_{\rm 1}) = \sqrt{\frac{2}{3}~\frac{G [M_{\rm Host}(<r_{\rm g}[r, \theta_{\rm 1}])]}{r_{\rm g}[r, \theta_{\rm 1}]}}. 
\end{equation}

\noindent Note that $r_{\rm g} = (r^2 + r^{2}_{\rm offset} - 2r r_{\rm offset}\cos{\theta_{\rm 1}})^{1/2}$, where $r_{\rm offset}=1~\rm kpc$ is the offset between the peak of stellar emission and the quasar position
in W2246--0526 and $\theta_{\rm 1}$ is the angle between the $r$ and $r_{\rm offset}$ vectors. The host-stellar component is assumed to follow a S\'ersic profile with index $n_{\rm Host} = 1.64$ and effective radius $R_e^{\rm Host} = 1.5$ kpc to match the morphology of the JWST F560W imaging. Its total mass $M_{\rm Host}^{\rm Total}$ is left as a free parameter. We modify appropriately equations \ref{intrisic spectrum}--\ref{ring spectrum} to carry out the angular integration in $\theta_{\rm 1}$ and compute the line velocity profile in each ring.

We show in Fig. \ref{offset_comparisons} that the $\sim$1 kpc offset to the center of the host galaxy would result in a nearly flat velocity dispersion profile for the host galaxy. This contrasts with the centrally decreasing profile we would expect in the absence of an offset. Assuming the outmost velocity dispersion profile points in Fig. 2 are dominated by the host galaxy, the best-fit host mass is $M_{\rm Host}^{\rm Total}$ $=$ 3.8$_{-0.7}^{+0.8}$ $\times$ 10$^{10}$ $\rm M_{\odot}$ (see Fig. \ref{offset_host}), which is consistent with the lower bound to the stellar mass obtained from the pPXF fit to the NIRSPEC and MIRI observations. Furthermore, a nearly flat velocity dispersion profile cannot account for the centrally enhanced velocity dispersion, and still requires an SMBH with a mass of {$5.9^{+1.9}_{-1.7}$ $\times$ 10$^{9}$ $\rm M_{\odot}$} to account for the central velocity dispersion. This is nearly the same mass as found in Section \ref{dispersion fitting}. Note that, unfortunately, a fit of this scenario that uses all the observed data points and takes into account the potential offset in the mass centers is not possible in the framework we have developed, and would require detailed orbit simulations that are well beyond the scope of this work. For the same reasons, we note that the concept of a SoI becomes ill-defined in this scenario. However, this analysis shows that the potential offset between the SMBH position and the center of the host galaxy could account for at least part of the discrepancy between our dynamical host mass and the stellar mass estimates.

\section{Nuclear mass contributions}\label{enclosed mass}

In Section \ref{Host galaxy mass issues}, we discussed possible explanations for the discrepancy between the total dynamical host mass ($M_{\rm Host}^{\rm Total} = $~1.3$^{+0.6}_{-0.5}$ $\times$ $\rm 10^{10}~\rm M_{\odot}$) and  the total stellar mass range (2--30 $\times$ 10$^{10}~\rm M_{\odot}$) estimated by JWST observations and \cite{2018Sci...362.1034D} (see Section \ref{stellar mass}). Importantly, this discrepancy does not affect the robustness of our dynamical SMBH mass measurement. Here we focus on the enclosed mass profiles within the region directly probed by our ALMA [C~{\sc ii}] and CO (12-11) dispersion profile (up to 1.5 kpc/$0.23''$), which are directly relevant for constraining the mass of the SMBH.

Fig.~\ref{mass_budget} presents the enclosed mass as a function of radius for different components: (i) the dynamical SMBH and the host from our preferred model (Section~\ref{joint_disperison_fitting} and Table~\ref{All_models_fitting}); (ii) the stellar mass derived from the JWST/MIRI F560W light profile ($R_e = 1.5\pm0.13$ kpc, $n=1.64\pm0.35$, with a total stellar mass of $2\times10^{10}\,\mathrm{M}_{\odot}$); and (iii) the dust‑based gas mass for optically thick dust cases based on our high‑resolution ALMA dust continuum observations (Section \ref{dust mass}). Within 1.5 kpc/$0.23''$, the enclosed stellar mass profile is comparable to that of the dynamical host, but the enclosed gas mass profile lies significantly above the dynamical host profile. This mismatch may raise a concern that our dynamical model underestimates the host mass, which would then lead to an overestimate of the SMBH mass and SoI size. However, as we demonstrate below, simply increasing the total host mass in the dynamical modelling cannot reproduce the observed velocity dispersion profile. A more plausible explanation is that the ALMA high-resolution [C~{\sc ii}] and CO (12-11) dispersion profile do not fully trace all nuclear gas components due to the projection effect. In particular, a substantial gas component rotates in or near the plane of the sky would contribute little to the line‑of‑sight velocity dispersion (see Section~\ref{Rotation}). Such a component could be present in the gas emission but remain kinematically invisible in our dispersion measurements.

To test whether a larger host mass could be accommodated, we repeat the dynamical modelling with an informative prior on the total host mass. Specifically, we fix the total host mass with $4.5\times10^{10}\,\mathrm{M}_{\odot}$, which is the sum of the dust‑based total gas mass ($2.5\times10^{10}\,\mathrm{M}_{\odot}$, scaled to match the observed nuclear gas mass profile assuming the JWST/MIRI Sérsic light distribution) and the JWST stellar mass ($2\times10^{10}\,\mathrm{M}_{\odot}$). The host mass profile is kept fixed to the JWST/MIRI Sérsic parameters. This forced fit yields a smaller SMBH mass of $3.3\pm0.4 \times10^{9}\,\mathrm{M}_{\odot}$, however, provides a very poor fit ($\chi^2$ = 40.44, $\chi^2_{\rm d.o.f}$ = 4.04; also see Table \ref{All_models_fitting}) to the observed dispersion profile (Fig. \ref{fix_host45}) as the model systematically overpredicts the dispersion at large radii (see the bottom panel of Fig. \ref{fix_host45}).

If we assume the best-fit host dynamical model underestimate the host mass, and we replace our best-fit host dynamical model with a host mass profile built from the combined dust-gas mass and JWST stellar mass estimates, as shown in Fig.~\ref{mass_budget}, the black hole SoI diameter could be as small as $765_{-78}^{+85}$ pc 
($0.117_{-0.012}^{+0.013}$$^{\prime\prime}$) 
corresponding optical thick dust for dust-gas mass estimates (see Section \ref{dust mass}). Even in this extreme scenario the SoI is still resolved by our observations.

\section{Radiation pressure and turbulence} \label{offset-stellar}
Throughout the article we have only considered gravitational origins to the observed [C{~\sc ii}] and CO (12-11) kinematics in W2246--0526. However, given the extreme luminosity of the source, it is reasonable to consider whether radiation pressure could be adding a significant additional force that we are not taking into account. Additionally, \cite{2025ApJ...989..230V} has shown the existence of powerful ionized gas outflows in W2246--0526, and their interaction with the ISM could be adding turbulence to the [C{~\sc ii}] kinematics. Below we discuss each case separately and why they are unlikely to play a significant role in the observed dynamics.

\subsection{Radiation Pressure}

Radiation pressure from the photons emitted by the accretion disk would act opposite to the gravitational pull of the SMBH and the host galaxy. If significant but not dominant, radiation pressure could lead to an underestimate of the SMBH and host masses. If it is dominant, however, it could lead to the gas being unbound to the system.

As shown in \cite{2000ApJ...533..631K} (see their Section 9.4), for optically thin gas surrounding a luminous quasar, the ratio of radiative to gravitational acceleration $g_{\rm rad}/g_{\rm G}$ can be significantly greater than unity, as the radiation reaches all gas particles and effectively transfers momentum through bound-bound and bound-free transitions (see section 2.9 of \cite{2008NewAR..52..257N}), implying gas would be unbound. However, the efficiency of radiative acceleration is greatly reduced if the surrounding gas is optically thick. In this scenario, the dominant mechanism for radiation pressure becomes the absorption of ionizing photons by neutral hydrogen, which are quickly absorbed and do not reach most gas particles. Radiation pressure in such a scenario is only important in the illuminated surface of the gas. A detailed treatment of this can be seen in Section 5.9.2 of \cite{2013peag.book.....N}. 

As discussed in the main text, \cite{2024A&A...682A.166F} showed using a multitude of far-IR line ratios that its nuclear gas density should have a $N_{\rm H}$ of at least $10^{24}~\rm cm^{-2}$, Compton-thick in the nuclear region of W2246-0526.
Therefore, we may only expect radiation pressure to be significant in the innermost regions of the quasar, within the torus-like scales (10s of pc assuming equation 1 of \cite{2008ApJ...685..160N}), but not to reach the scales of 100s pc where we measure the gas dynamics through [C~{\sc ii}] and CO (12-11). A similar scenario could occur with radiation pressure on the dust grains, which would be dominated by absorption from the UV/optical emission of the accretion disk. However, as the large amount of dust in the nuclear region, associated to the high gas column densities discussed above and to a torus covering fraction close to unity (see discussion in \cite{2015ApJ...805...90T}), the majority of that radiation is unlikely to reach dust in sufficiently large scales so as to affect the gas dynamics measured through [C~{\sc ii}] and CO (12-11).

Note, radiation pressure is likely important in the innermost regions of the quasar and likely contributes to the launch of nuclear winds that may turn into the large scale [O~{\sc iii}] and dusty outflows observed in this source (see \cite{2025ApJ...989..230V,2025arXiv251009870V}).

\subsection{Turbulence}\label{turbulence}

A contribution from turbulence to our dispersion profile could result in the masses of both the SMBH and the host being misestimated. However, there are some reasons to think that these are not dominant here.

From an empirical point of view, if the turbulence was dominant, it would be difficult to explain the consistency between the dynamical estimate of the SMBH mass and that obtained from the broad Mg~{\sc ii} emission line using the single-epoch method, as these are independent from each other while we acknowledge the latter has large errorbars. It could explain why our host mass estimate is seemingly inconsistent with that estimated from other indicators although in Section \ref{Host galaxy mass issues} we discuss other plausible reasons for the mismatch.

If turbulence was adding a sub-dominant contribution to the dispersion, then the SMBH and host masses could be both overestimated. The effects would be stronger, however, for the host mass, as it comes from scales where the gravitational pull is weaker, and hence the gravitationally driven dispersion would be smaller. This would not considerably affect our conclusions regarding the SMBH mass, but would increase the tension between the different host mass estimates as discussed in Section \ref{Host galaxy mass issues}.

To explore the potential effects of turbulence within beam-like scales and spatial regions  dominated by the SMBH gravity of the combined dispersion profile of [C{~\sc ii}] and CO (12-11) in SMBH mass estimate (see Section \ref{dispersion fitting}), we also model the moment 0 of the CO (12-11) emission using a S\'ersic profile with the public code PySerisc \cite{Pasha2023}, since turbulence dominated regions tend to have very flat distributions. For example, systems with turbulent gas driven by mergers or other external factors typically has S\'ersic indices of ~0.5-1 \cite{2014ApJ...781...11E}, while central gas turbulence driven by AGN outflows is observed to be distributed on a disk that has a S\'ersic index of 1 within a few hundred pc \cite{2024A&A...684A..24P}. Furthermore, strong AGN-feedback could result in a sharp decrease in the central molecular gas concentration and flat or inverted radial profiles \cite{2024A&A...689A.347G}. In contrast, dispersion dominated gas tends to have much larger S\'ersic index (canonically closer to 4).
Using the CO (12-11) data cube obtained with the robust = 0.5 weighting (as close to uniform weighting as we can go to fit the source morphology given the SNR of the observations) we find a S\'ersic index of 3.69$^{+0.55}_{-0.50}$ (see Fig. \ref{LF_co}), which strongly supports the gas being dispersion supported. A consistent index of 3.72$^{+0.45}_{-0.39}$ (see Fig. \ref{LF_cii}) is found when we use the [C~{\sc ii}] data cube with uniform-weighting imaging, which has a slightly lower spatial resolution than CO with robust of 0.5 (geometric average angular resolution: 0.069$^{\prime\prime}$ vs 0.058$^{\prime\prime}$).

Therefore, for all these reasons, we consider it is unlikely that turbulence is affecting the dynamical measurement of the SMBH mass presented in Section \ref{dispersion fitting}. Hence, neglecting its contribution does not qualitatively affect our conclusions.

\clearpage

\begin{figure} [htp!]
\centering
\includegraphics[width=1.\textwidth]
    {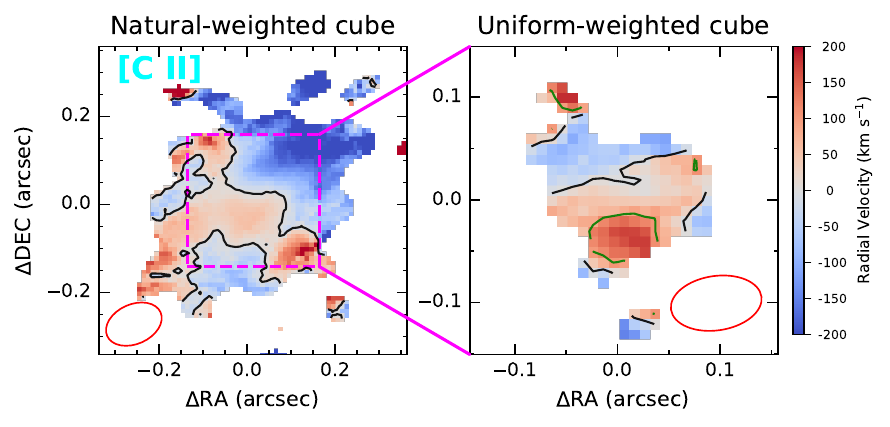}
\includegraphics[width=1.\textwidth]
{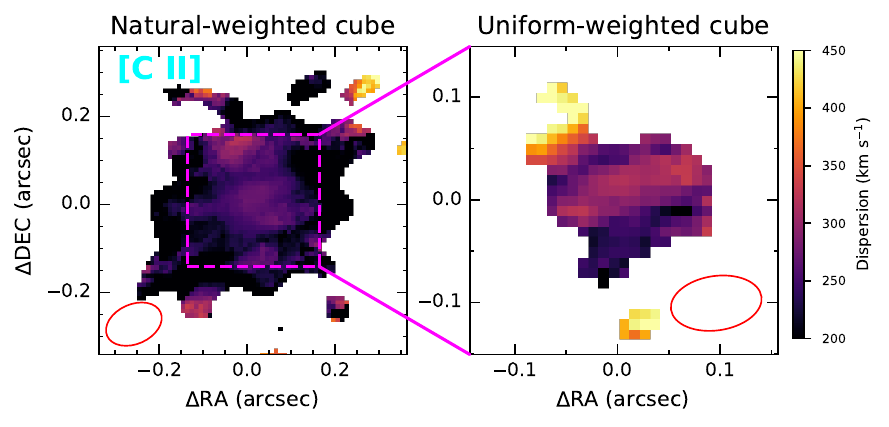}
    
\end{figure}

\afterpage{
  \begin{figure}[htp!]
    \caption[]{The first and second moment maps (referred as M1 and M2 maps) collapsed from the two data cubes of [C{~\sc ii}] (see Table \ref{cube_info}). The synthesized beams are shown as an ellipse in the lower left corner of each map. All the moments are obtained using the continuum-subtracted cubes and clipping at 1$\sigma$. We only show pixels detected at $\ge$ 3$\sigma$ detection in the zeroth moment (see Fig. \ref{M0}). The top two panels use contours to highlight velocity structures along three different directions, and locating at the south direction, respectively (see Section \ref{channels maps} for details).} 
\label{cii_M1_M2}
  \end{figure}
}

\clearpage
\clearpage
\begin{figure} [htp!]
\centering
\includegraphics[width=1\textwidth]
    {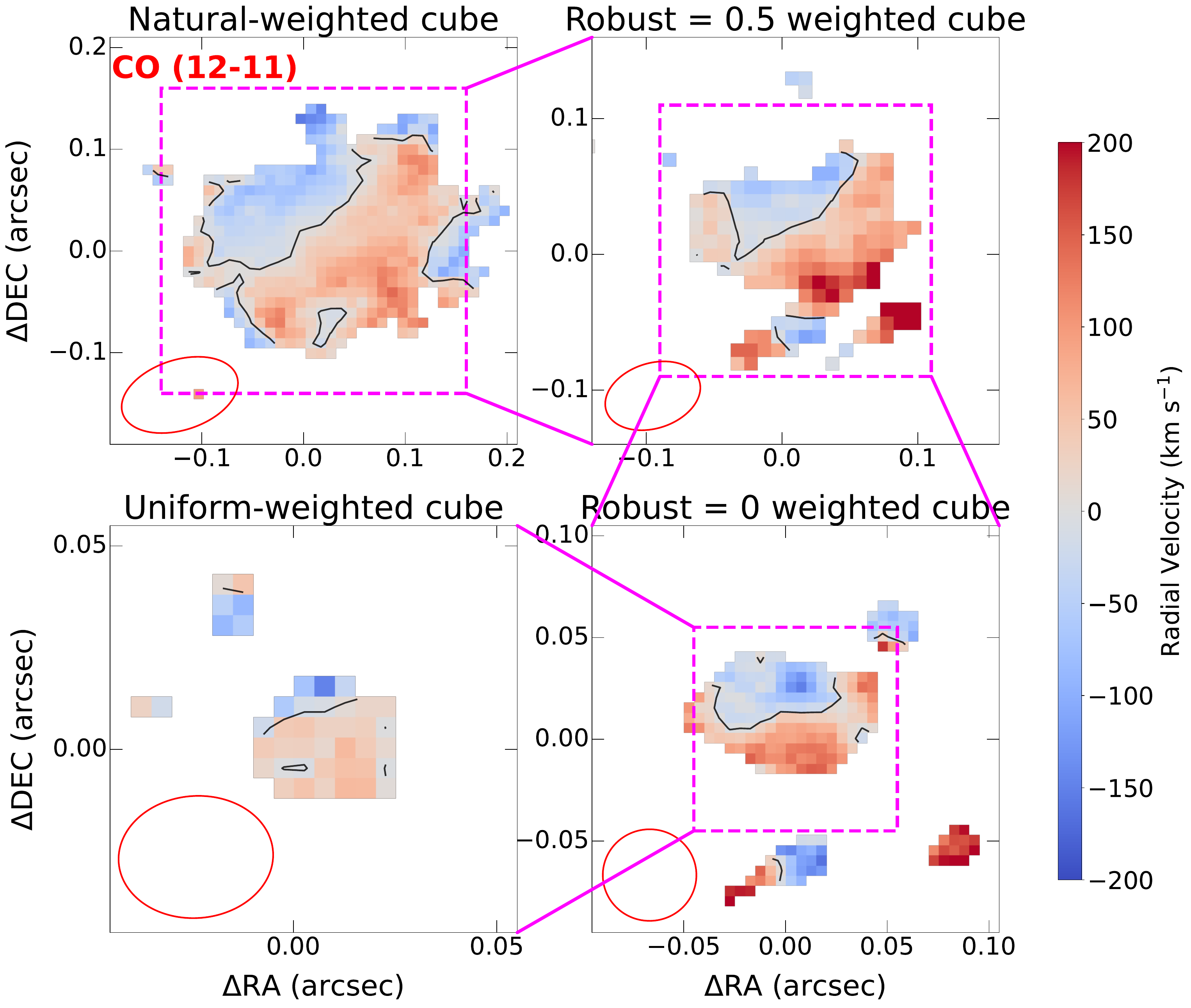}
  \caption[]{Same as Figure \ref{cii_M1_M2}, but for the four data cubes of CO (12-11) M1 maps. See details in Sections \ref{ALMA_data_reduction} and 9.1.}\label{CO_M1}
\end{figure}

\begin{figure} [htp!]
\centering
\includegraphics[width=1\textwidth]
    {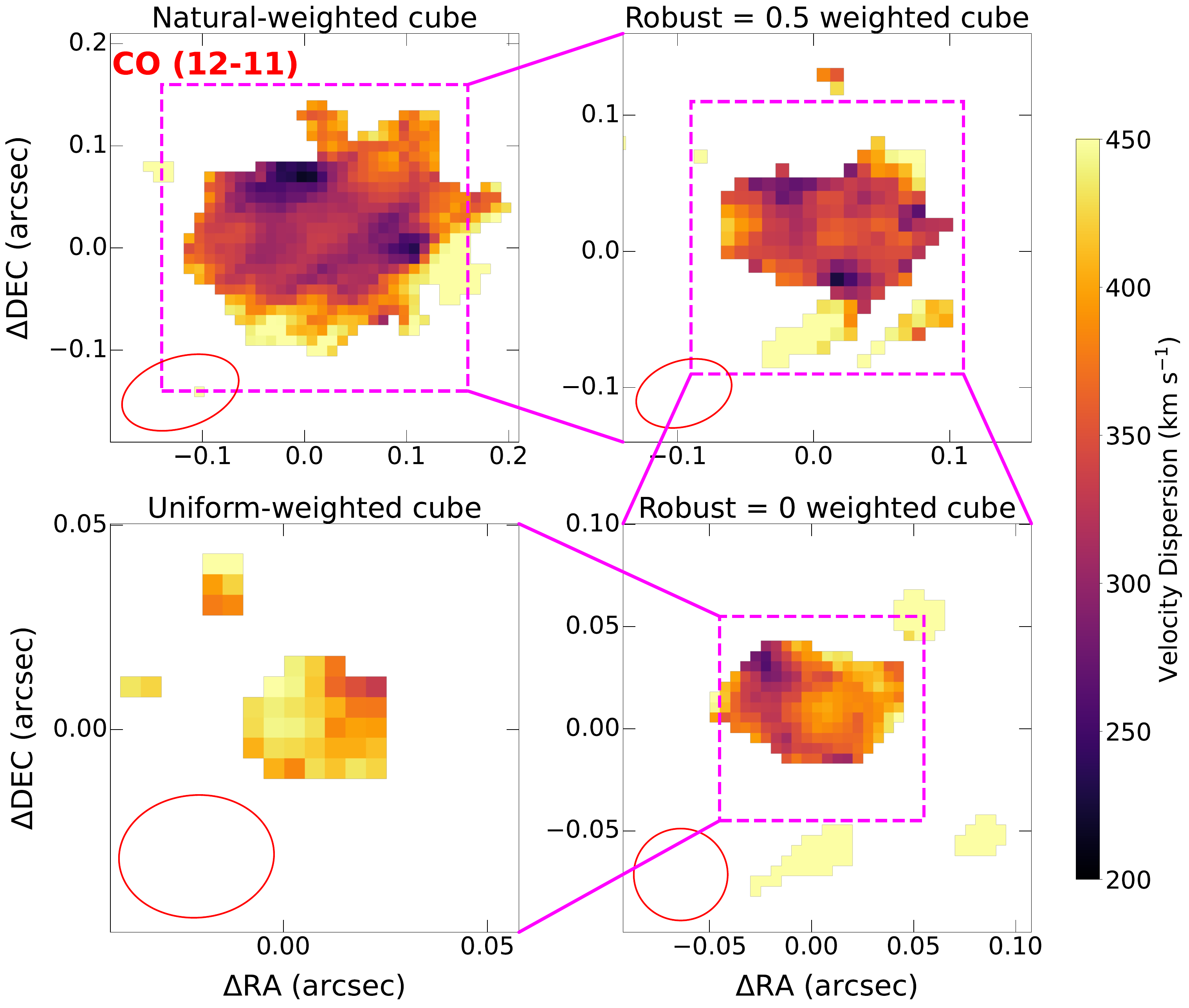}
\caption[]{Same as Figure \ref{cii_M1_M2}, but for the four data cubes of CO (12-11) M2 maps. See details in Sections \ref{ALMA_data_reduction} and 9.1.}\label{CO_M2}
\end{figure}

\begin{figure*}[h]
\centering
	\includegraphics[width=1.\columnwidth] {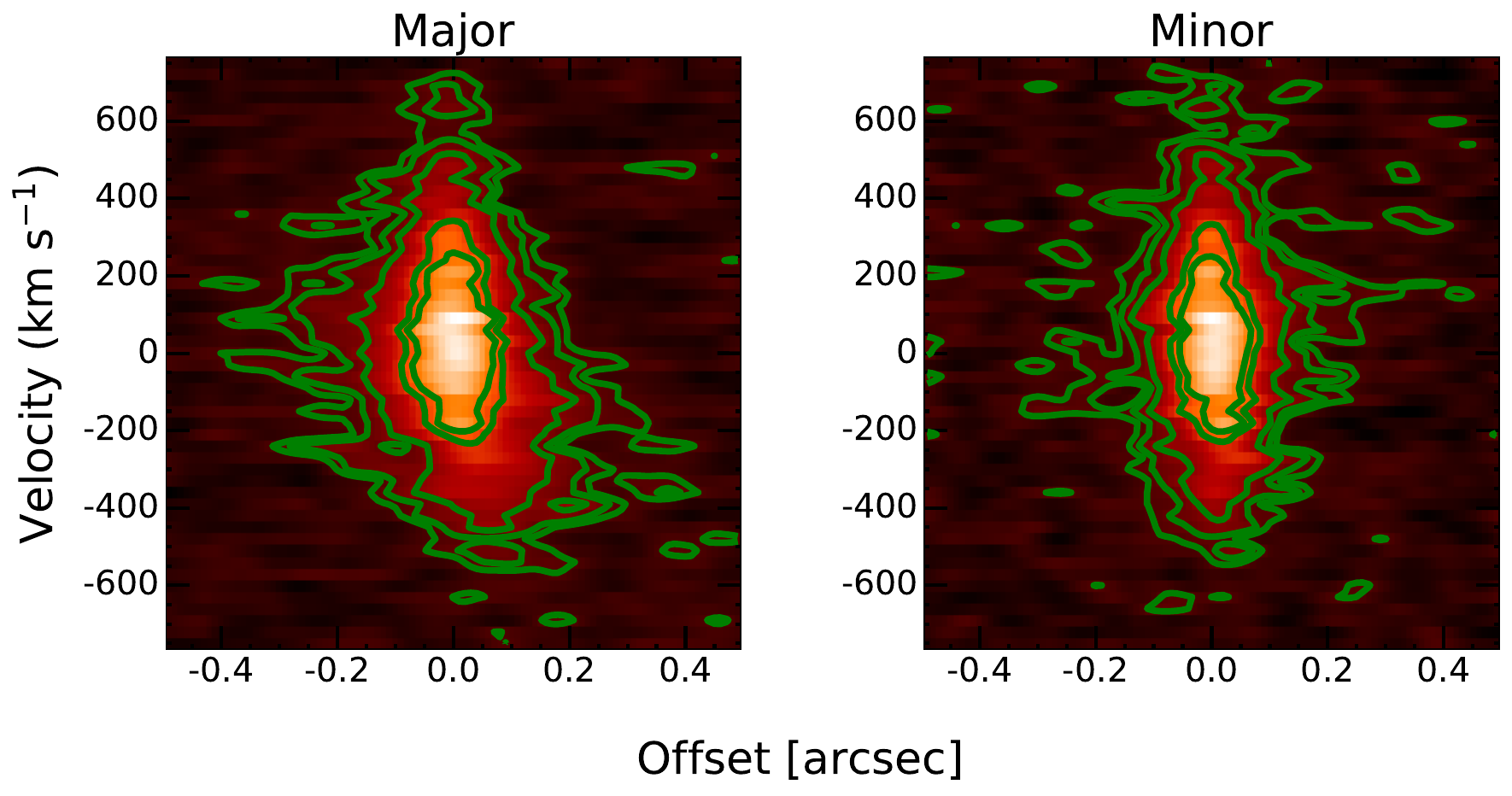}
	\caption{PV diagrams of the [C{~\sc ii}] emission in the natural-weighted cube along the major axis (PA = 138$^{\circ}$, $\it left$) and minor axis (PA = 228$^{\circ}$, $\it right$), where the axis directions are based on the 2D elliptical Gaussian fits to the integrated intensity (zeroth moment) map discussed in Section \ref{moments}. These diagrams are extracted over a 10-pixel (one beam size) and 13-pixel extraction width, respectively, where this ratio corresponds to the axis ratio determined from the 2D Gaussian fitting. This ensures we capture the same kinematic structures along both axes. The green contours are drawn at (2, 3, 5, 10, 12)$\sigma$ .}
\label{fig:PV}
\end{figure*} 

\begin{figure*}[h]
\centering
	\includegraphics[width=1.\columnwidth]{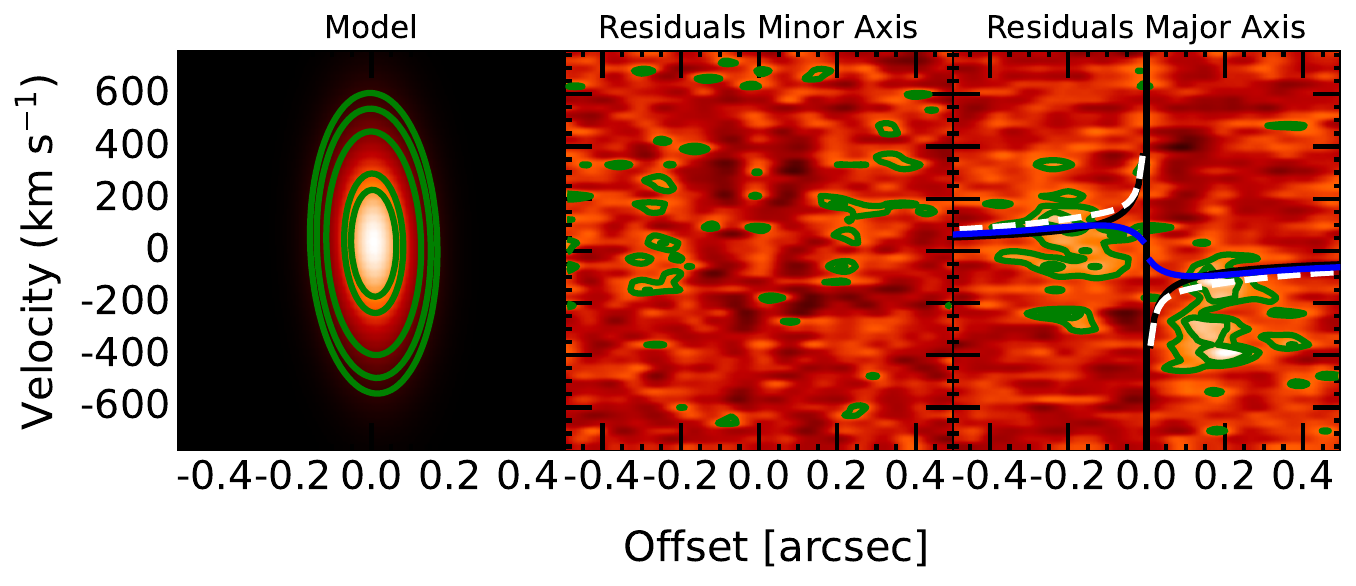}
	\caption{The 2D Gaussian model to the minor axis PV diagram ($\it left$), the residual emission after subtracting this model along the minor axis ($\it middle$) and along the major axis ($\it right$). Green contours are drawn at (2, 3, 5, 10, 12)$\sigma$. The black and blue lines in the right panel show the rotation curves predicted from
the SMBH ($M_{\rm BH} = 4.0 \times 10^{9}~\rm M_{\odot}$ \cite{2018ApJ...868...15T}) and the stellar distribution ($M_{\star} = 2.5 \times 10^{11} \rm M_{\odot}$ within a 2$^{\prime\prime}$ radius, assuming an effective radius $r_{e}$ = 1 kpc and Sersic index $n$ = 3 \cite{2018Sci...362.1034D}), respectively. The white dashed lines represent the sum of the two components.} 
\label{fig:2D}
\end{figure*} 

\begin{figure*}[h]
\centering
	\includegraphics[width=1.\columnwidth]
     {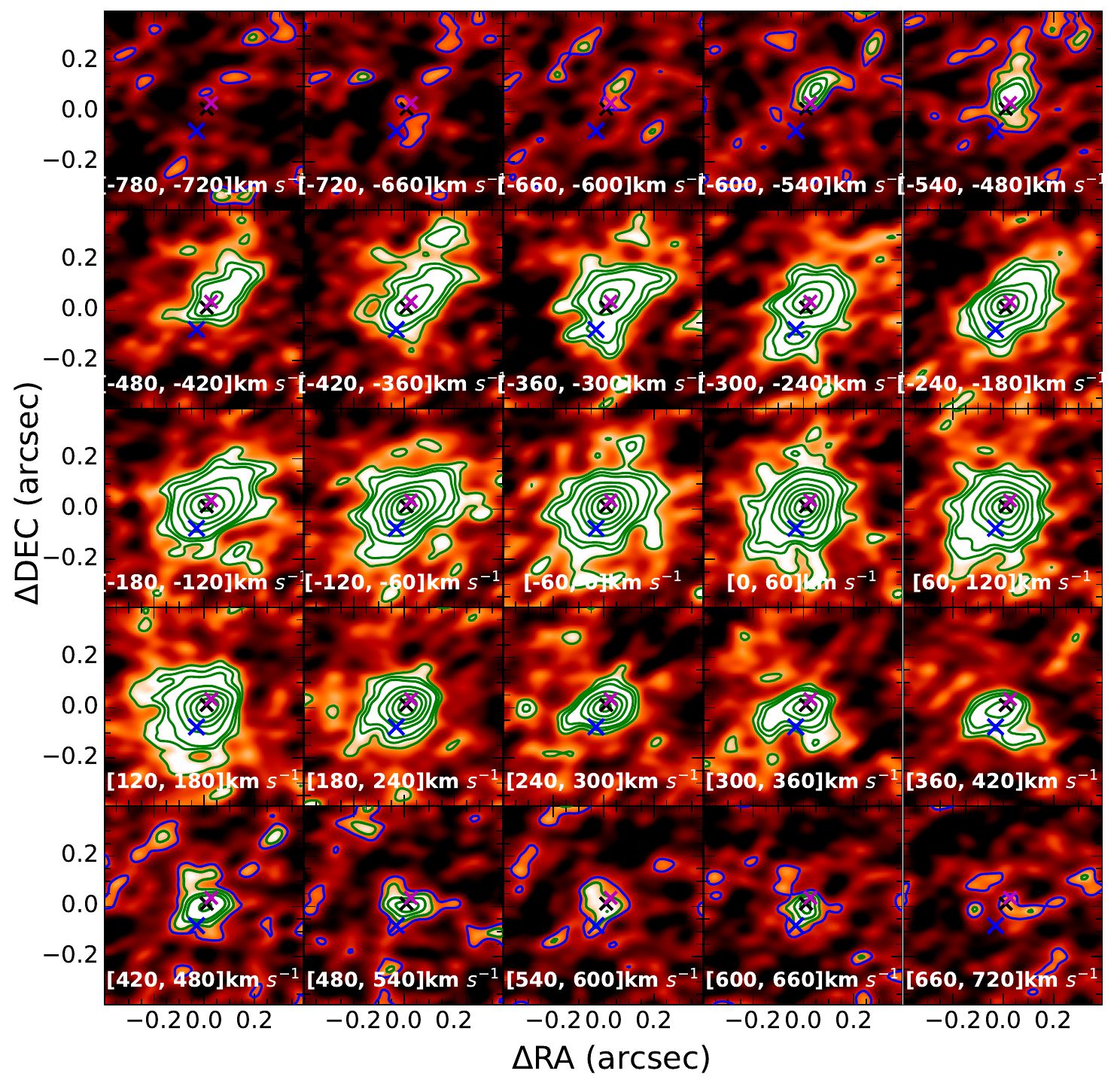}
	\caption{Channel maps in bins of 60 $\rm km~s^{-1}$ derived from [C~{\sc ii}] natural-weighted cube. In each map, the overlaid (blue) green contours show the levels of ((2) 3, 4, 5, 8, 11, 14, 17, 20)$\sigma$. The black crosses mark the position of [C~{\sc ii}] central peak of the integrated emission (M0) map, which coincide with the JWST quasar position (magenta crosses) on JWST/MIRI F160W image within the uncertainties (see Section \ref{JWST observations} for details), while blue crosses mark the position of JWST stellar-light center.}
\label{fig:cii_channel_maps}
\end{figure*}

\begin{figure*}[h]
\centering
	\includegraphics[width=1\columnwidth]
     {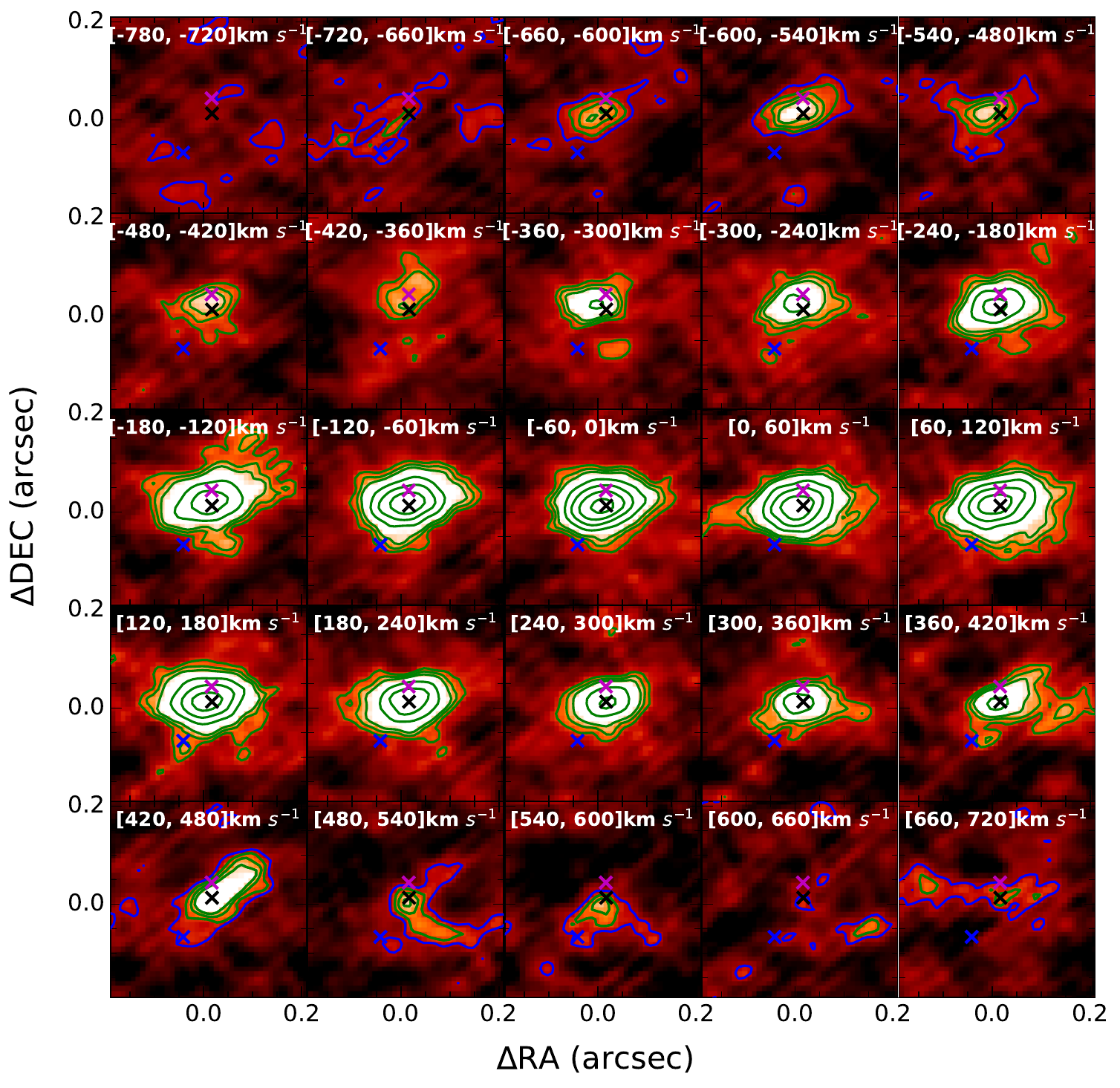}
	\caption{Same as Figure \ref{fig:cii_channel_maps}, but for the CO (12-11) natural-weighted cube.}
\label{CO_channel_maps}
\end{figure*} 

\begin{figure}[h]
\centering
 \includegraphics[width=1\columnwidth]{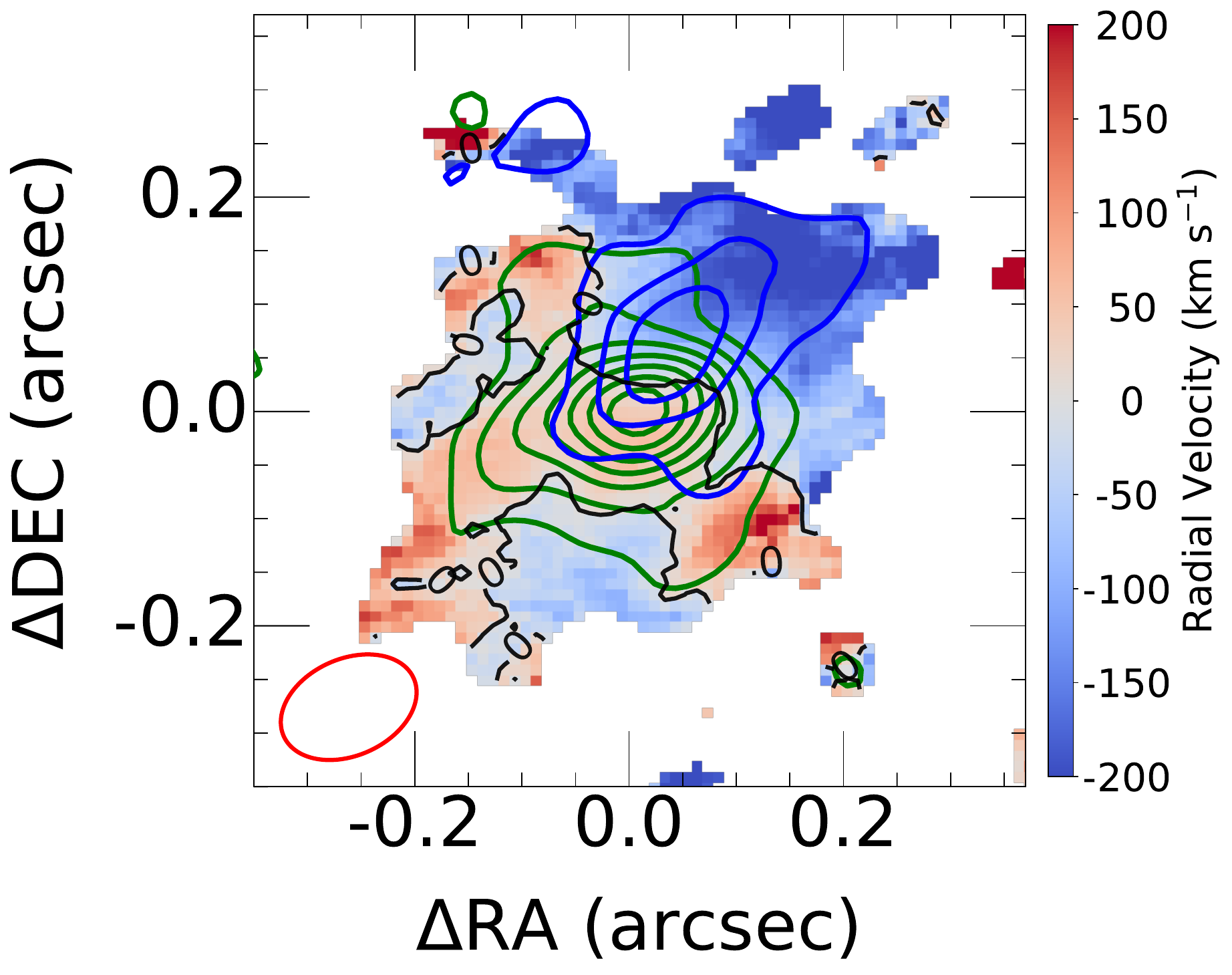}
	\caption{First moment overlays with the [C{~\sc ii}] emission for the natural-weighted cube at the high velocity, where the blue contours illustrate emission within [-660, -420] $\rm km~s^{-1}$ while green contours are for [240, 720] $\rm km~s^{-1}$ . The contours are plotted at (3, 5, 7, 9, 11, 13, 15)$\sigma$ with $\sigma$ = 0.025 (0.039) Jy $\rm km~s^{-1} beam^{-1}$ for the former (latter). Black contours at 0 $\rm km~s^{-1}$ highlight the velocity gradients along three directions.}
\label{fig: outflows with the first moment}
\end{figure} 

\begin{figure*}
\centering
\includegraphics[width=0.45\textwidth]{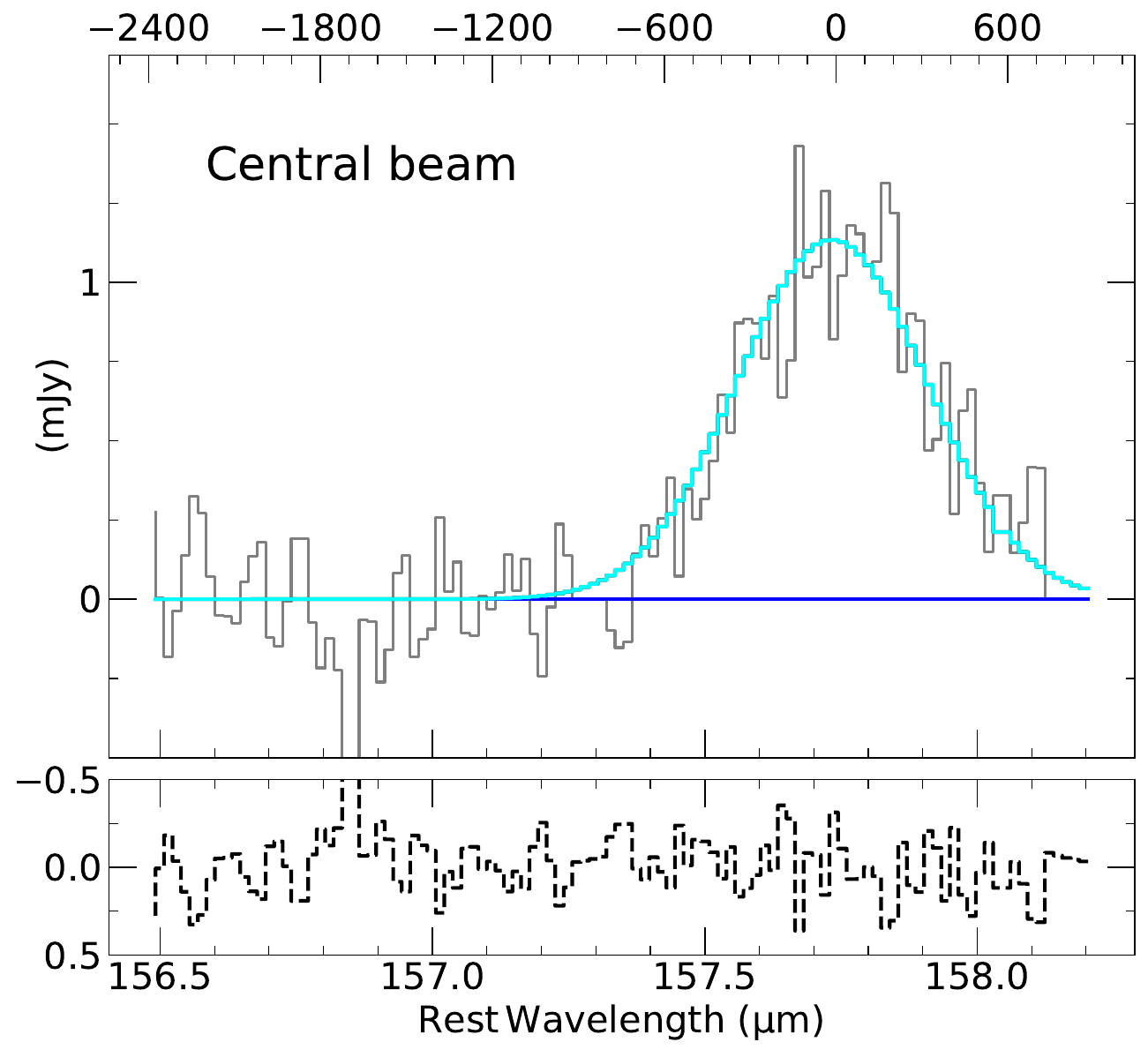}
\includegraphics[width=0.45\textwidth]{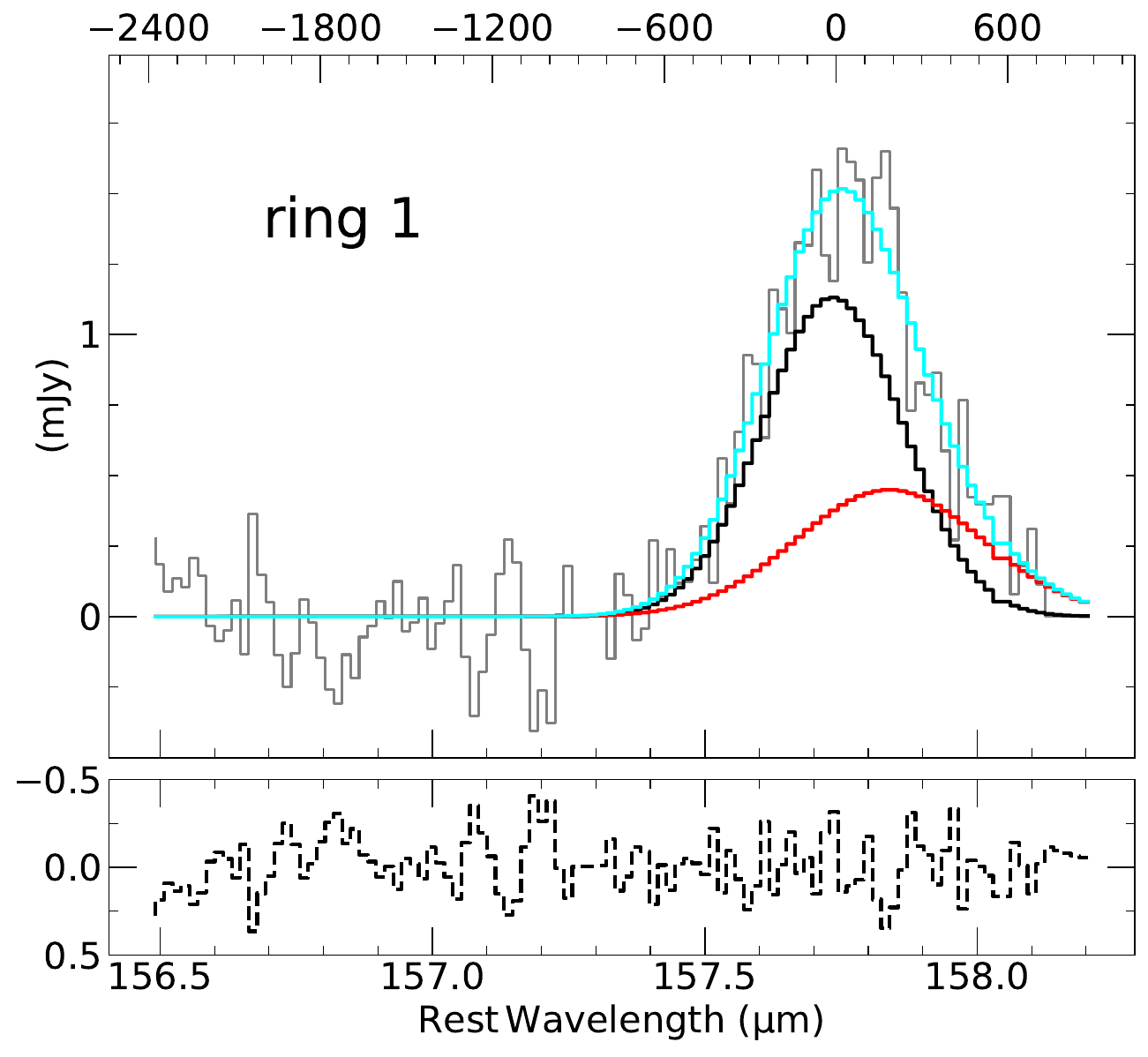}
\caption{
The integrated spectral fitting of the central beam and $U_{\rm R1}$ for [C{~\sc ii}] uniform-weighted data cube. The grey, black, and red solid lines are the observed data, fitted host component and red outflows, respectively. The cyan solid line
sums the two components in the right panel, while the observed data in the left panel can be well fitted by the host component alone. The top scale of each plot is in velocity space. The black dashed lines are the residuals.}
\label{cii_ring_fitting_uniform}

\end{figure*} 

\begin{figure*}
\centering
\includegraphics[width=0.45\textwidth]{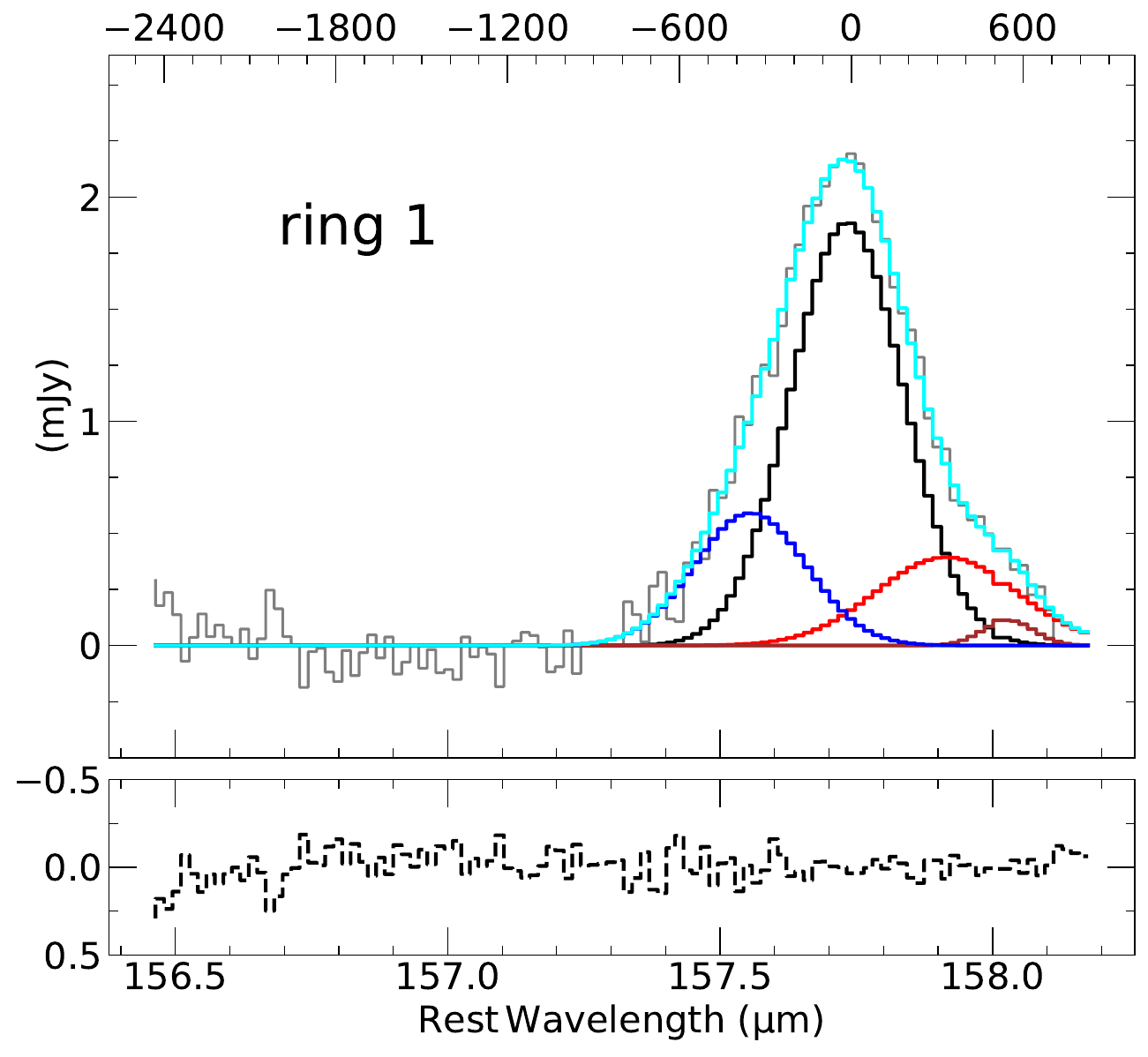}
\includegraphics[width=0.45\textwidth]{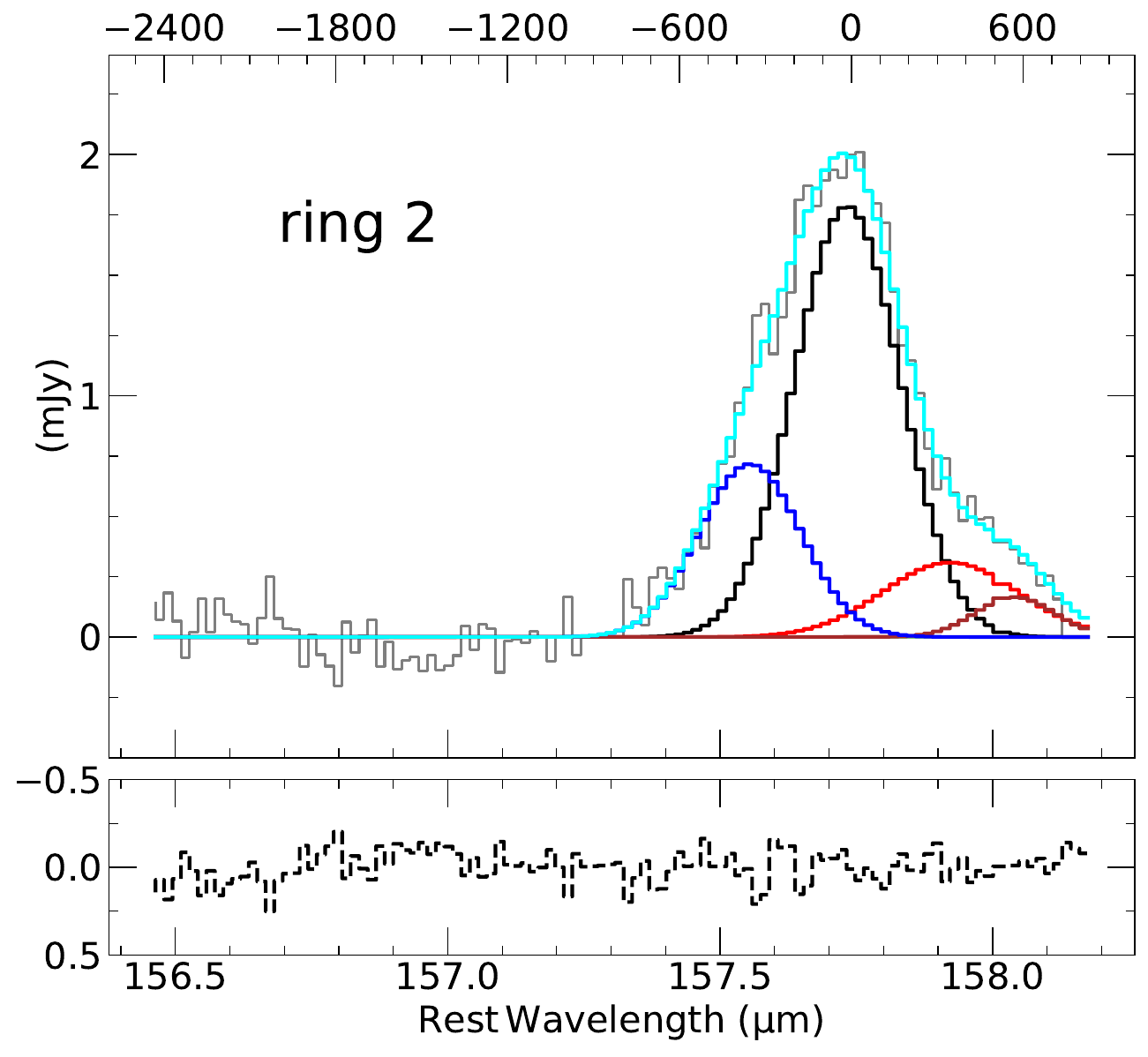}\\
\includegraphics[width=0.45\textwidth]{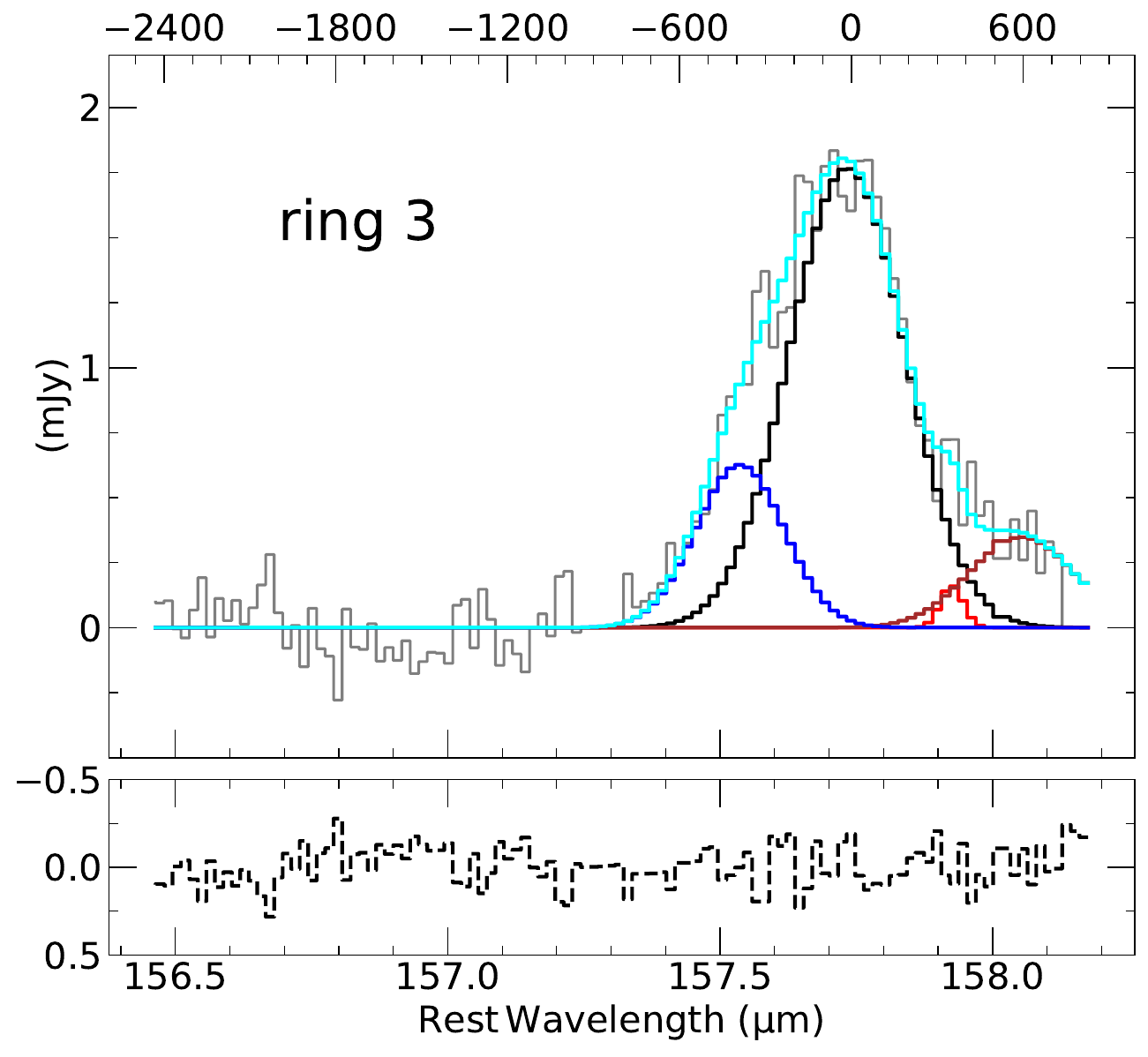}
\includegraphics[width=0.45\textwidth]{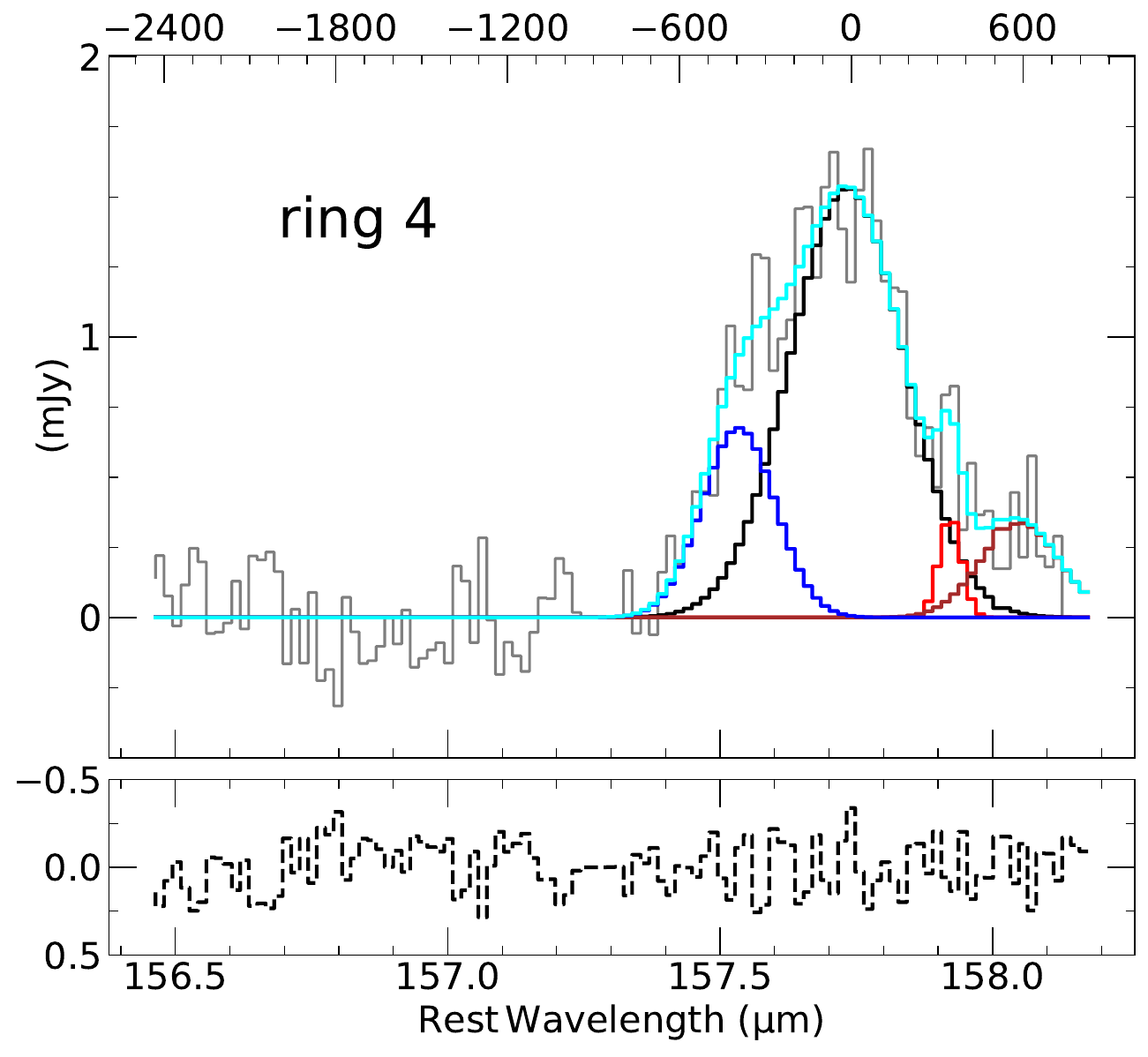}
\includegraphics[width=0.45\textwidth]{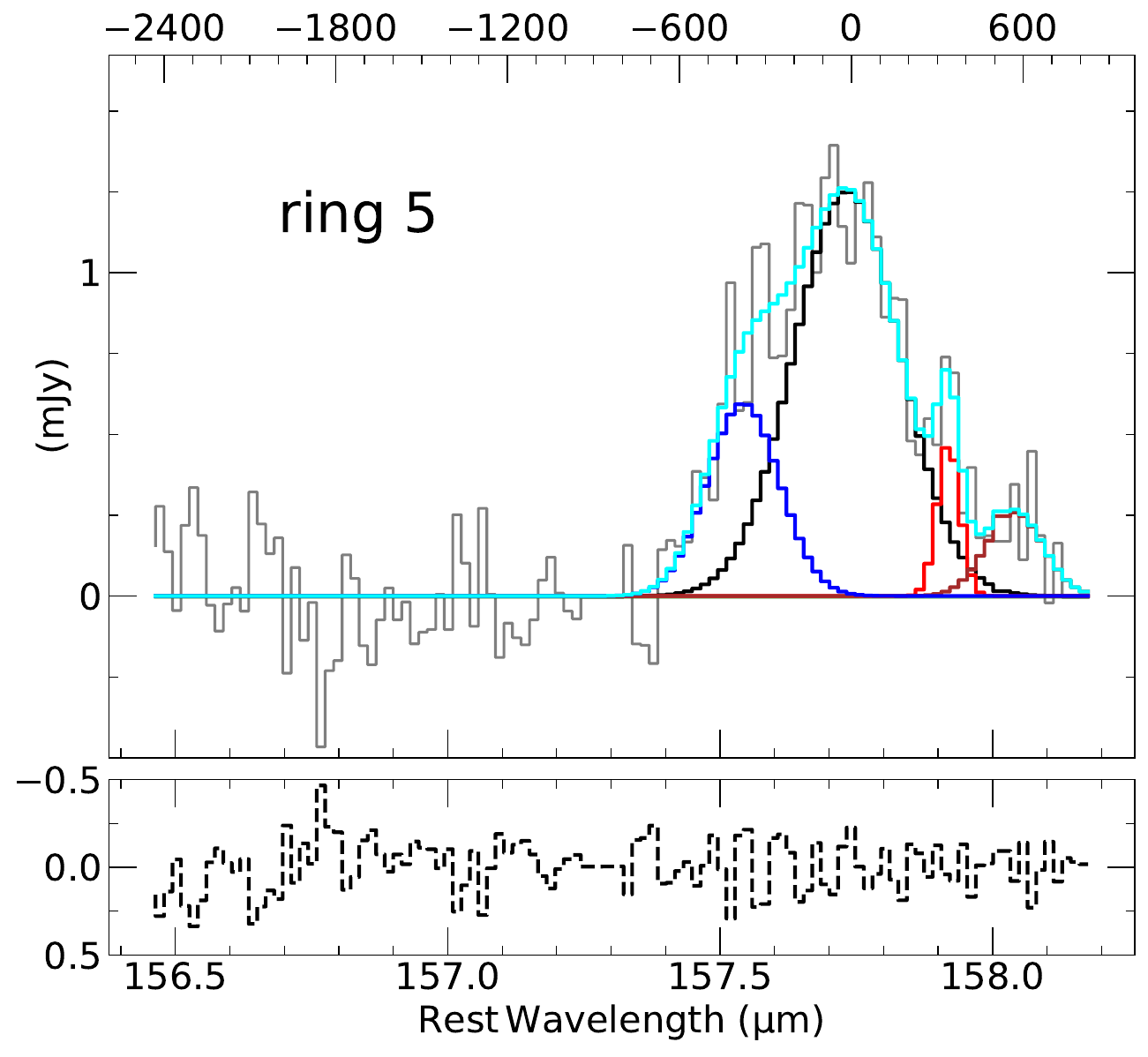}
 \caption{Same as Figure \ref{cii_ring_fitting_uniform} but for spectral fitting of rings $N_{\rm R1}$, $N_{\rm R2}$, $N_{\rm R3}$, $N_{\rm R4}$ and $N_{\rm R5}$ for [C{~\sc ii}] natural-weighted cube. The grey, black, red/brown, and blue solid lines are the observed data, fitted host component, low-/high-velocity red-outflows, and blue-outflows, respectively. The cyan solid lines sum the four total components.}
\label{cii_ring_fitting_natutal}
\end{figure*} 

\begin{figure*}
\centering
\includegraphics[width=0.5\textwidth]{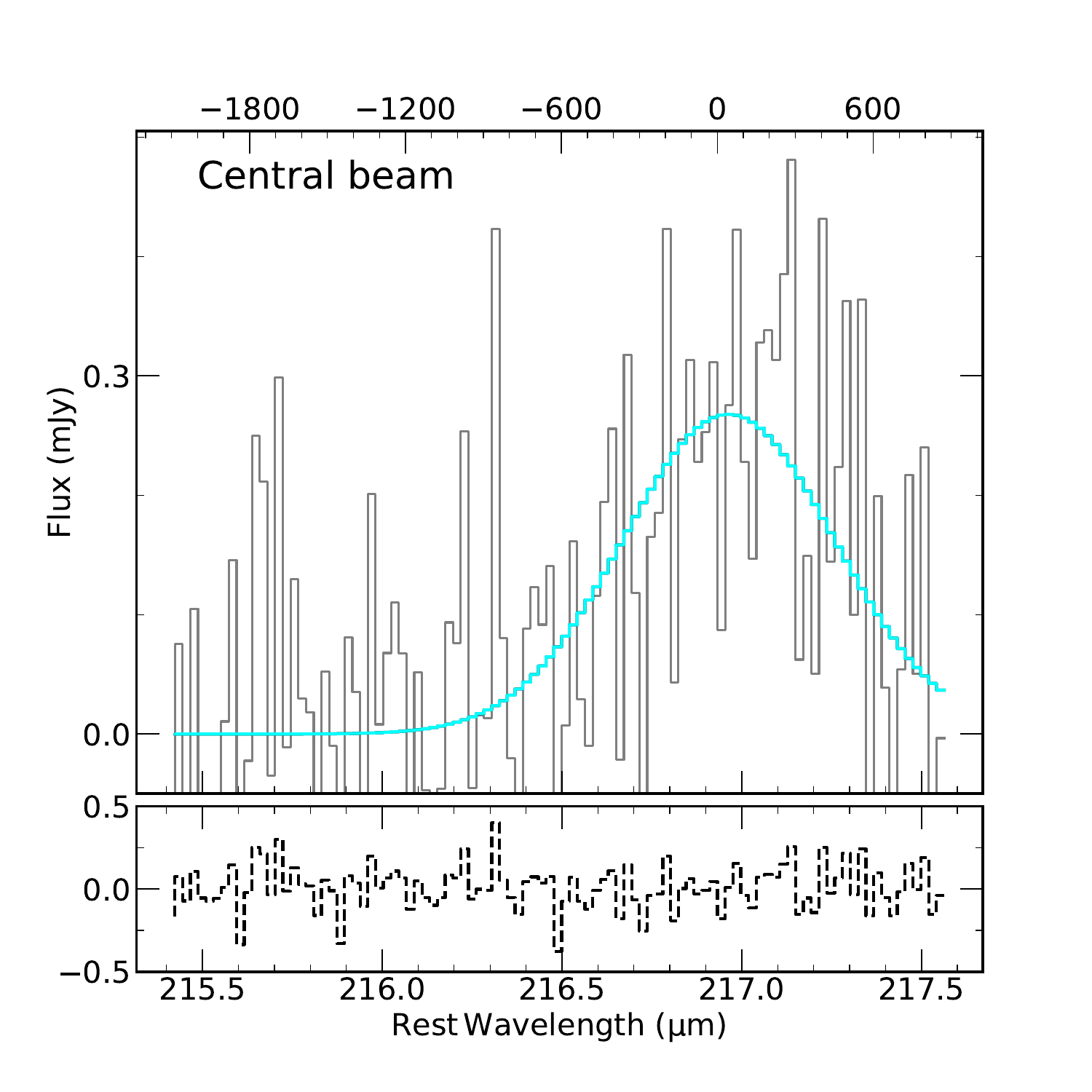}\\
\includegraphics[width=0.45\textwidth]{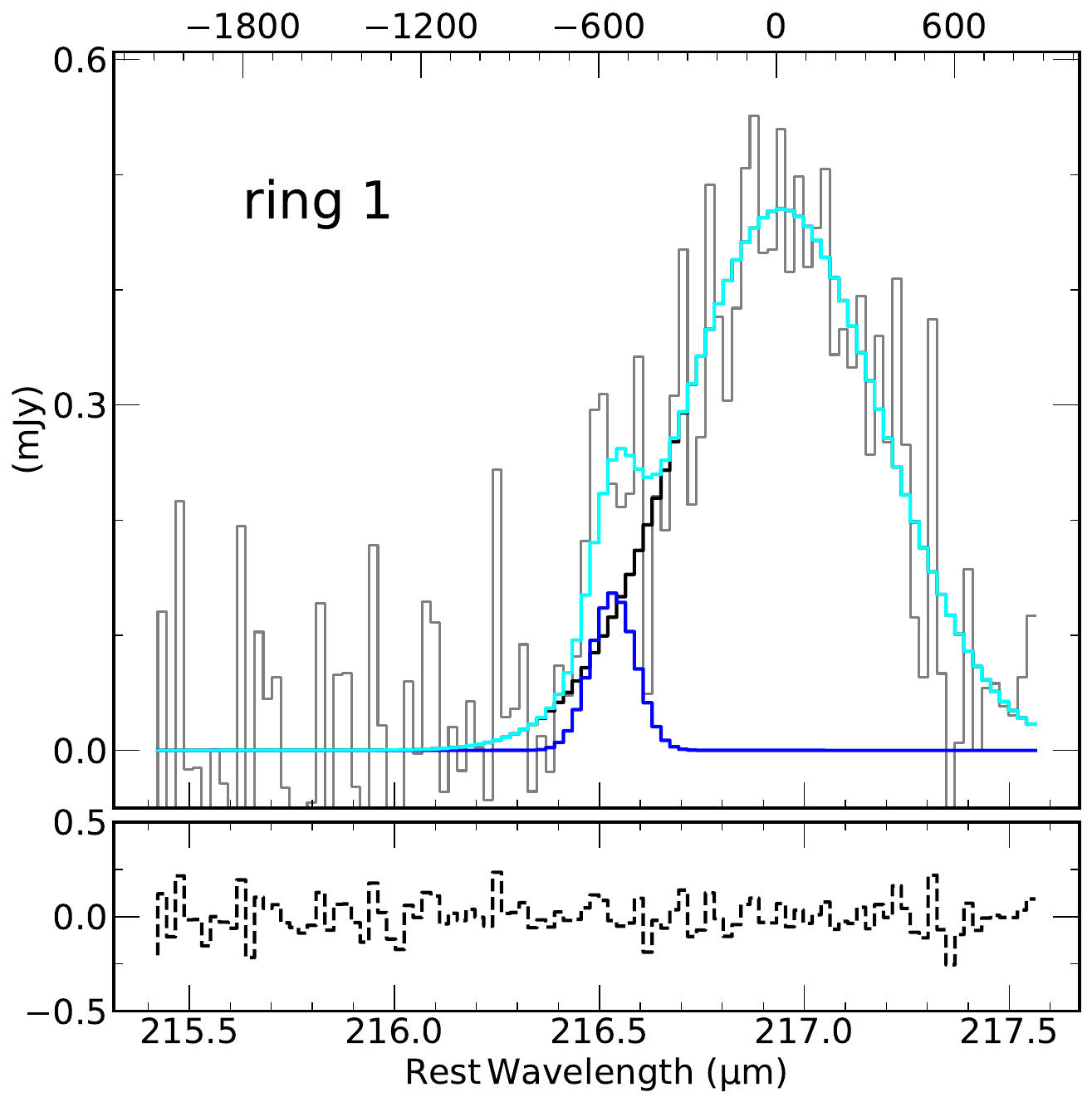}
\includegraphics[width=0.45\textwidth]{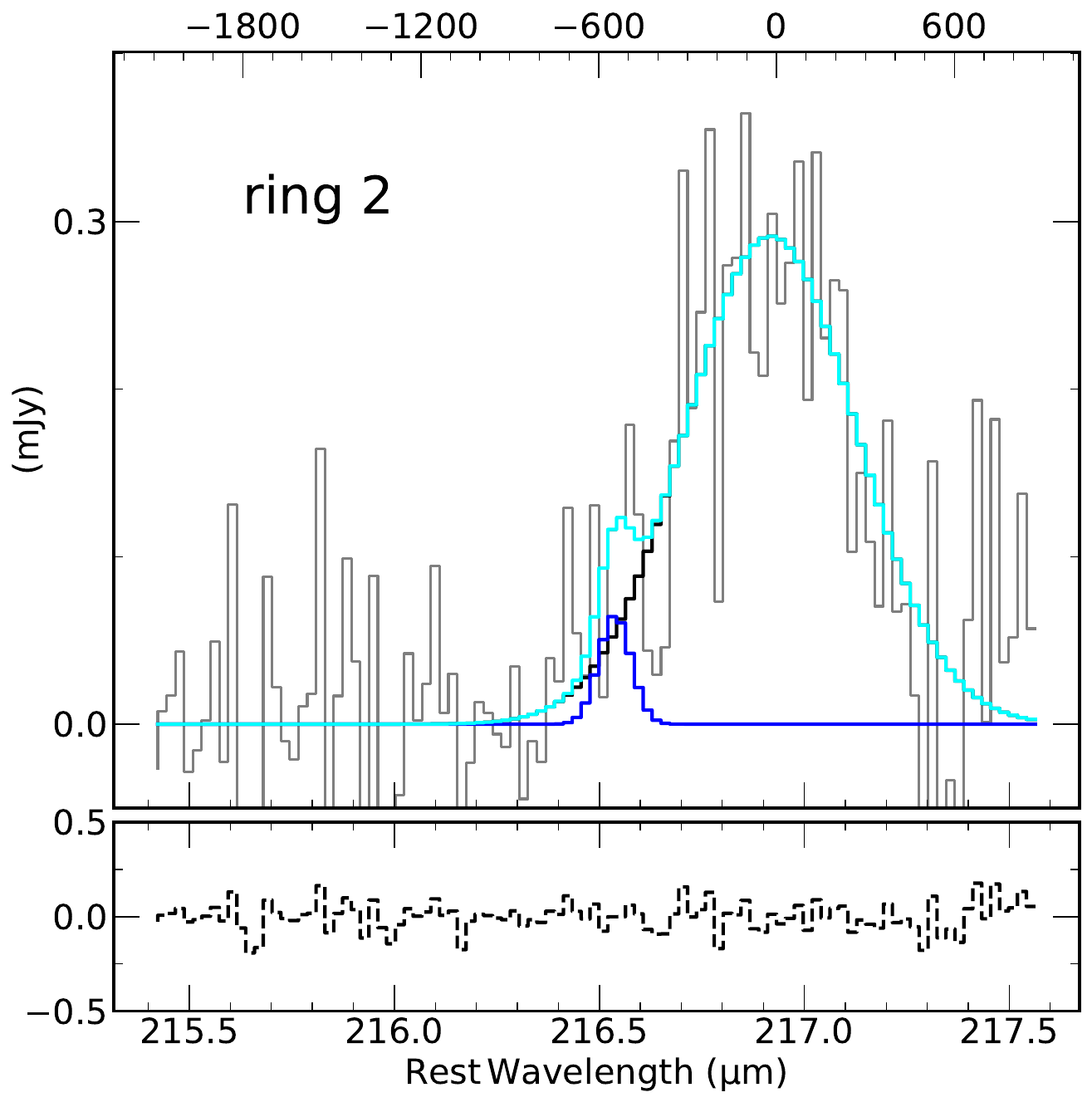}
 \caption{Same as Figure \ref{cii_ring_fitting_uniform} but for spectral fitting for CO (12-11) emission line for its central beam created from the data cube with uniform-weighting (upper panel), two ring spectra created from the data cube with robust = 0 weighting (bottom: $R0_{\rm R1}$ and $R0_{\rm R2}$). See discussion in Section \ref{spectra_fitting}.}
\label{CO_ring_fitting1}
\end{figure*}

\begin{figure*}
\centering
\includegraphics[width=0.5\textwidth]{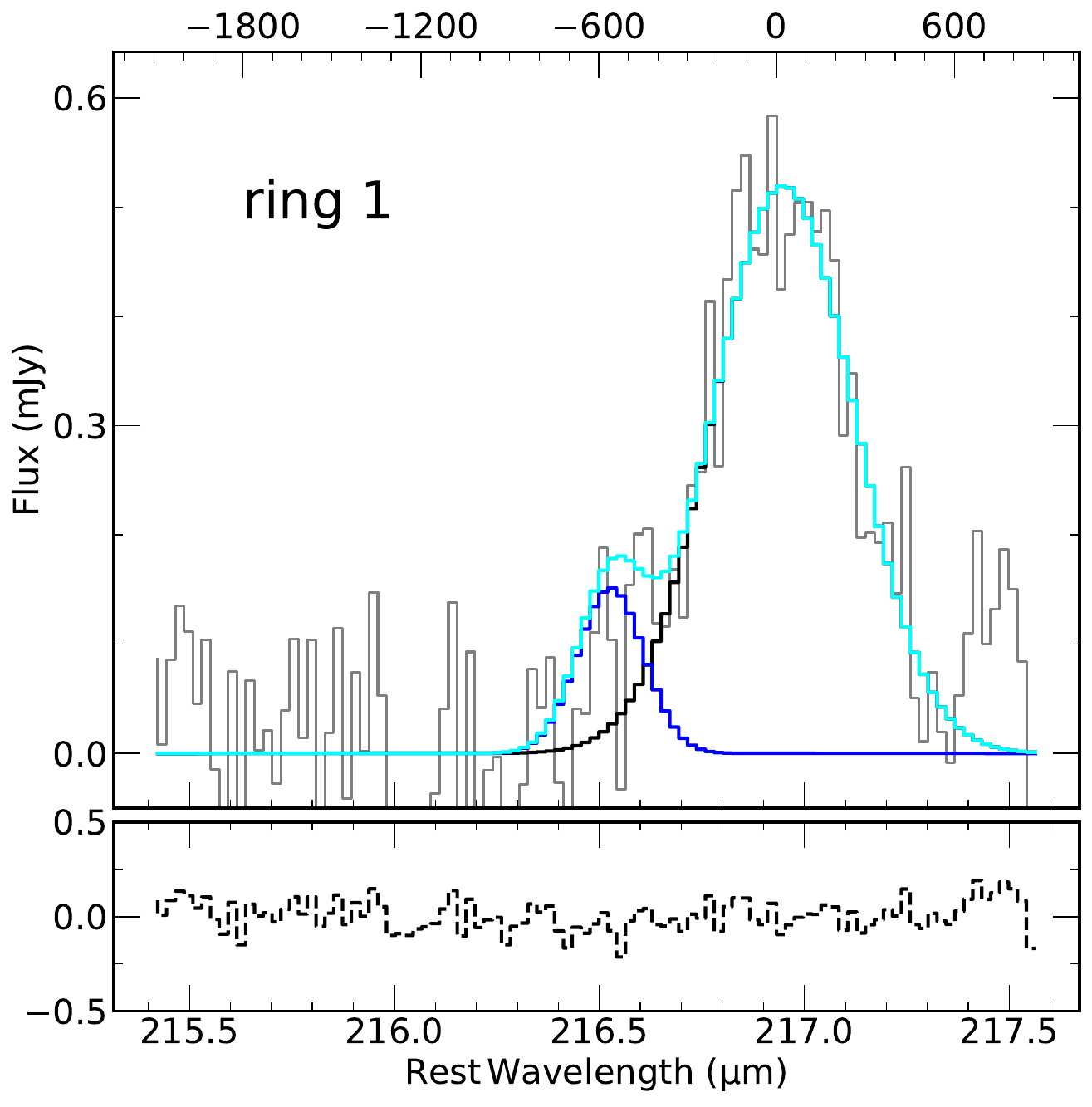}\\
\includegraphics[width=0.45\textwidth]{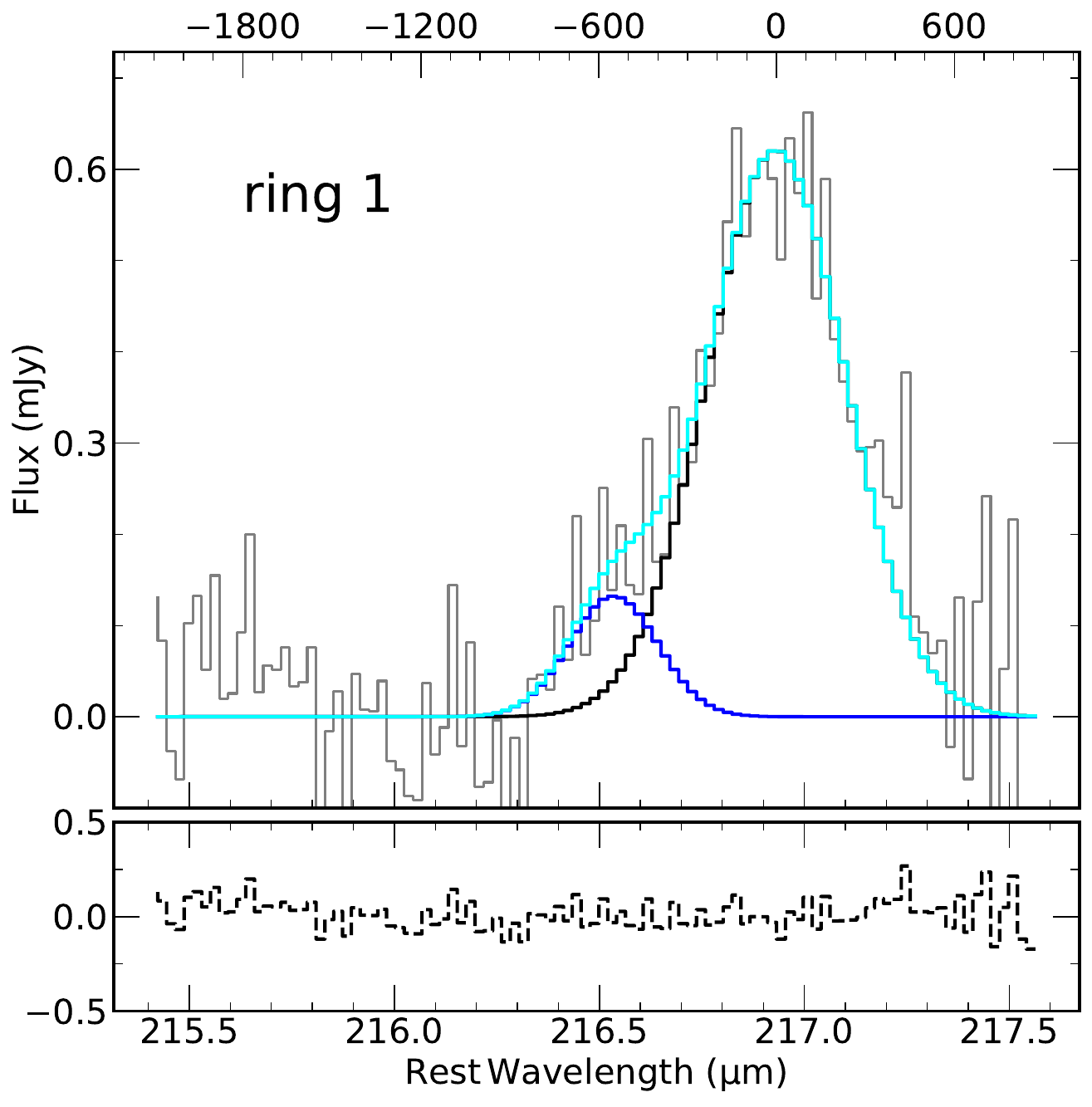}
\includegraphics[width=0.45\textwidth]{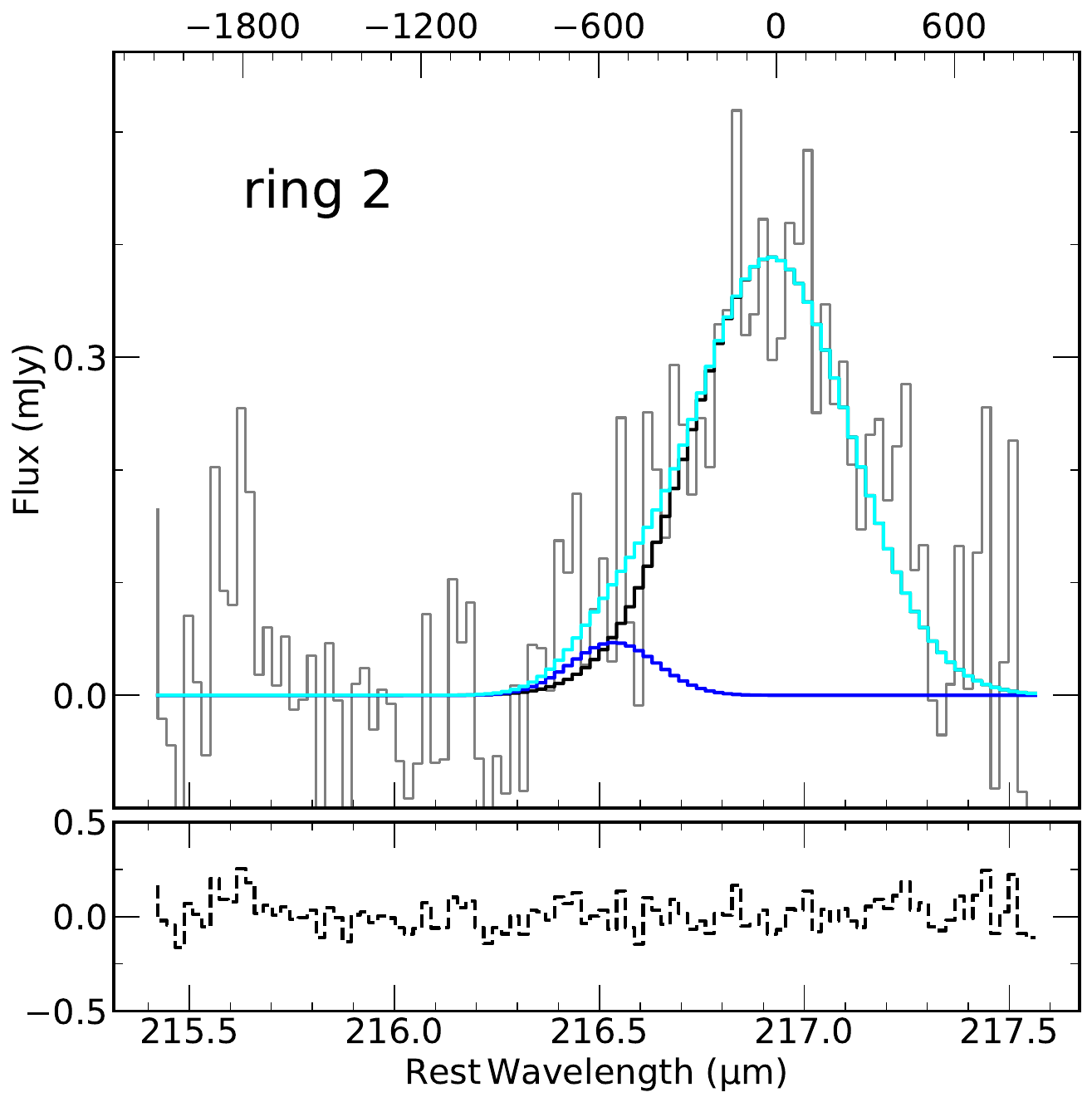}
 \caption{Same as Figure \ref{cii_ring_fitting_uniform} but for spectral fitting for CO (12-11) emission line for its one ring spectrum created from data cube with robust = 0.5 (upper panel: $R0.5_{\rm R1}$), two ring spectra created from the data cube with natural-weighting (bottom: $NN_{\rm R1}$ and $NN_{\rm R2}$). See discussion in Section \ref{spectra_fitting}.}
\label{CO_ring_fitting2}
\end{figure*}



\begin{figure*}
\centering
\includegraphics[width=1\textwidth]{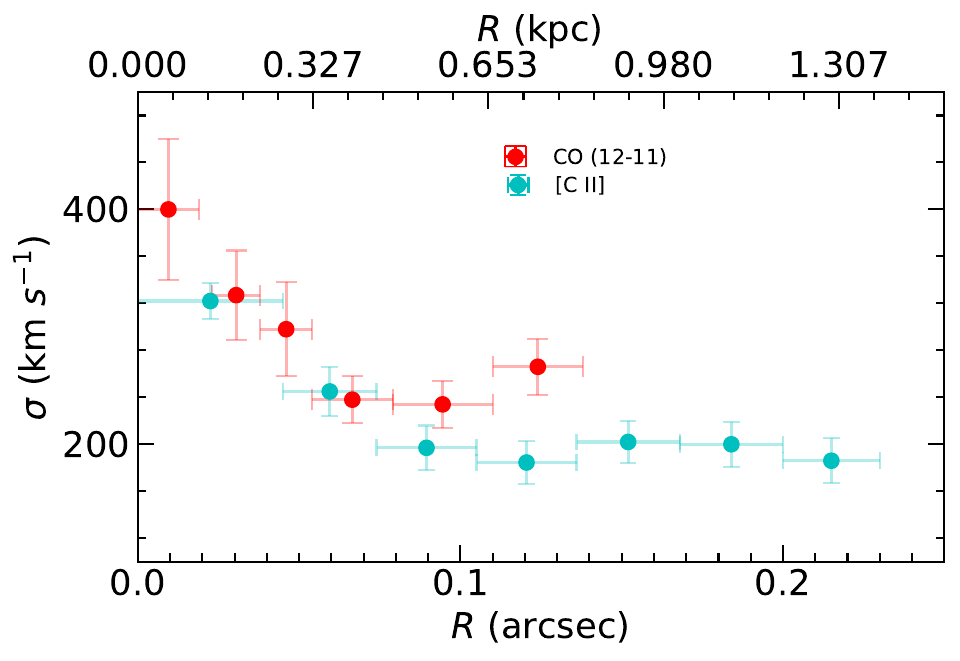}
 \caption{The full combined dispersion profile of [C{~\sc ii}] and CO (12‑11) derived from the best-fit central Guassian component in concentric rings, highlighting that the dispersion radial profile of both emission lines is highly consistent out to a radius of 520 pc (0.08$^{\prime\prime}$) and starts diverging at larger angular scales, where heating likely starts being dominated by photodissociation regions (PDRs) for [C{~\sc ii}] and shocks by CO (12-11).}
\label{full combined dispersion profile}
\end{figure*} 

	

	


\begin{figure}[htbp!]
\centering
 \includegraphics[width=1.0\columnwidth]{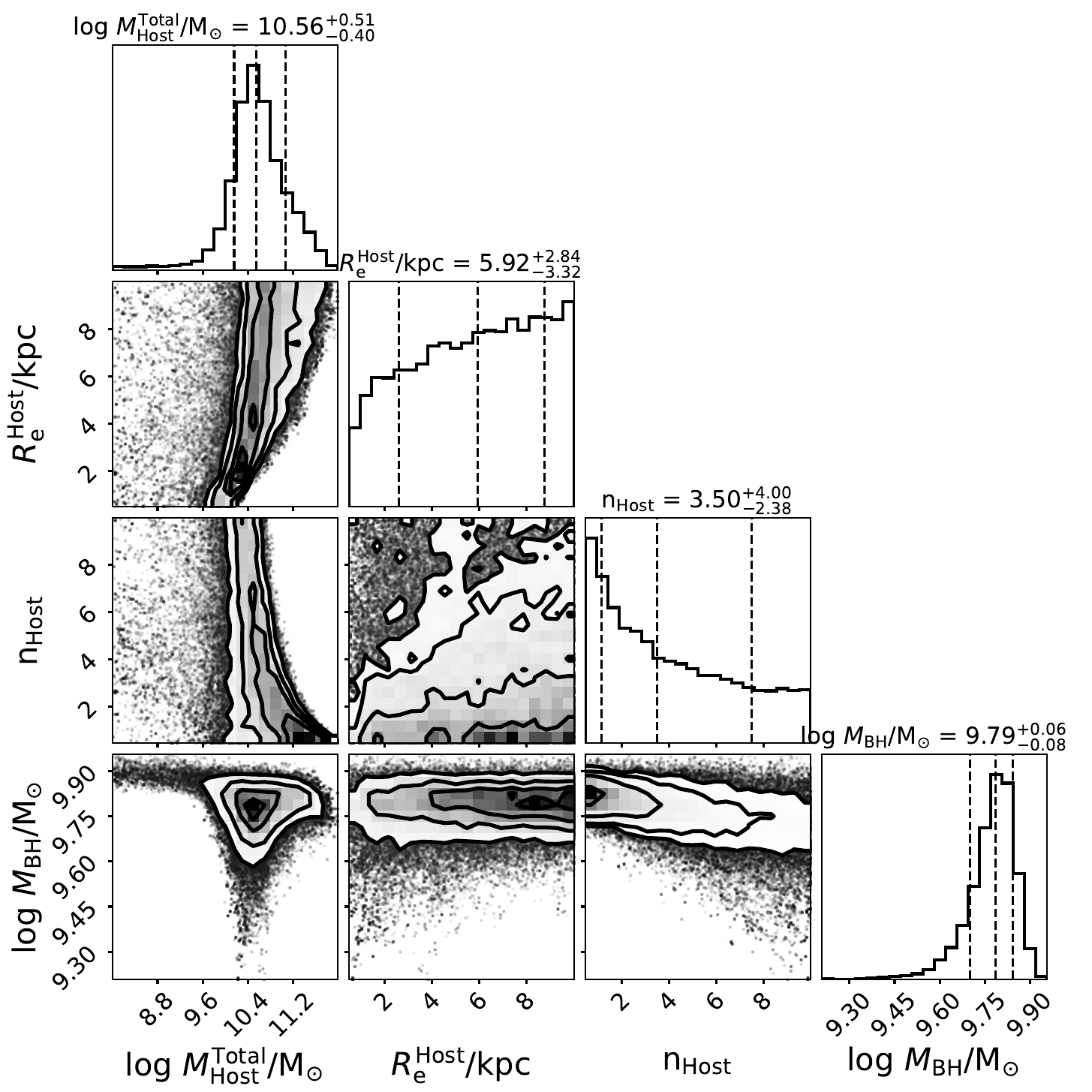}
 	\caption{Corner plots of the resulting posterior distributions of all four fitted parameters ($M_{\rm BH}$, $M_{\rm Host}^{\rm Total}$, $n_{\rm Host}$ and $R_e^{\rm Host}$) considering the gravity from black hole plus host galaxy for combined CO (12-11) + [C{~\sc ii}] dispersion profile. See discussion in Section \ref{dispersion fitting}.}
\label{fig:BH_host_combined_dis}
\end{figure} 

\begin{figure}[htbp!]
\centering
 \includegraphics[width=1.0\columnwidth]{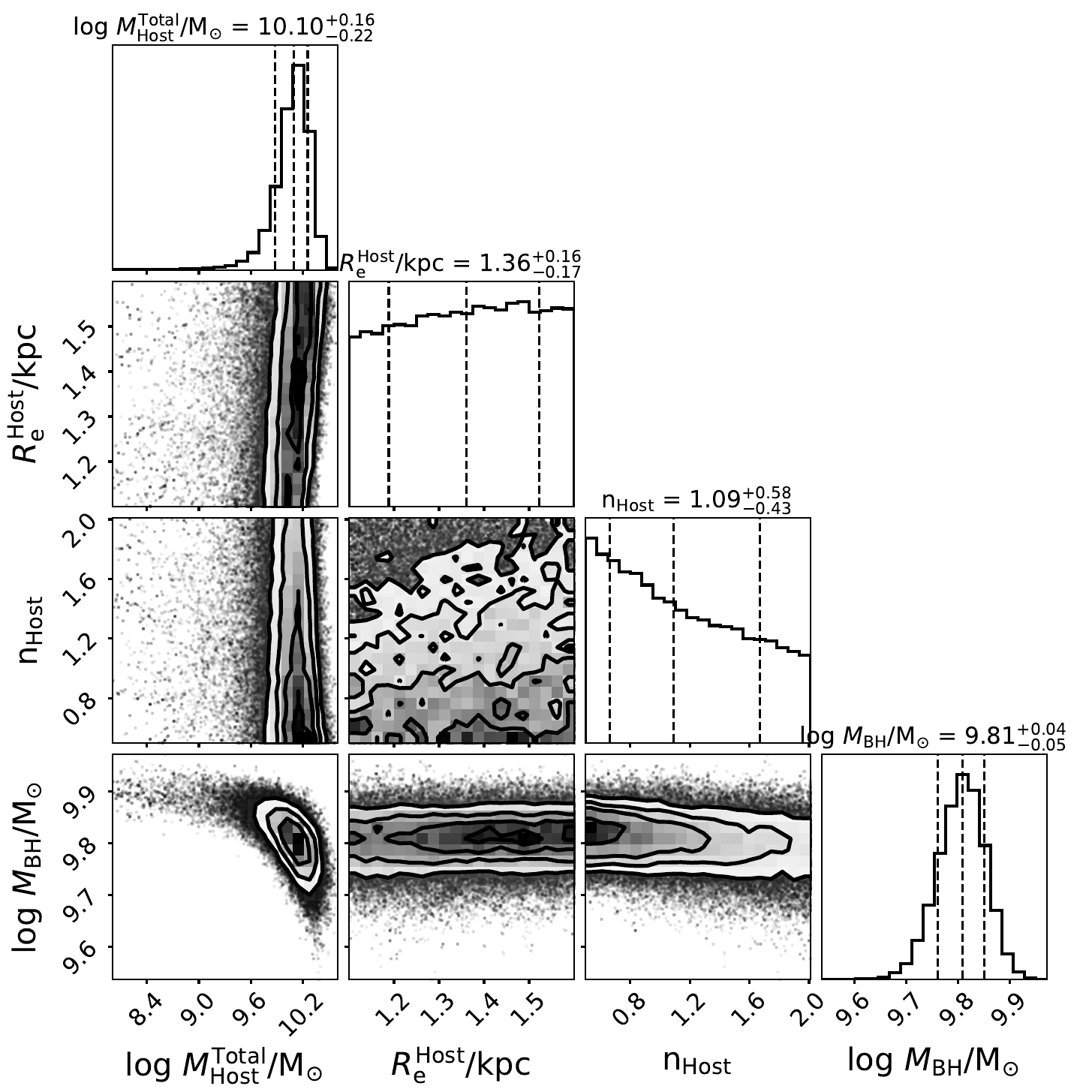}
	\caption{Same as Figure \ref{fig:BH_host_combined_dis}, but here the host parameters of S\'ersic profile index $n_{\rm Host}$ and effective radius $R_{e}^{\rm Host}$ are only allowed to change in the ranges of [0.5, 2] and [1.1, 1.6] kpc, respectively. See discussion in Section \ref{dispersion fitting}.}
\label{fig: BH_host_combined_fixed_dis}
\end{figure} 


\begin{figure}[htbp!]
\centering
 \includegraphics[width=1\columnwidth]{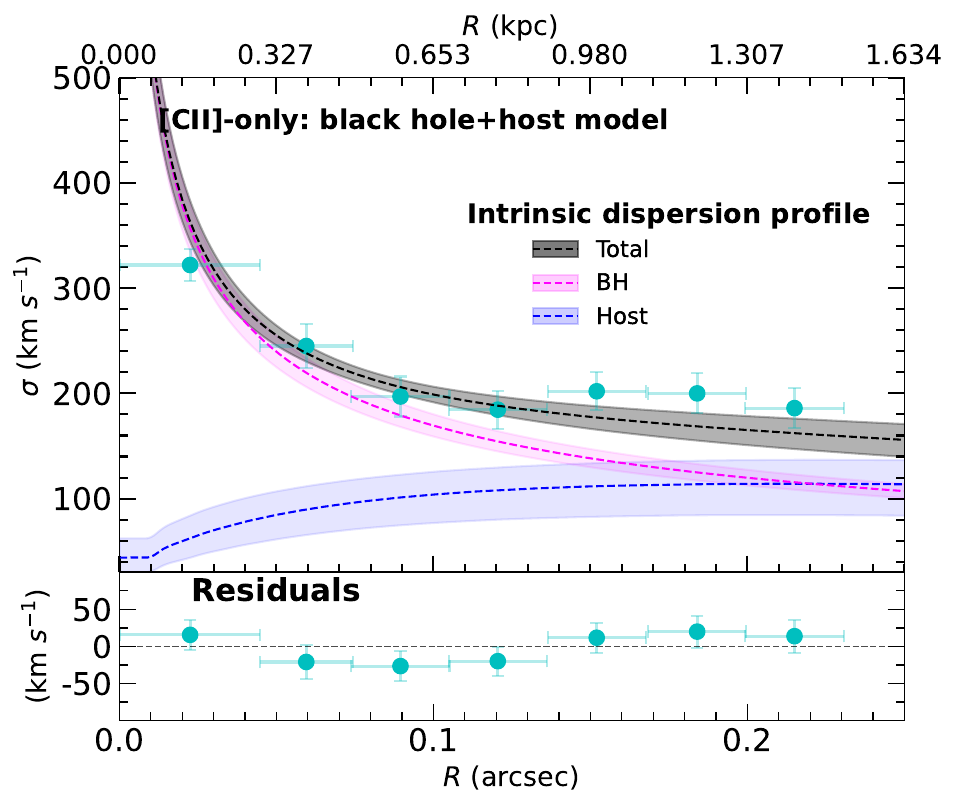}
 	\caption{$\it Top$: The best-fit intrinsic models for black hole and host components after accounting for the beam smoothing effect to the observed [C{~\sc ii}]-only dispersion profile with the shaded regions showing the $1\sigma$ confidence intervals, by considering the gravity from black hole plus host galaxy with host parameters of S\'ersic profile index $n_{\rm Host}$ and effective radius $R_{e}^{\rm Host}$ being only allowed to change in the ranges of [0.5, 2] and [1.1, 1.6] kpc. $\it Bottom$: the residuals between the best-fit total model and the observed data points with considering the beam smoothing effect. See discussions in Section \ref{cii_disperison_fitting}, and Figure \ref{model_plot_only_cii_dis} for the distribution of each model parameter.}
\label{model_plot_only_cii}
\end{figure} 

\begin{figure}[htbp!]
\centering
 \includegraphics[width=1.0\columnwidth]{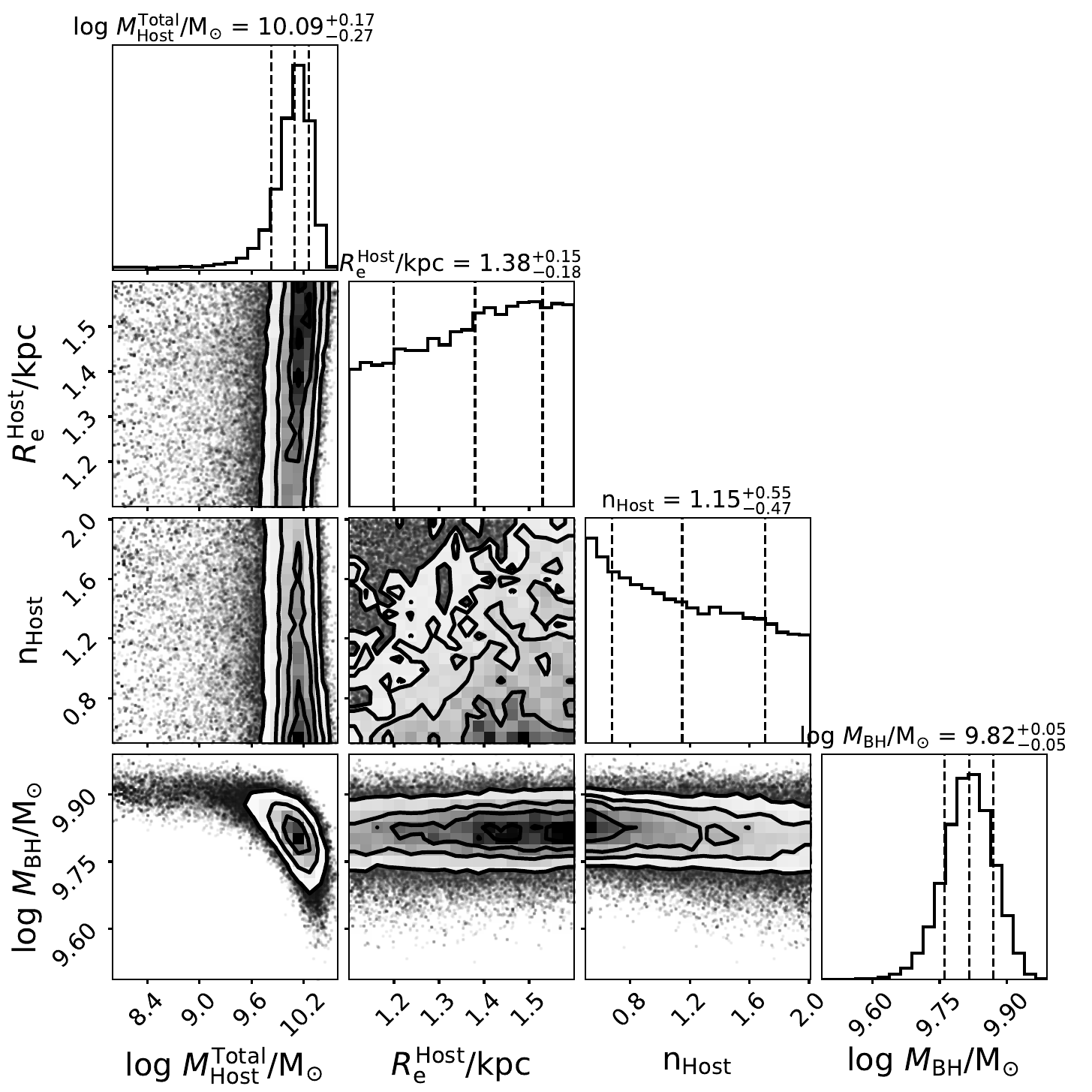}
	\caption{Same as Figure \ref{fig: BH_host_combined_fixed_dis} but for [C{~\sc ii}]-only dispersion profile. See discussion in Section \ref{cii_disperison_fitting}.}
\label{model_plot_only_cii_dis}
\end{figure} 

\begin{figure}[htbp!]
\centering
 \includegraphics[width=0.8\columnwidth]{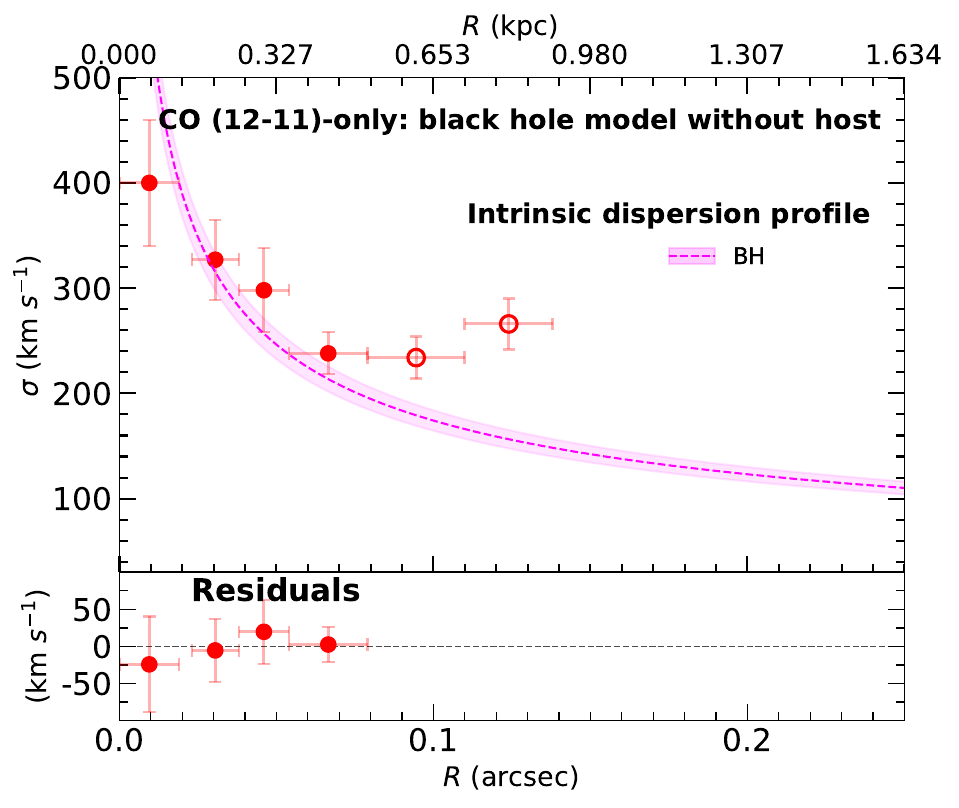}
 \includegraphics[width=0.5\textwidth]{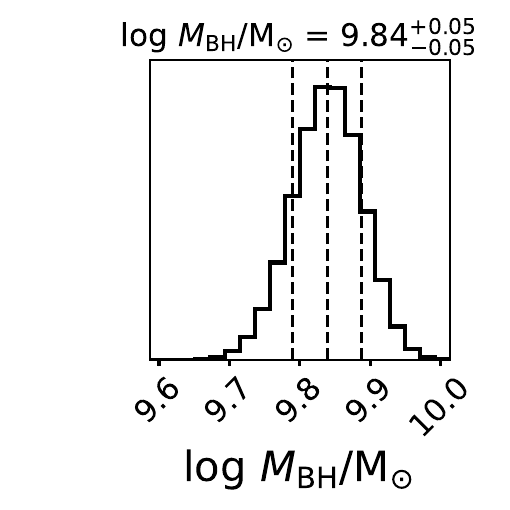}
 	\caption{$\it Top$: Same as Figure \ref{model_plot_only_cii}, but for observed CO (12-11)-only dispersion profile with only considering the
gravity from black hole. The CO (12-11) outermost two data points (the red open circles) is masked in the fitting (see Section \ref{spectra_fitting} for details). $\it Bottom$: Same as Figure \ref{fig: BH_host_combined_fixed_dis} but for CO (12-11)-only dispersion profile with only considering the black hole gravity. See Section \ref{co_disperison_fitting} for details.}
\label{model_plot_only_co}
\end{figure} 



\begin{figure*}
\centering
\includegraphics[width=0.9\textwidth]
{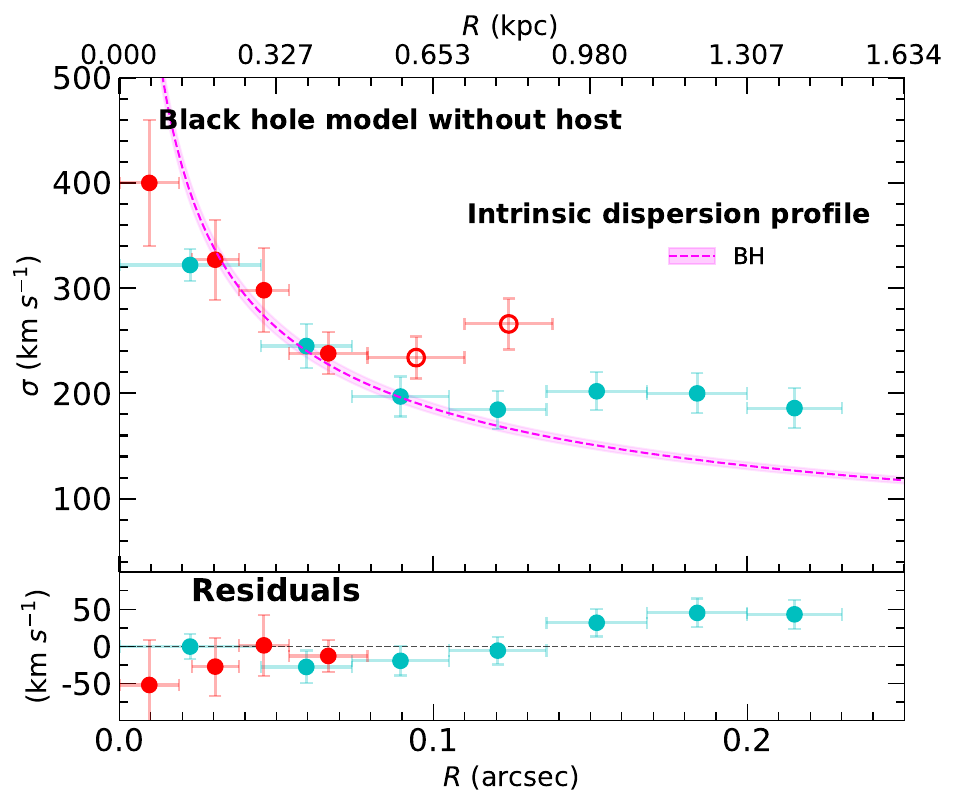}
\includegraphics[width=0.5\textwidth]{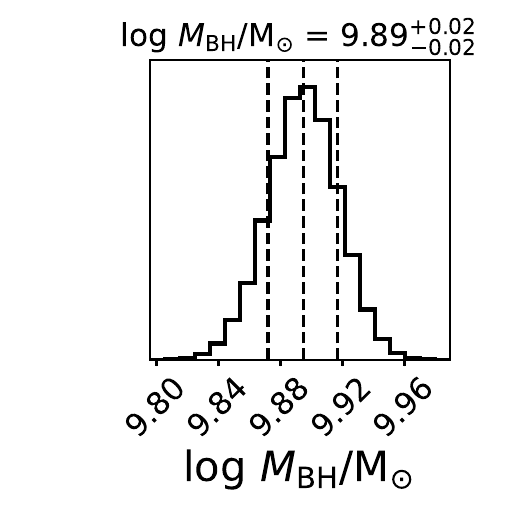}
\caption{$\it Top$: Same as Fig. \ref{model_plot_only_cii}, but for only considering the gravity of SMBH for the combined [C~{\sc ii}] and CO (12-11) dispersion profile. The CO (12-11) outermost two data points (the red open circles) is masked in the fitting; $\it Bottom$: Same as Fig. \ref{fig: BH_host_combined_fixed_dis}, but for only considering the gravity of SMBH. See Section 11.2 for details.}
\label{BH_only} 
\end{figure*}

\begin{figure*}
\centering
\includegraphics[width=1\textwidth]{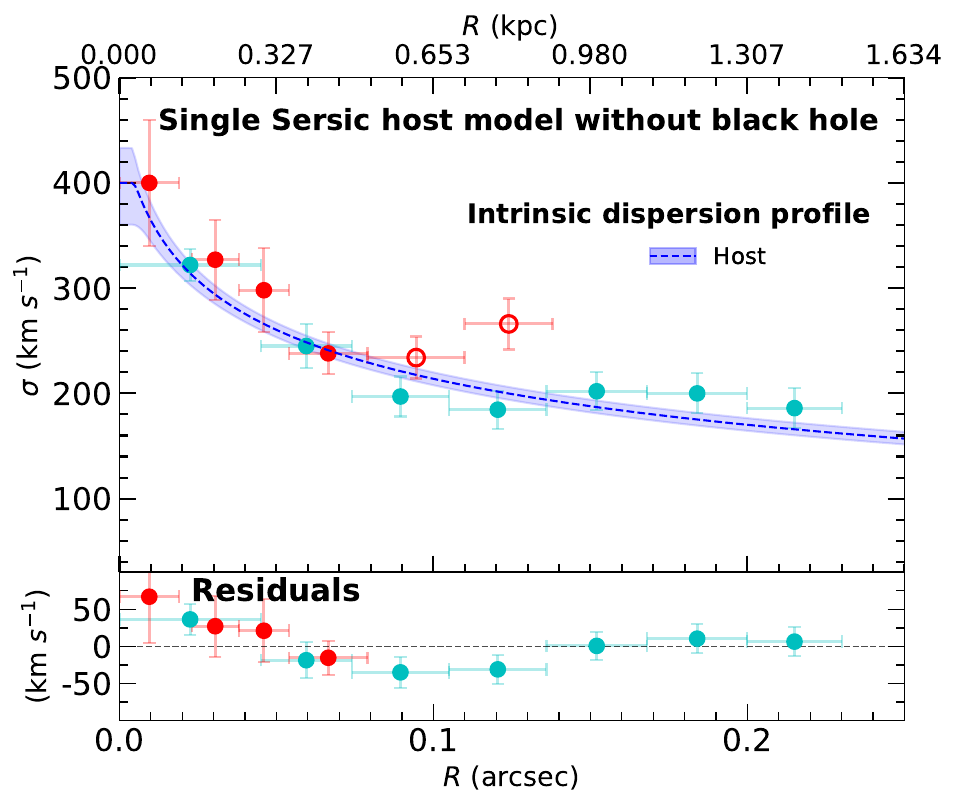}
\caption{Same as Fig. \ref{model_plot_only_cii}, but for the combined dispersion profile of [C~{\sc ii}] and CO (12-11) assuming no SMBH and only considering the gravity from the host galaxy for a single S\'ersic profile. The CO (12-11) outermost two data points (the red open circles) is masked in the fitting. see discussions in Section \ref{Host-only Model} and Fig. \ref{only_host_single_dis} for the distribution of each model parameter.}
\label{only_host_single} 
\end{figure*}

\begin{figure*}
\centering
\includegraphics[width=1\textwidth]{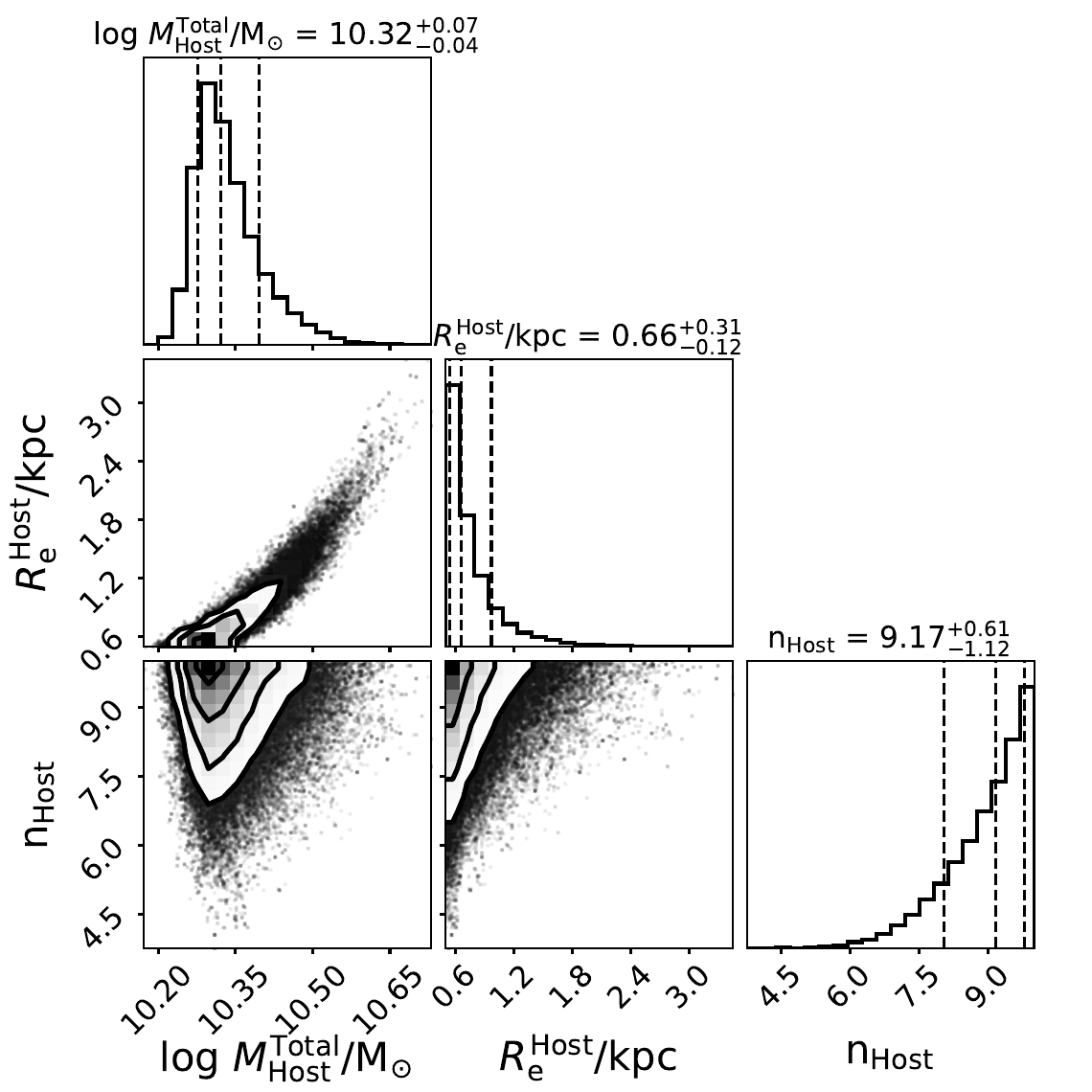}\\
\caption{Same as Fig. \ref{fig: BH_host_combined_fixed_dis}, but for only considering the gravity from the host galaxy with a single S\'ersic profile. See Section \ref{Host-only Model} for details.}
\label{only_host_single_dis} 
\end{figure*}

\begin{figure*}
\centering
\includegraphics[width=1\textwidth]{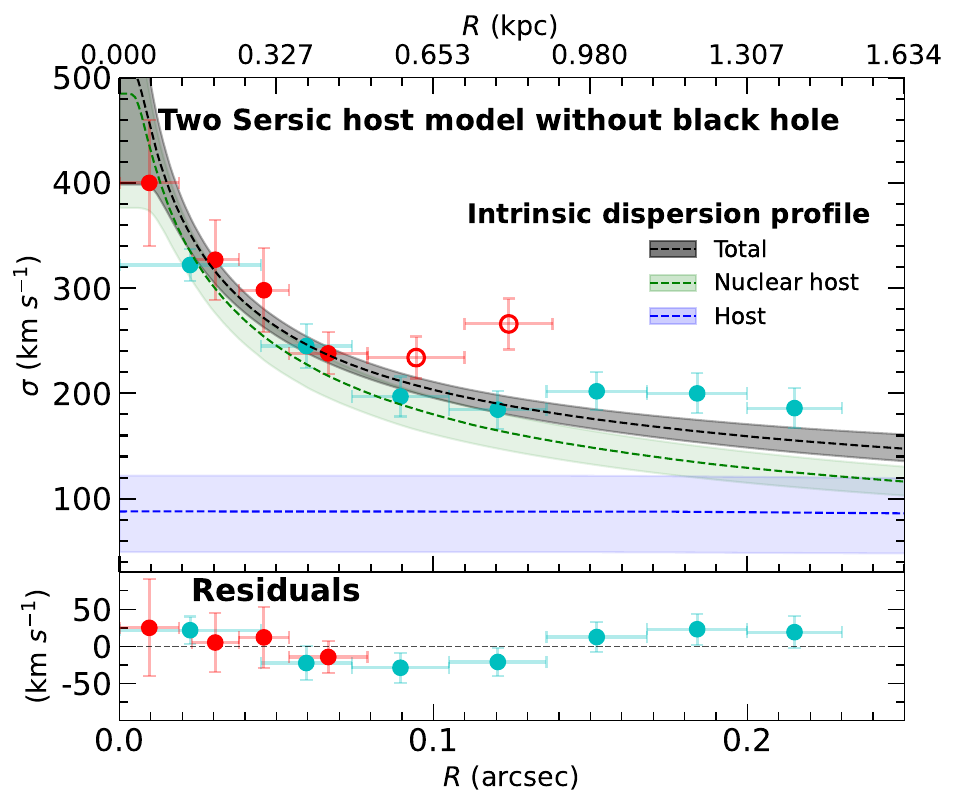}
\caption{Same as Fig. \ref{model_plot_only_cii}, but for the combined dispersion profile of [C~{\sc ii}] and CO (12-11) assuming no SMBH and only considering the gravity from the host galaxy for double S\'ersic profile. The CO (12-11) outermost two data points (the red open circles) is masked in the fitting. See Section \ref{Host-only Model}, and Fig. \ref{only_host_double_dis} for the distribution of each model parameter.}
\label{only_host_double} 
\end{figure*}

\begin{figure*}
\centering
\includegraphics[width=1.\textwidth]{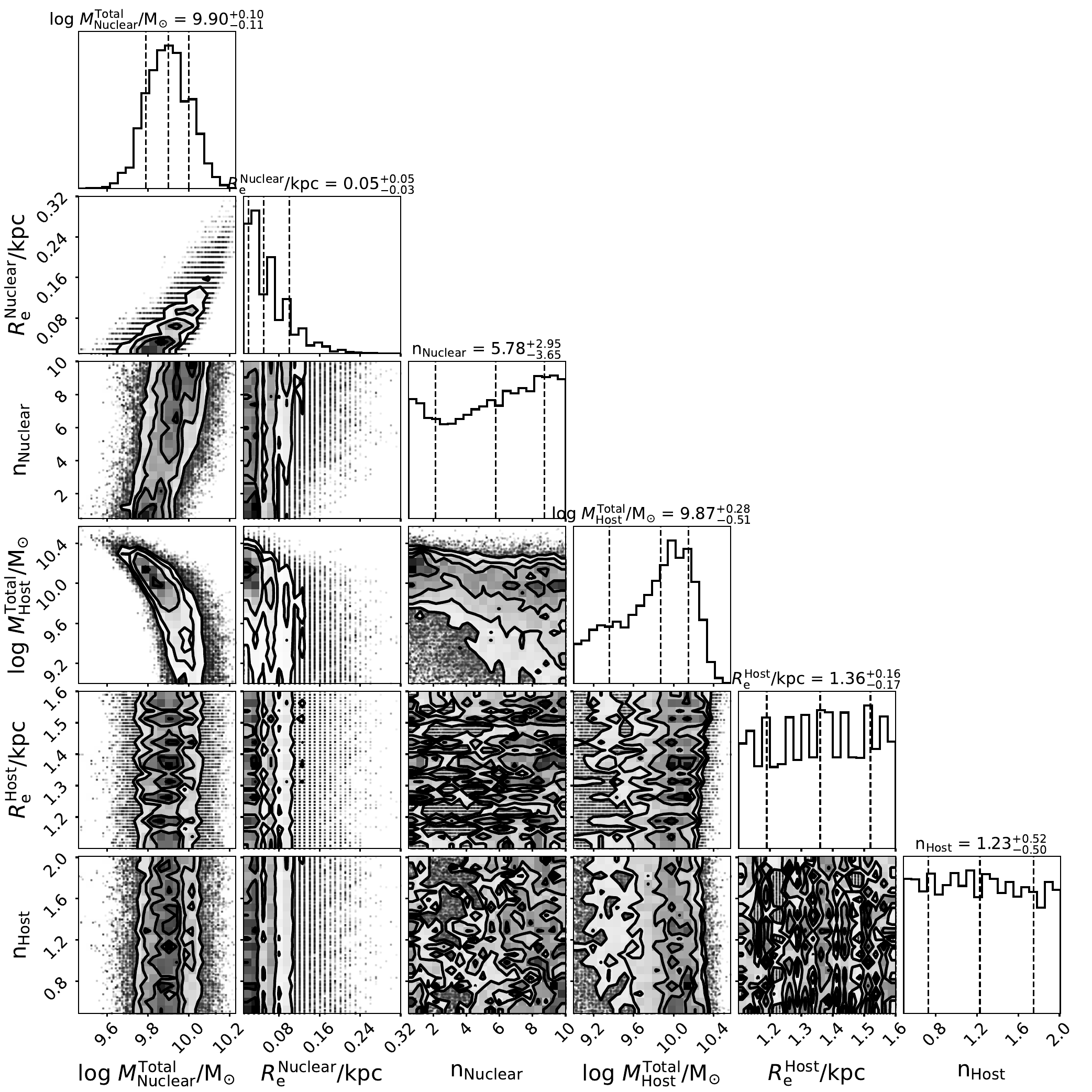}
\caption{Same as Figure \ref{fig: BH_host_combined_fixed_dis}, but for only considering the gravity from the host galaxy with double S\'ersic profile.}
\label{only_host_double_dis} 
\end{figure*}

\begin{figure*}
\centering
\includegraphics[width=1\textwidth]
{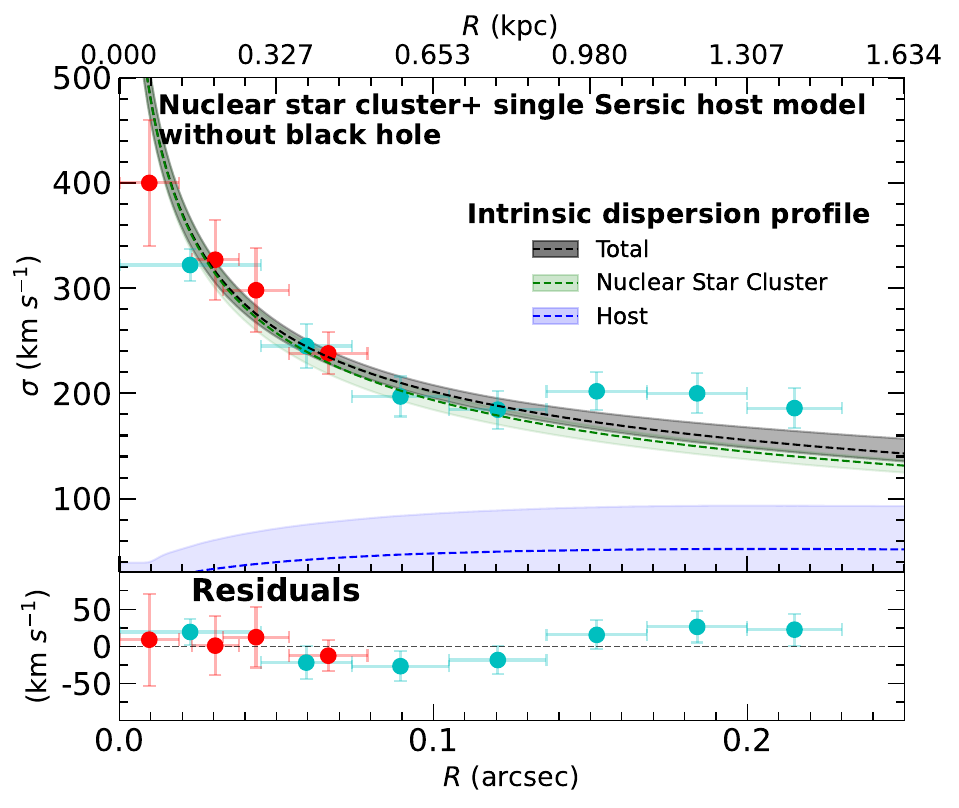}
\caption{Same as Fig. \ref{model_plot_only_cii} for considering the gravity from a nuclear star cluster plus host galaxy. The CO (12-11) outermost two data points (the red open circles) is masked in the fitting. See Sections \ref{Nuclear star cluster} for details, and Fig. \ref{star_cluster_corner} for the distribution of each model parameter.}
\label{star_cluster} 
\end{figure*}

\begin{figure*}
\centering
\includegraphics[width=1.0\textwidth]{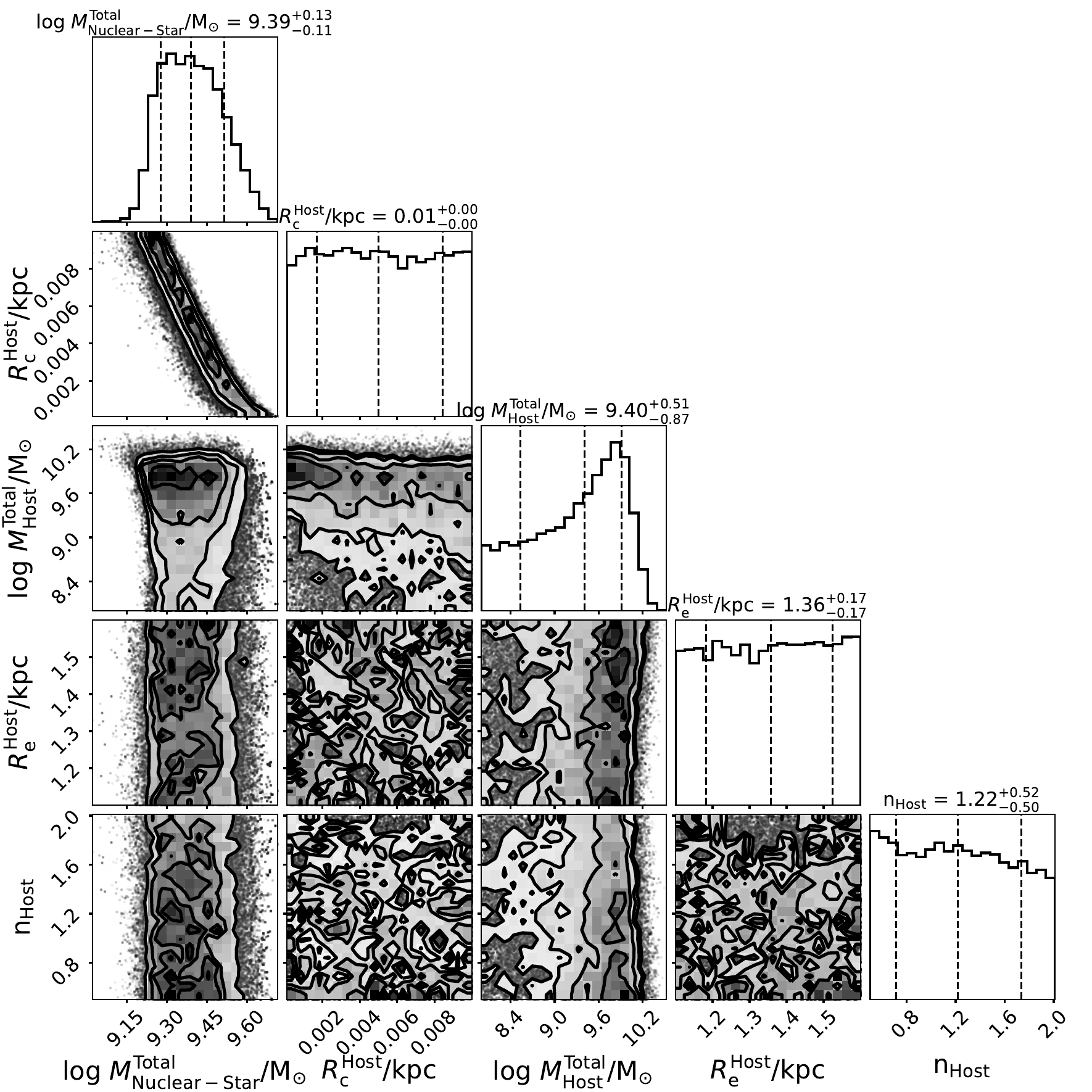}
\caption{Same as Fig. \ref{fig: BH_host_combined_fixed_dis}, but for considering the gravity from a nuclear star cluster plus host galaxy. See Section \ref{Nuclear star cluster} for details.}
\label{star_cluster_corner} 
\end{figure*}

\begin{figure*}
\centering
\includegraphics[width=1\textwidth]
{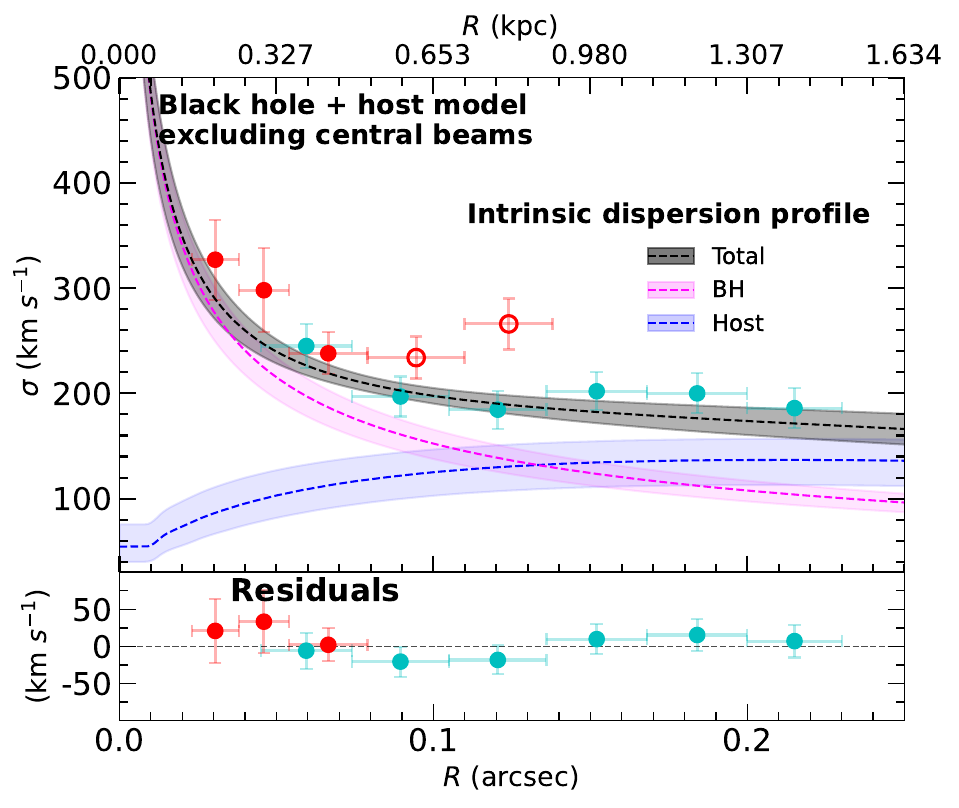}
\caption{Same as Fig. \ref{model_plot_only_cii}, but for excluding the central beam data points for the combined dispersion profile of [C~{\sc ii}] and CO (12-11), and considering the gravity from the black hole and host galaxy with a single S\'ersic profile. The CO (12-11) outermost two data points (the red open circles) is masked in the fitting. See Section \ref{Importance of the innermost velocity dispersion measurement} for details, and Fig. \ref{no_central_beam_model_BH+host_dis} for the distribution of each model parameter.}
\label{no_central_beam_model_BH+host} 
\end{figure*}

\begin{figure}[htbp!]
\centering
 \includegraphics[width=1.0\columnwidth]{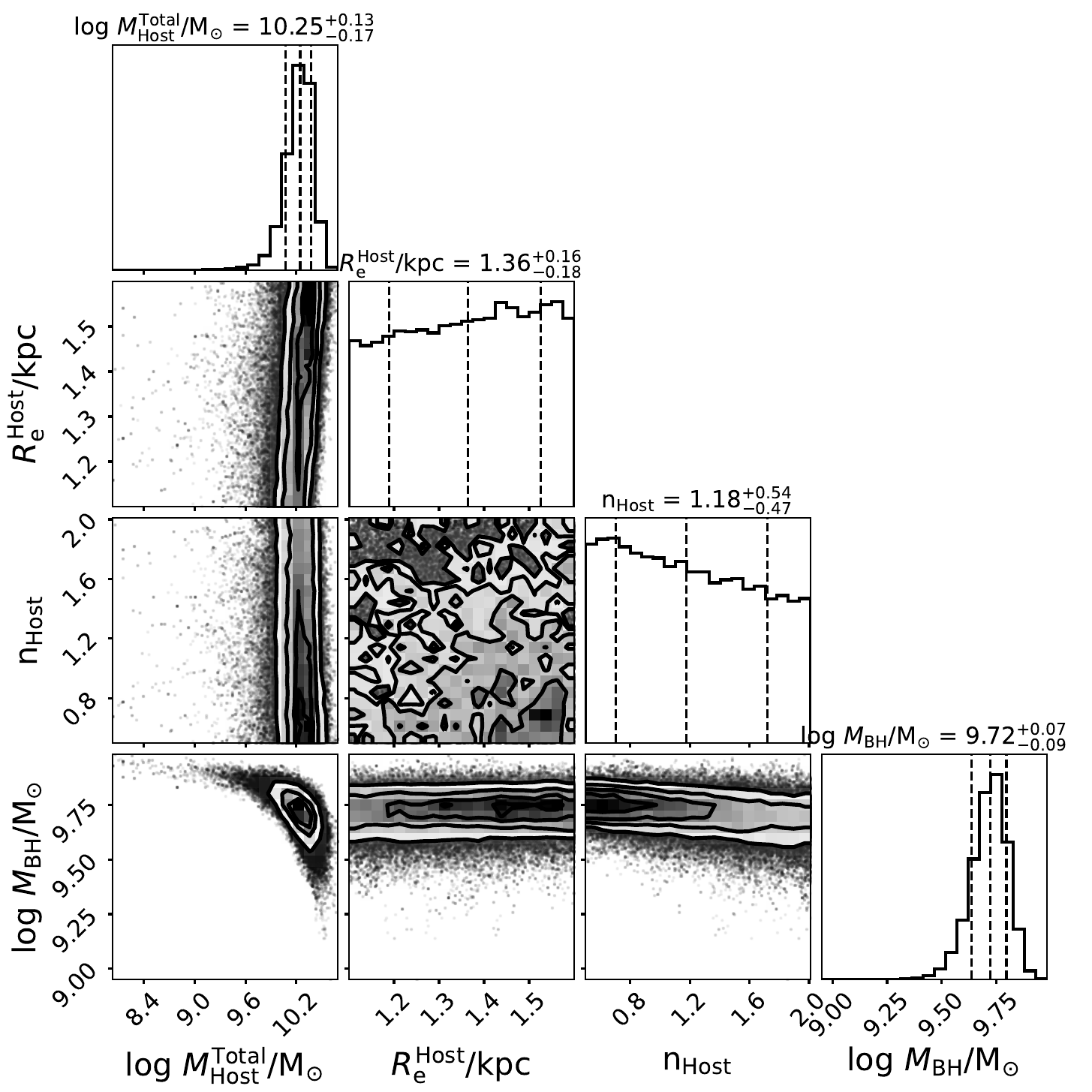}
	\caption{Same as Fig. \ref{fig: BH_host_combined_fixed_dis}, but for considering the gravity from the black hole and host
galaxy for excluding the central beam data points. See Section \ref{Importance of the innermost velocity dispersion measurement} for details.}
\label{no_central_beam_model_BH+host_dis}
\end{figure} 

\begin{figure*}
\centering
\includegraphics[width=1\textwidth]
{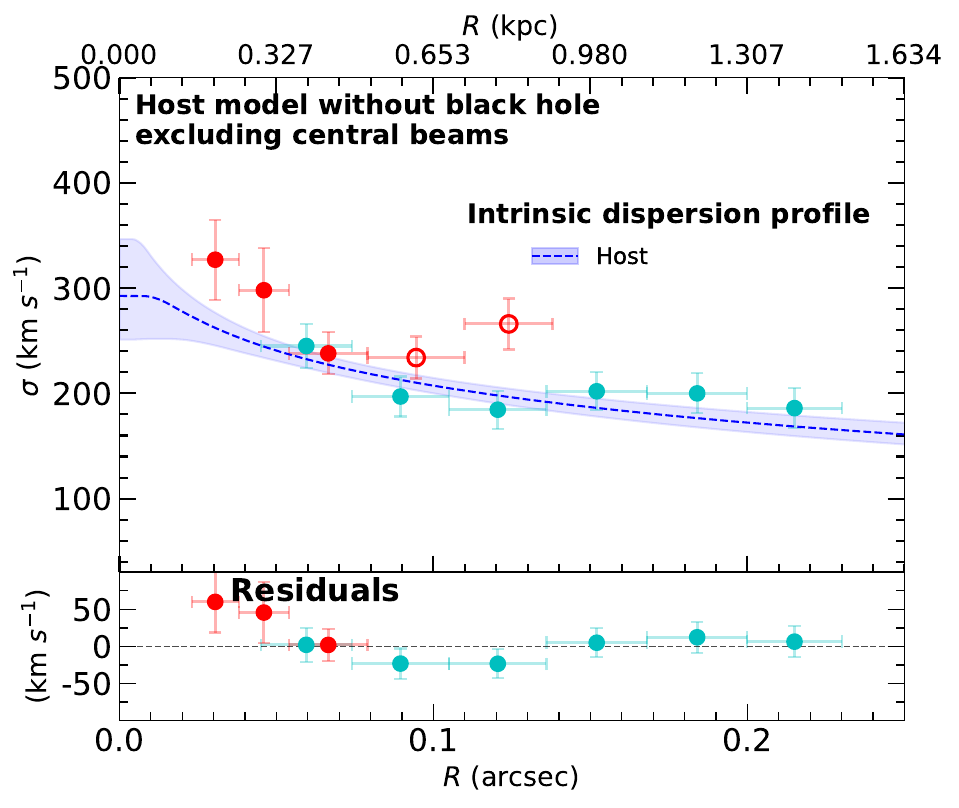}
\caption{Same as Fig. \ref{model_plot_only_cii}, but for excluding the central beam data points for the combined dispersion profile of [C~{\sc ii}] and CO (12-11), and considering the gravity only from the host galaxy with a single S\'ersic profile. The CO (12-11) outermost two data points (the red open circles) is masked in the fitting. See Section \ref{Importance of the innermost velocity dispersion measurement} for details, and Fig. \ref{no_central_beam_model_only_host_dis} for the distribution of each model parameter.}
\label{no_central_beam_model_only_host} 
\end{figure*}

\begin{figure}[htbp!]
\centering
 \includegraphics[width=1.0\columnwidth]{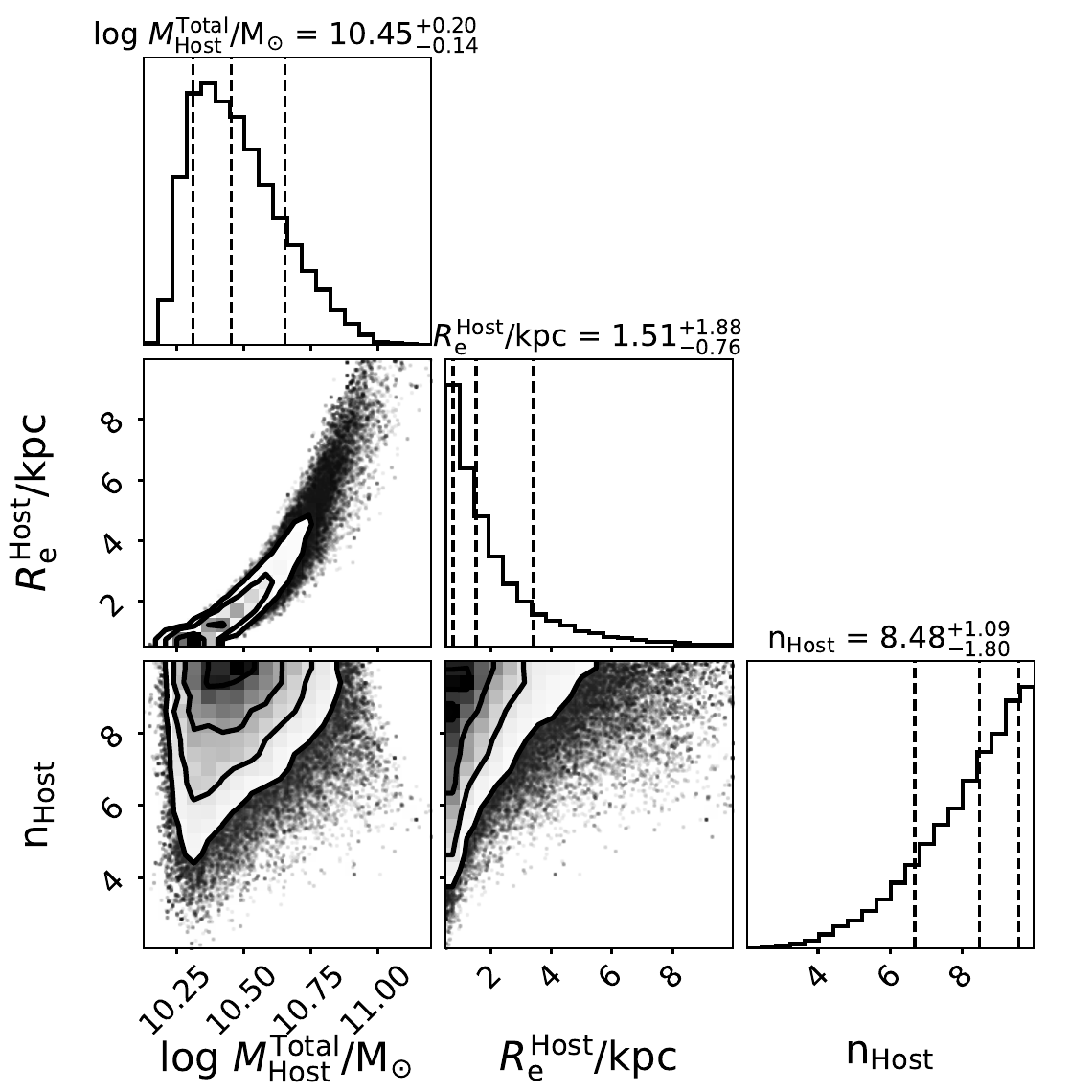}
	\caption{Same as Fig. \ref{fig: BH_host_combined_fixed_dis}, but for considering the gravity only from the host galaxy with a
single S\'ersic profile for excluding the central beam data points for the combined dispersion profile of [C~{\sc ii}] and CO (12-11). See Section \ref{Importance of the innermost velocity dispersion measurement} for details.}
\label{no_central_beam_model_only_host_dis}
\end{figure}

\begin{figure*}
\centering
\includegraphics[width=1\textwidth]
{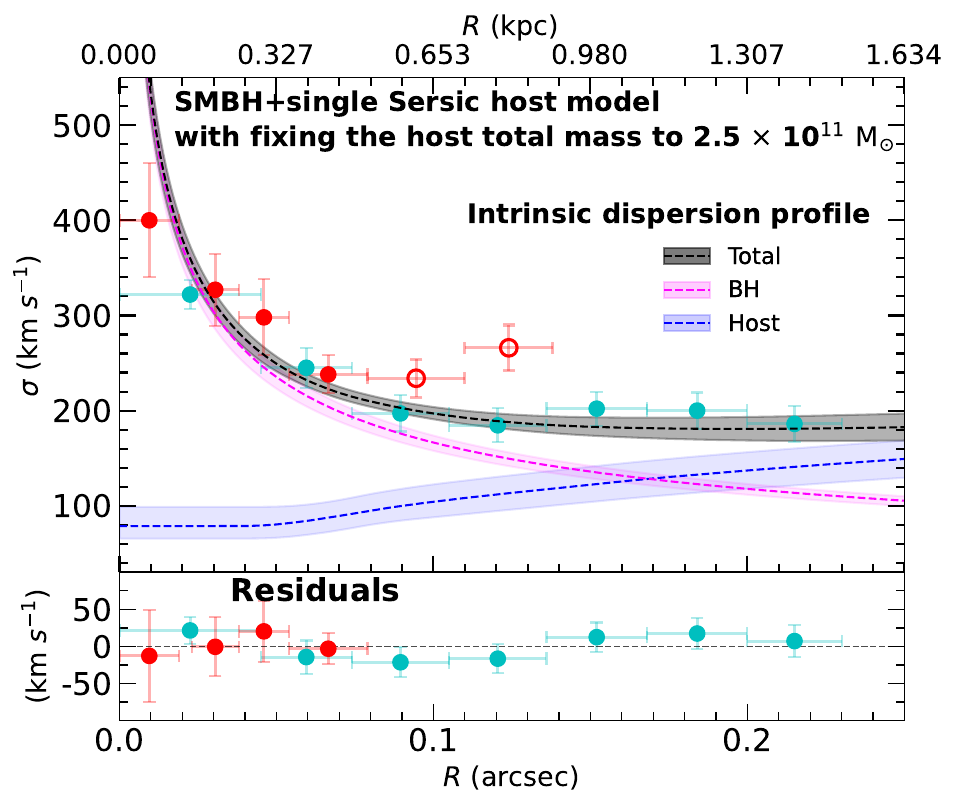}
\caption{Same as Fig. \ref{model_plot_only_cii}, but for considering the gravity from the black hole
and the host galaxy with a single S\'ersic profile and matching its total host mass to 2.5 $\times$ 10$^{11} \rm M_{\odot}$ for the combined dispersion profile of [C~{\sc ii}] and CO (12-11). The CO (12-11) outermost two data points (the red open circles) is masked in the fitting. See Section \ref{Host galaxy mass issues} for details, and Fig. \ref{fix_tanio_mass_corner} for the distribution of each model parameter.}
\label{BH_host_fix_host} 
\end{figure*}

\begin{figure*}
\centering
\includegraphics[width=1\textwidth]{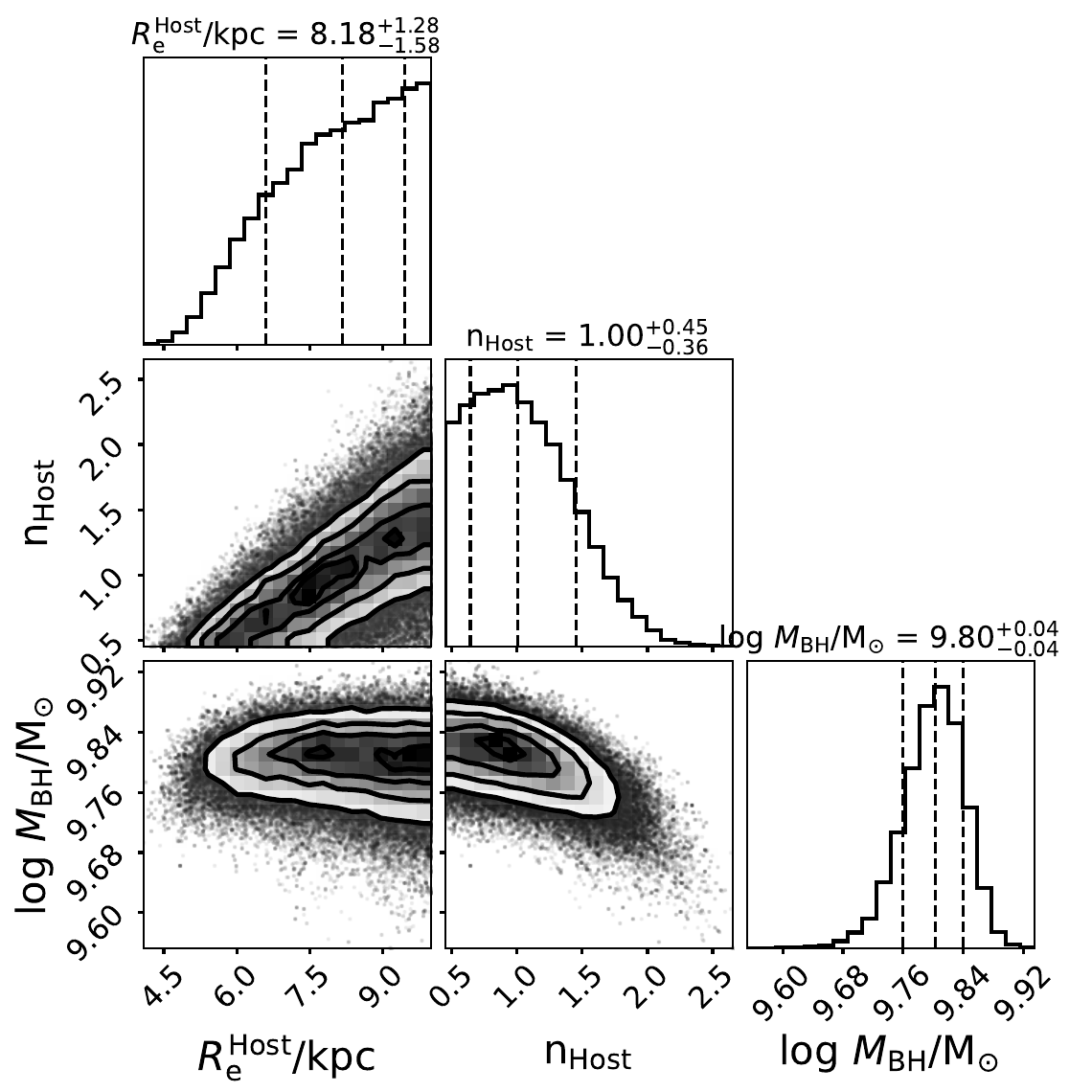}
\caption{Same as Fig. \ref{fig: BH_host_combined_fixed_dis}, but for considering the gravity from black hole plus the host galaxy with matching its total host mass to 2.5 $\times~ 10^{11}~\rm M_{\odot}$. See Section \ref{Host galaxy mass issues} for details.}
\label{fix_tanio_mass_corner} 
\end{figure*}

\begin{figure*}
\centering
\includegraphics[width=1\textwidth]
{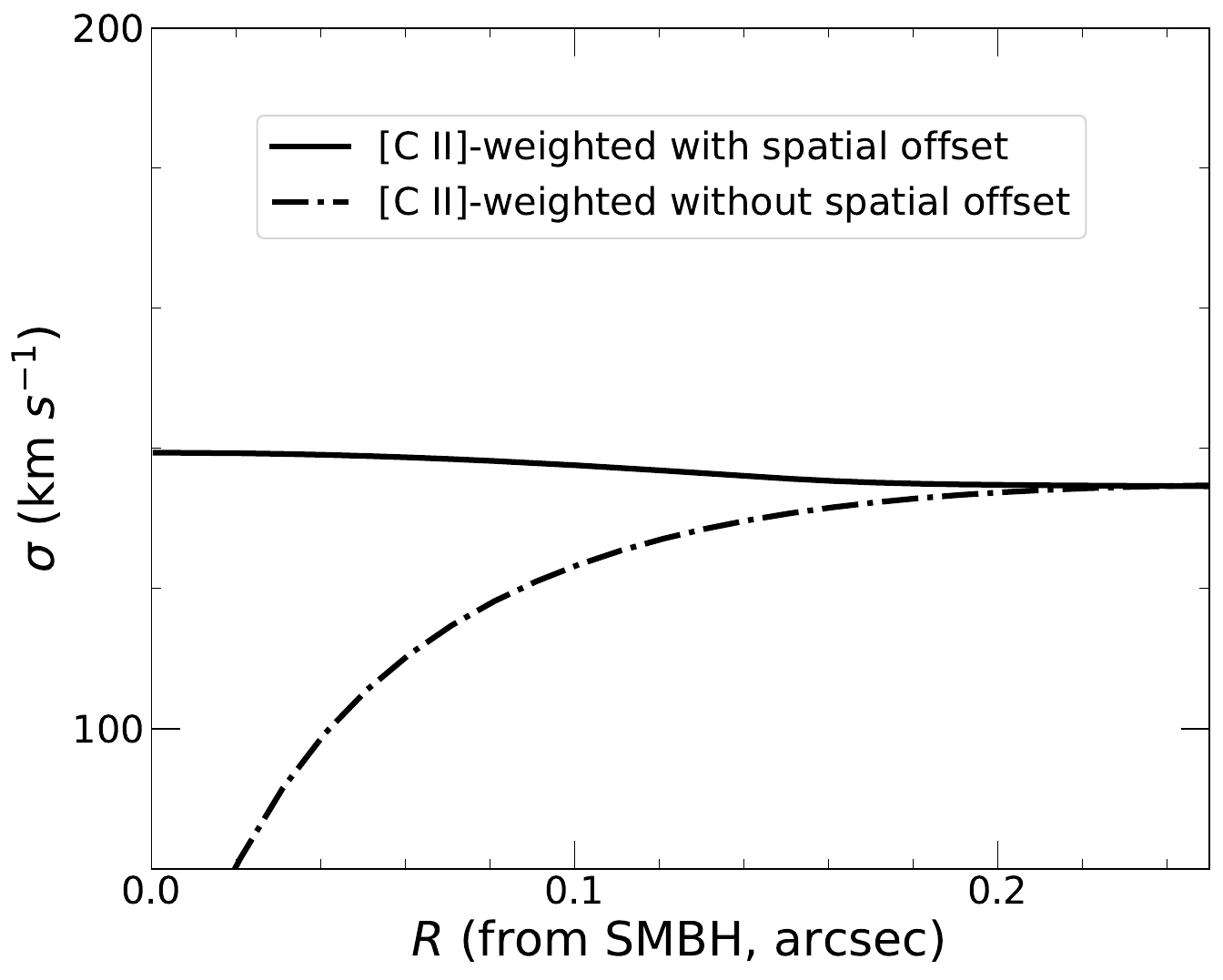}
\caption{The dispersion profile comparison between offset stellar mass and non-offset stellar mass with weighted by [C{~\sc ii}] emission and using a total stellar mass of 2 $\times$ 10$^{10}$ $\rm M_{\odot}$ with a stellar mass profile (S\'ersic profile index $n_{\rm Host}$ = 1.64 and effective radius $R_{e}^{\rm Host}$ = 1.5 kpc) derived from JWST estimates (see Sections \ref{JWST observations} and \ref{stellar mass}). Obviously, the offset-host galaxy would result in a nearly flat velocity dispersion profile, instead of the centrally decreasing profile.}
\label{offset_comparisons} 
\end{figure*}

\begin{figure*}
\centering
\includegraphics[width=0.8\textwidth]
{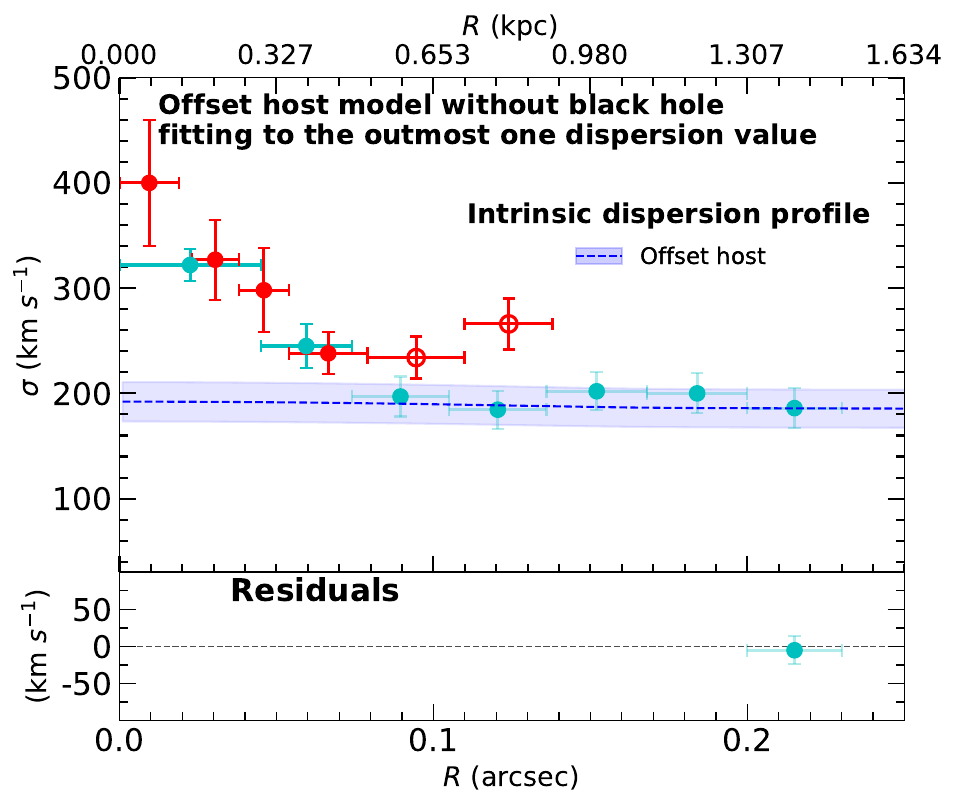}
\includegraphics[width=0.55\textwidth]{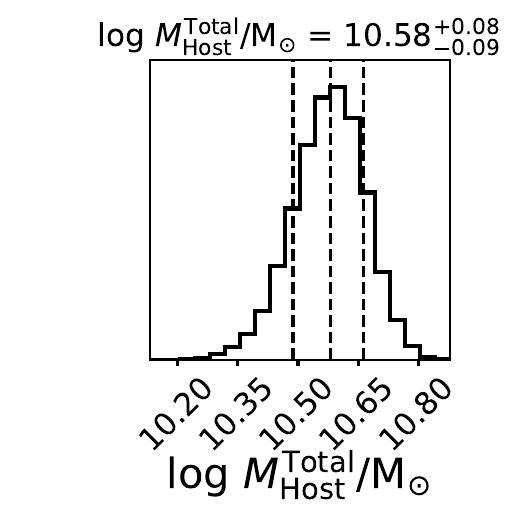}
\caption{$\it Top$: Same as Fig. \ref{model_plot_only_cii}, but for considering the gravity from host mass being offset from the SMBH position, and dominating the outermost dispersion value. Here $n_{\rm Host}$ and $R_{e}^{\rm Host}$ are fixed at 1.64 and 1.5 kpc from JWST analysis (see Section \ref{JWST observations}). $\it Bottom$: Distribution of the $M_{\rm Host}^{\rm Total}$ values. See Section \ref{Offset stellar distribution} for details.}
\label{offset_host} 
\end{figure*}

\begin{figure*} [hbp!]
\centering
\includegraphics[width=1.\columnwidth]{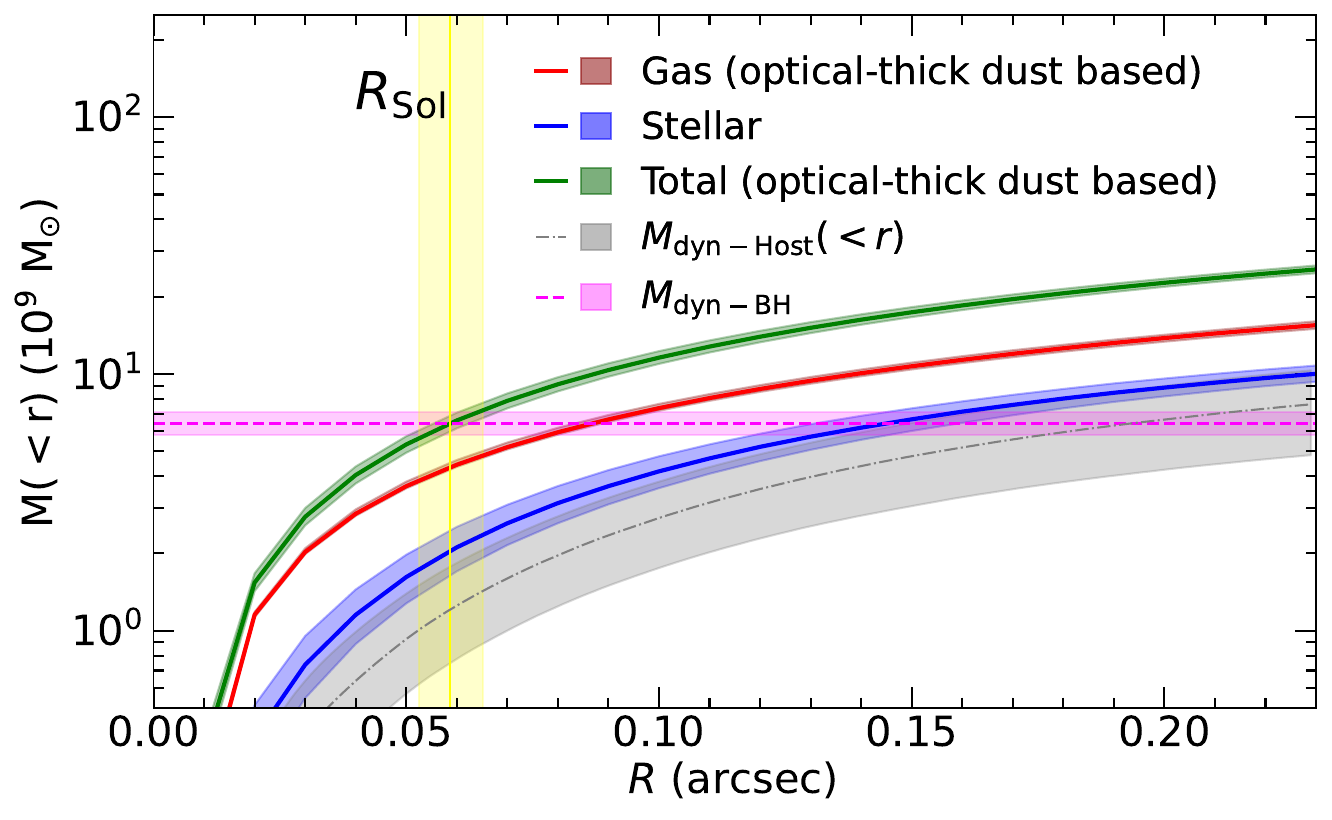}
\caption{\textbf{Enclosed mass as a function of radius for different mass contributions.} Shaded regions and lines represent the median and $1\sigma$ uncertainty intervals derived from MCMC posterior samples. Gas is dust-based gas masses inferred using optically thick (solid red) dust emission models (see Section \ref{dust mass}). Stellar is the stellar mass profile from JWST observations (solid blue, see Section \ref{stellar mass}). Total sums the dust-gas and stellar masses (green solid). $M_{\rm dyn-Host} (< r)$ (grey dash-dotted line and band) and $M_{\rm dyn-BH}$ (horizontal magenta dashed line and band) are the dynamical host galaxy and BH masses from our preferred model based on [C~{\sc ii}]+CO (12-11) kinematics (see Section \ref{joint_disperison_fitting}). $R_{\rm SoI}$ marks the SoI radius where the total enclosed mass (gas+stellar) equals the black hole mass. The horizontal axis is truncated at 0.23$''$ to focus on the inner region where the SoI lies.}
\label{mass_budget}
\end{figure*}


\begin{figure*}
\centering
\includegraphics[width=0.8\textwidth]
{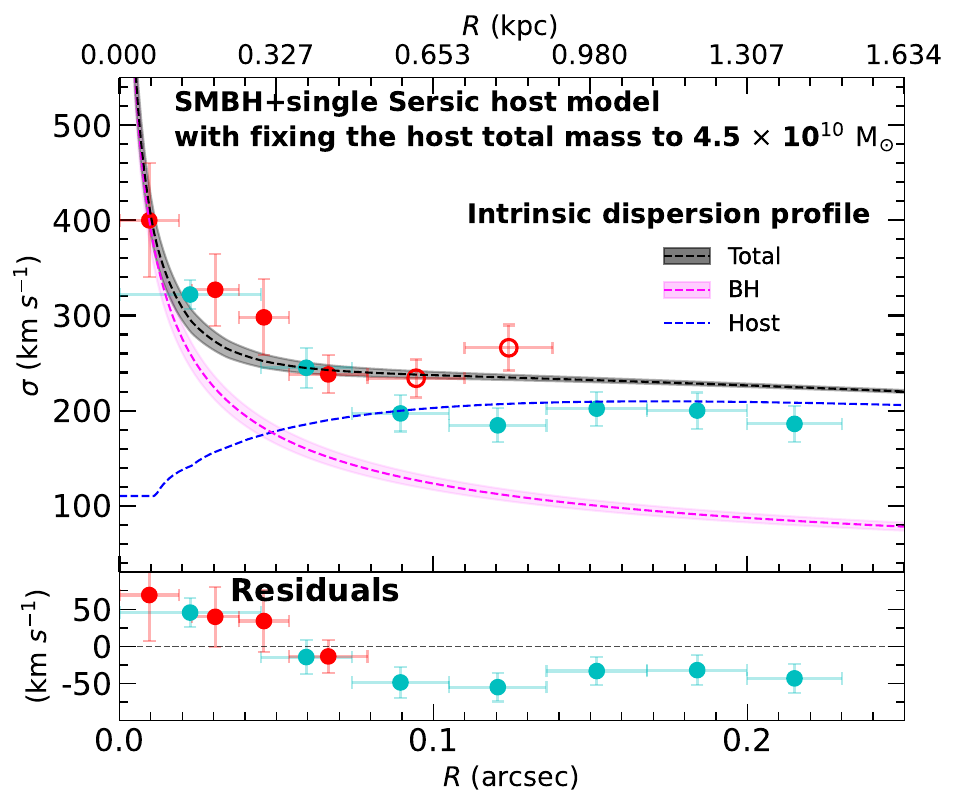}
\includegraphics[width=0.5\textwidth]{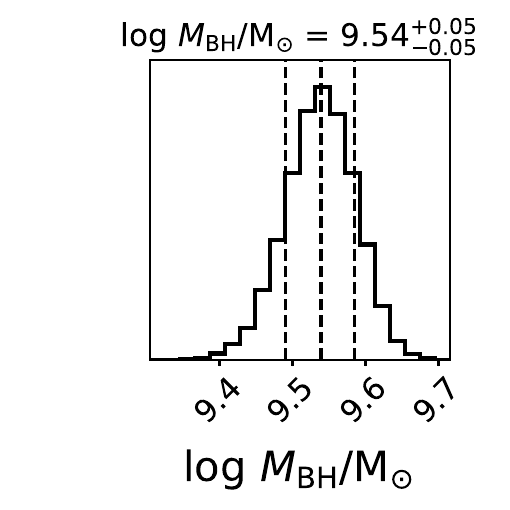}
\caption{$\it Top$: Same as Fig. \ref{model_plot_only_cii}, but for considering the gravity from the black hole and match the total host mass to 4.5 $\times$ 10$^{10}~ \rm M_{\odot}$ with $n_{\rm Host}$ and $R_{e}^{\rm Host}$ are fixed at 1.64 and 1.5 kpc from JWST analysis (see Section \ref{JWST observations}) for the combined dispersion profile of [C~{\sc ii}] and CO (12-11). The outermost two data points (shown as red open circles) of CO (12-11) is masked in the fitting. $\it Bottom$: Same as Fig. \ref{fig: BH_host_combined_fixed_dis}, but for considering
a fixed host with SMBH. See Section \ref{enclosed mass} for details.}
\label{fix_host45} 
\end{figure*}

\begin{figure*}
\centering
\includegraphics[width=1\textwidth]
{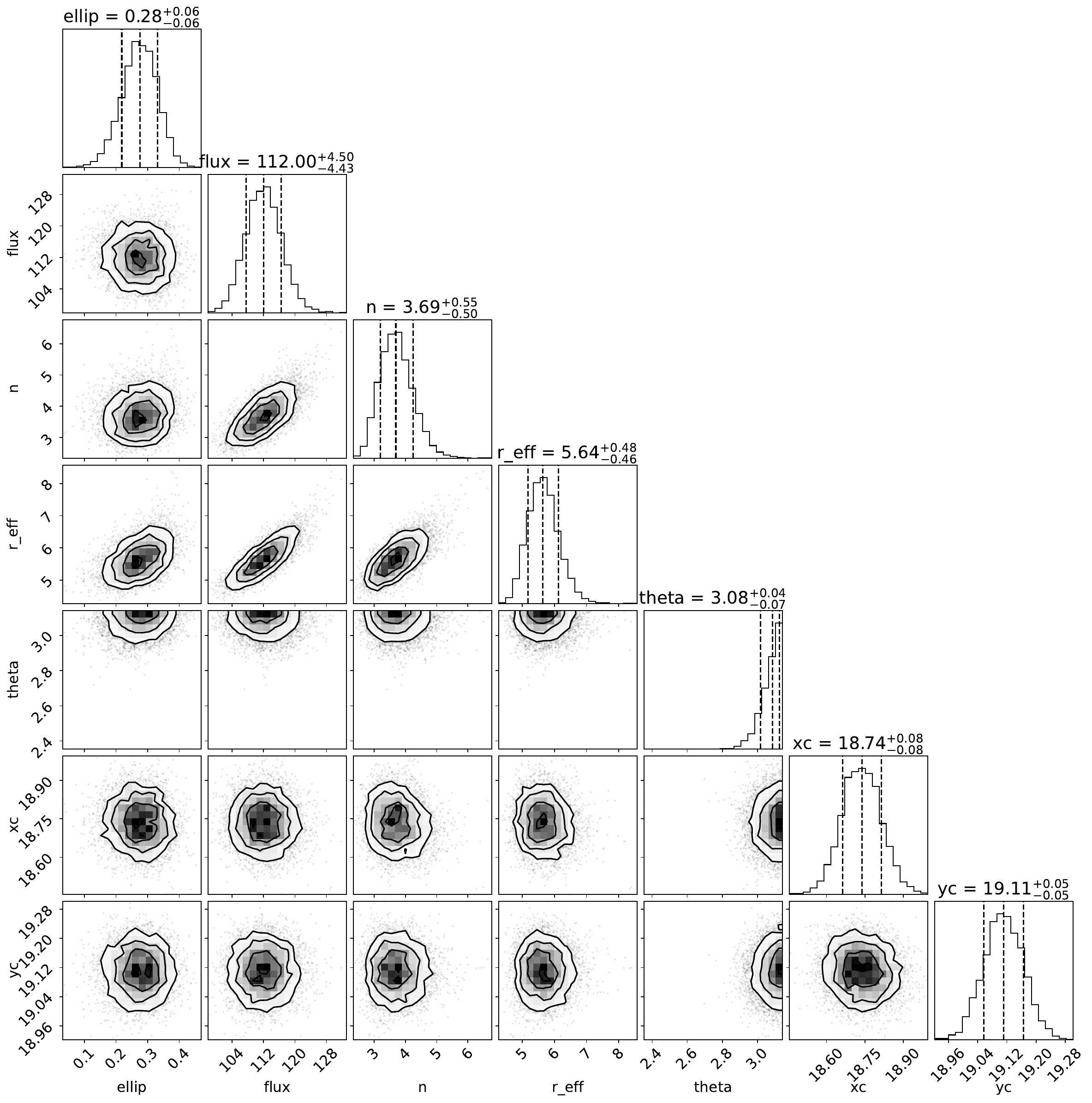}
\caption{Corner plot of the posterior distributions of the free parameters optimised in the PySerisc code fitting to the M0 map of the CO (12-11) data cube with robust = 0.5 with focusing on the central region of 0.4 $''\times$ 0.4 $''$ with 0.01$''$/pixel. Dashed lines
indicate the median value and 16th and 84th percentile range for each parameter (from top to bottom and left to right: ellip, ellipticity of the profile; flux, total flux (mJy/beam); n, S\'ersic index; r$_{\rm eff}$ (in pixel unit), the effective (half-light) radius; theta, rotation angle; xc and yc: x and y position of the center in pixel coordinate. See Section \ref{turbulence} for details.}
\label{LF_co} 
\end{figure*}

\begin{figure*}
\centering
\includegraphics[width=1\textwidth]{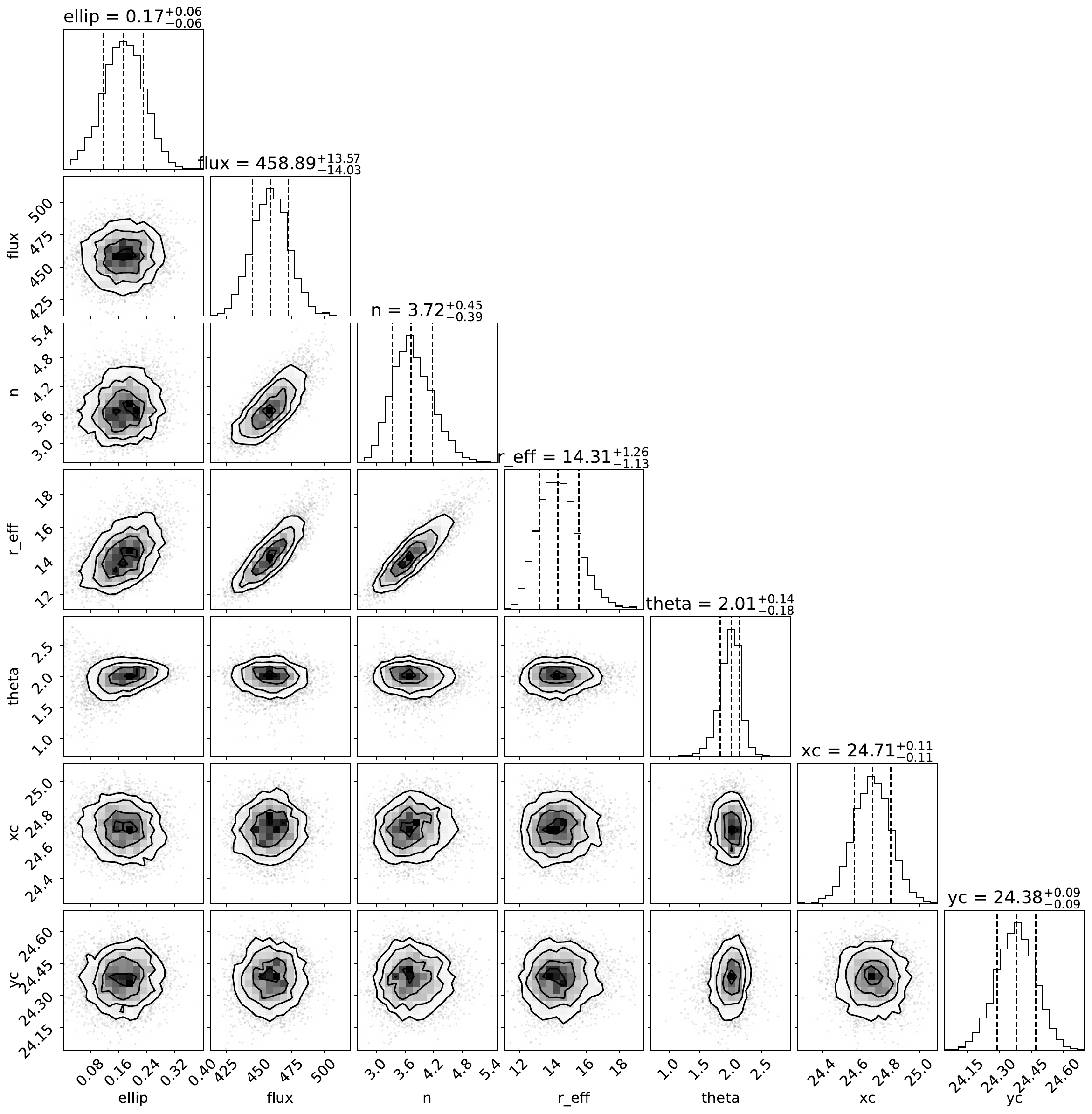}
\caption{As same Fig. \ref{LF_co}, but for the M0 map of the [C~{\sc ii}] data cube with uniform-weighting with focusing on the central region of 0.5 $''\times$ 0.5 $''$ with 0.01$''$/pixel. See Section \ref{turbulence} for details.}
\label{LF_cii} 
\end{figure*}

\clearpage 

\begin{table}[htb!]
\centering
\caption{\normalsize \textbf{Properties of the different data cubes used in this work. U and N indicate the uniform-weighting (maximize the spatial resolution at cost of the sensitivity)
and natural-weighting line data cubes (maximize the sensitivity at cost of the spatial resolution), while R0 and R0.5 indicate the intermediate CO (12-11) data cubes with robust = 0 and 0.5 weighting imaging, respectively.} \\} \label{cube_info}
\setlength{\tabcolsep}{0.1in} 
\begin{tabular}{llcc}
\hline
Data cube & Resolution (FWHM) & Geometric resolution (FWHM) & Pixel size \\
\hline
\hline
[C{~\sc ii}] & & & \\
\hline

robust = -2 (U)  & $0.089'' \times 0.053''$, PA = -82 deg & $0.069''$ & $0.010''$ \\
robust = 2 (N)  & $0.130'' \times 0.090''$, PA = -66 deg & $0.100''$  & $0.010''$ \\
\hline
CO (12-11) & & & \\
\hline
robust = -2 (U)   & $0.038'' \times 0.030''$, PA = -86 deg  & $0.034''$ & $0.005''$ \\
robust = 0  (R0)   & $0.046'' \times 0.035''$, PA = -89 deg & $0.040''$  & $0.005''$ \\
robust = 0.5 (R0.5)  & $0.072'' \times 0.048''$, PA = 84 deg & $0.058''$ & $0.010''$ \\
robust = 2 (N)    & $0.118'' \times 0.069''$, PA = -72 deg & $0.090''$  & $0.010''$ \\
\hline
\end{tabular}
\end{table}

\begin{table}[htb!]
\centering
\caption{\normalsize \textbf{The continuum-weighted average dust temperature and dust-based gas mass for each ring estimated under the optical thick and optical thin scenarios. CB stands for central beam. U and N indicate whether the continuum comes from the uniform-weighting or natural-weighting continuum maps, respectively.}\\}\label{dust_con_T_gas_mass}
\setlength{\tabcolsep}{0.26in}
\begin{tabular}{cccccc}
\hline
& & \multicolumn{2}{c}{\textit{Optically thick}} & \multicolumn{2}{c}{\textit{Optically thin}} \\

\textbf{Rings} & \it{$\mathbf{r_{\rm out}}$} & \textbf{$\mathbf{\langle T \rangle}$}   & \textbf{$\mathbf{M_{\rm gas}}$} & \textbf{$\mathbf{\langle T \rangle}$}  & \textbf{$\mathbf{M_{\rm gas}}$}    \\

& ($^{\prime\prime}$) & (K)  & (10$^{9}$ $\rm M_{\odot}$) &  (K) & (10$^{9}$ $\rm M_{\odot}$)  \\
\hline
\hline
CB (U) & 0.045 & 251 & 2.06 & 349&  1.35  \\

1 (U) & 0.074 & 114   & 1.89& 131 &1.53   \\
\hline
1 (N) & 0.11& 91 & 2.15 & 99 &  1.84  \\

2 (N) & 0.14& 80 & 1.97 &85   & 1.75 \\

3 (N) & 0.17& 72 & 1.89&76 &  1.73   \\

4 (N) & 0.20& 66 &1.82 &69 & 1.71   \\

5 (N) &0.23& 62 &1.82 & 64  & 1.75   \\
\hline
\end{tabular}
\end{table}

\begin{table}[htb!]
\centering
\caption{\normalsize \textbf{Best-fit spectral parameters for each elliptical ring of [C~{\sc ii}]. The outer radius is along the major-axis for each ring. U and N indicate the uniform-weighting and natural-weighting [C~{\sc ii}] line data cubes, respectively. CB stands for central beam. Numbers in parentheses indicate the uncertainties.}\\}\label{cii_fitting_table}
\setlength{\tabcolsep}{0.06in}
\begin{tabular}{cccccccccc}
\hline
\textbf{Rings} & \textbf{$\mathbf{r_{\rm out}}$} & \textbf{$\mathbf{\sigma_{\rm host}}$}  & \multicolumn{2}{c}{\textbf{$\mathbf{V_{\rm red}}$}}  & \textbf{$\mathbf{V_{\rm blue}}$} & \multicolumn{2}{c}{\textbf{$\mathbf{\sigma_{\rm red}}$}}  & $\mathbf{\sigma_{\rm blue}}$ & \textbf{$\mathbf{\chi^{2}_{\rm d.o.f.}}$}  \\

 &  &  &  $V_{\rm red1}$ & $V_{\rm red2}$ &  & $\sigma_{\rm red1}$ & $\sigma_{\rm red2}$&  &  \\

& ($^{\prime\prime}$) & ($\rm km~s^{-1}$)  & ($\rm km~s^{-1}$) & ($\rm km~s^{-1}$)  & ($\rm km~s^{-1}$)& ($\rm km~s^{-1}$) & ($\rm km~s^{-1}$) & ($\rm km~s^{-1}$) &   \\
\hline
\hline
CB (U) & 0.045$^{\prime\prime}$& 322 (15) & -- & --   & --  & --   & -- & -- & 1.37  \\

1 (U) & 0.074$^{\prime\prime}$ & 245 (21)   &201 (10) & --   & -- &  326 (40)  & --   &--  &0.89    \\
\hline
1 (N) & 0.11$^{\prime\prime}$ & 197 (22)  &346 (88) & 562 (26) & --337 (32)  & 246 (42)&90 (24)  & 196 (28)  & 0.62  \\

2 (N) & 0.14$^{\prime\prime}$ & 185 (20)   &360 (21)&586 (18)  & --339 (18)  & 235 (36)&137 (18)  & 179 (19)  & 0.98  \\

3 (N) & 0.17$^{\prime\prime}$ & 202 (18)  &366 (22) &609 (5)  & --374 (20)  & 40 (45)& 183 (50)  & 154 (16)  & 1.16  \\

4 (N) & 0.20$^{\prime\prime}$ & 200 (18)   &360 (22)& 604 (3)  & --378 (16)  & 40 (20)&137 (50)  & 128 (17) & 1.20  \\

5 (N) & 0.23$^{\prime\prime}$& 186 (20)   & 354 (40) & 585 (32)  & --363 (24)  & 40 (10)&100 (26)   & 126 (20)  & 0.89   \\
\hline
\end{tabular}
\end{table}

\begin{table}[htb!]
\centering
\caption{\normalsize \textbf{Best-fit spectral parameters for each elliptical ring of CO (12-11). The outer radius is along the major-axis for each ring. U and N indicate the uniform-weighting and natural-weighting CO (12-11) line data cubes, respectively. R0 and R0.5 indicate the intermediate CO (12-11) data cubes with robust = 0 and 0.5 weighting, respectively.
CB stands for central beam. Numbers in parentheses indicate the uncertainties.}\\}\label{co_ring_fitting_table}
\setlength{\tabcolsep}{0.06in}
\begin{tabular}{cccccc}
\hline
\textbf{Rings} & \textbf{$\mathbf{r_{\rm out}}$} & \textbf{$\mathbf{\sigma_{\rm host}}$} & \textbf{$\mathbf{V_{\rm blue}}$} & \textbf{$\mathbf{\sigma_{\rm blue}}$} & \textbf{$\mathbf{\chi^{2}_{\rm d.o.f.}}$} \\
& ($^{\prime\prime}$) & ($\rm km~s^{-1}$) & ($\rm km~s^{-1}$) & ($\rm km~s^{-1}$) & \\
\hline
\hline
CB (U) & 0.019$^{\prime\prime}$ & 400 (60) & -- & -- & 0.99 \\
\hline
1 (R0) & 0.038$^{\prime\prime}$ & 327 (38) & -535 (110) & 75 (45) & 0.76 \\
2 (R0) & 0.054$^{\prime\prime}$ & 298 (40) & -530 (80) & 55 (35) & 0.67 \\
\hline
1 (R0.5) & 0.079$^{\prime\prime}$ & 238 (20) & -544 (94) & 112 (32) & 1.26 \\
\hline
1 (N) & 0.110$^{\prime\prime}$ & 234 (20) & -534 (75) & 150 (30) & 0.98 \\
2 (N) & 0.138$^{\prime\prime}$ & 266 (24) & -533 (70) & 145 (40) & 0.84 \\
\hline
\end{tabular}
\end{table}

\begin{table}[ht!]
	\centering
	\caption{\normalsize \textbf{Best-fit parameters derived from the dynamical modeling fitting to the observed dispersion data points from  [C~{\sc ii}] and CO (12-11) (unless otherwise specified, only central four CO (12-11) data points are used, see details in Section \ref{spectra_fitting}), for different adopted models in this work. For the combined dispersion profile,
    $^{\star}$ indicates the $\chi^2$/$\chi^2_{d.o.f}$ calculated only for the central eight data points, while $^{a}$ ($^{b}$) indicates the corresponding model only fits to the most inner (outer) data point. $^{\#}$ means the host mass profile follows S\'ersic profile index $n_{\rm Host}$ = 1.64 and effective radius $R_{e}^{\rm Host}$ = 1.5 kpc derived from JWST estimates (see Section \ref{JWST observations}).}\\}\label{All_models_fitting}
\setlength{\tabcolsep}{0.06in}
\begin{tabular}{lcccccc} 
\hline
\textbf{Models} & $\mathbf{M_{\rm BH}}$  & $\mathbf{M_{\rm Nuclear-Host}^{\rm Total}}$  & $\mathbf{M_{\rm Host}^{\rm Total}}$ & $\mathbf{\chi^2}$ & $\mathbf{\chi^2_{d.o.f}}$ & Discussion \\
& (10$^{9}~ \rm M_{\odot}$) & (10$^{10}~ \rm M_{\odot}$) & (10$^{10}~ \rm M_{\odot}$)& & & Section \\
\hline
\hline
\textbf{SMBH+host} & &  &  & & &  \\
Preferred model & $6.5^{+0.6}_{-0.7}$ & -- &  $1.3^{+0.6}_{-0.5}$ & 7.63& 1.09 & Main text, \ref{joint_disperison_fitting}\\
Only [C~{\sc ii}] 7 data points & $6.6^{+0.8}_{-0.7}$ & -- & $1.2\pm0.6$ & 7.20 & 2.40 & Main text, \ref{cii_disperison_fitting}\\
Fixed $M_{\rm Host}^{\rm Total} = 2.5 \times 10^{11} \rm M_{\odot}$
 & $6.3 \pm 0.6$ & -- & -- &6.39 & 0.80 & \ref{Host galaxy mass issues}\\
\hline
\textbf{SMBH-only} &   &  &  &  & & \\
All points & $7.8 \pm 0.4$  & -- &  --&  18.56 & 1.86 & \ref{joint_disperison_fitting}\\
 &  &  &  & (4.54)$^{\star}$ & (0.65)$^{\star}$ &\\

Only CO 4 data points  &  $6.9\pm0.8$ & -- &-- & 0.44 & 0.15& Main text, \ref{co_disperison_fitting}\\
 Central point$^{a}$  & $5.9^{+1.9}_{-1.7}$  & -- &-- & -- & --& \ref{Offset stellar distribution}\\
 Fixed $M_{\rm Host}^{\rm Total} = 4.5 \times 10^{10} \rm M_{\odot}$$^{\#}$
 & $3.5\pm0.4$ & -- & -- &40.44 & 4.04 & \ref{enclosed mass}\\
\hline
\textbf{Host-only} & &  &  & & &\\
Single S\'ersic & --  & -- & $2.1^{+0.4}_{-0.2}$ & 16.08  & 2.01 & \ref{Host-only Model}\\
Double S\'ersic & -- & $0.8^{+0.2}_{-0.2}$ & $0.7^{+0.7}_{-0.5}$ & 10.72 & 2.14 & \ref{Host-only Model}\\
NSC & -- & $0.3 \pm 0.1$ & $0.3^{+0.6}_{-0.2}$ & 10.42 & 1.74 & \ref{Nuclear star cluster}\\
Offset-host$^{b}$  & -- & -- & $3.8^{+0.8}_{-0.7}$ & -- & --& \ref{Offset stellar distribution}\\
\hline
\end{tabular}
\end{table}

\begin{table}[htb!]
	\centering
	\caption{\normalsize \textbf{Best-fit parameters derived from the dynamical modeling by excluding the central dispersion data points (9 data points in total), for different adopted models.}\\}\label{All_models_fitting_no_central_beam}
\setlength{\tabcolsep}{0.13in}
\begin{tabular}{lcccccc} 
\hline
\textbf{Models} & $\mathbf{M_{\rm BH}}$  & $\mathbf{M_{\rm Nuclear-Host}^{\rm Total}}$  & $\mathbf{M_{\rm Host}^{\rm Total}}$ & $\mathbf{\chi^2}$ & $\mathbf{\chi^2_{d.o.f}}$ & Discussion\\

& (10$^{9}~ \rm M_{\odot}$) &  (10$^{10}~ \rm M_{\odot}$) & (10$^{10}~ \rm M_{\odot}$)&   & & Section \\
\hline
\hline
\textbf{SMBH+host} & &  &  & &  \\
Preferred model
 & $5.2^{+1.0}_{-0.9}$ & -- & $1.8\pm0.6$ &  4.38 & 0.88 & \ref{Importance of the innermost velocity dispersion measurement} \\
\hline
\textbf{Host-only} & &  &  & &  \\
Single S\'ersic & -- & --  & $2.8^{+1.6}_{-0.8}$ &  7.60& 1.27 & \ref{Importance of the innermost velocity dispersion measurement}\\
\hline
\end{tabular}
\end{table}

\begin{table}[htb!]
	\centering
	\caption{\normalsize \textbf{Summarizing the priors and free parameters used for our preferred and alternative models in this work for the combined dispersion profile of [C~{\sc ii}] and CO (12-11), as well as the reasons that alternative models cannot provide the satisfactories than the SMBH interpretation for the observed central velocity dispersion
rise shown in Fig. \ref{fig: dispersion_profile}. $^{a}$ stands for only nine data points are used after excluding the two central beam data points.}\\}\label{All_models_summary}
\setlength{\tabcolsep}{0.18in}
\begin{tabular}{lccc} 
\hline
\textbf{Models} & \multicolumn{2}{c}{\textbf{$\mathbf{Parameters}$}} & \textbf{Rejected reasons} \\
 &\textbf{Informative priors} &\textbf{Non-informative priors} & \textbf{Section} \\
\hline
\hline
\textbf{SMBH+host} &  &  &   \\
Preferred model& $n_{\rm Host}$, $R_{\rm e}^{\rm Host}$  & $\rm M_{\rm BH}$, $\rm M_{\rm Host}^{\rm Total}$&\\
\hline
\textbf{Host-only} &  & &\\
Single S\'ersic & --&  $\rm M_{\rm Host}^{\rm Total}$, $n_{\rm Host}$,  $R_{\rm e}^{\rm Host}$ & (1), \ref{Host-only Model}\\
Double S\'ersic & $n_{\rm Host}$, $R_{\rm e}^{\rm Host}$ & $\rm M^{\rm Total}_{\rm Nuclear-Host}$, $n_{\rm Nuclear}$, $R_{\rm e}^{\rm Nulcear}$; $\rm M_{\rm Host}^{\rm Total}$ & (2), \ref{Host-only Model} \\
NSC & $n_{\rm Host}$, $R_{\rm e}^{\rm Host}$ & $\rm M_{\rm Nuclear-Star}^{\rm Total}$, $R_{\rm c}^{\rm Nulcear-Star}$; $\rm M_{\rm Host}^{\rm Total}$ &  (3), \ref{Nuclear star cluster}\\
\hline
\hline
\textbf{SMBH+host}$^{a}$ &  &  &   \\
Preferred model& $n_{\rm Host}$, $R_{\rm e}^{\rm Host}$ & $\rm M_{\rm BH}$, $\rm M_{\rm Host}^{\rm Total}$ &\\
\hline
\textbf{Host-only}$^{a}$ &  & &\\
Single S\'ersic & --& $\rm M_{\rm Host}^{\rm Total}$, $n_{\rm Host}$, $R_{\rm e}^{\rm Host}$& (4), \ref{Importance of the innermost velocity dispersion measurement}\\
\hline      
\end{tabular}
\normalsize{
\textbf{Notes:} (1) Worse fitting and an unrealistically compact host.; (2) Worse fitting and an unrealistically compact nuclear host; (3) Worse fitting and an unrealistically massive nuclear star cluster; (4)  Worse fitting and an unrealistically compact host.}
\end{table}

\clearpage

\clearpage

\end{document}